\documentclass[%
 reprint,
superscriptaddress,
groupedaddress,
 amsmath,amssymb,
 aps,
]{revtex4-2}

\usepackage{graphicx}% Include figure files
\usepackage{dcolumn}% Align table columns on decimal point
\usepackage{bm}% bold math

\newcommand{\rulesep}{\unskip\ \hrule\ } % SBK
\usepackage{booktabs, multirow} % SBK

\usepackage{color}
\usepackage{float}
\usepackage[letterpaper, portrait, margin=0.8in]{geometry}
\usepackage[utf8]{inputenc}
\usepackage{graphicx}
\usepackage{subcaption}
\usepackage{bibunits}
\usepackage{bigints}
\graphicspath{{Images_new/}}
\usepackage{placeins}
\usepackage{afterpage}

\usepackage[pdftex,breaklinks,colorlinks,
citecolor=blue,
urlcolor=blue,
pdfsubject={LaTeX}]{hyperref}
\begin{document}

% \preprint{APS/123-QED}

% \title{Interphase Boundary Segregation in Quaternary alloys: A Generalized Multicomponent Phase-field Model for Chemical Effects}

\title{Phase Boundary Segregation in Multicomponent Alloys: A Diffuse-Interface Thermodynamic Model}
% \title{Thermodynamics of Phase Boundary Segregation in Quaternary Alloys: A Diffuse Interface Model}

\author{Sourabh B Kadambi}
% \affiliation{%
% Department of Materials Science and Engineering, North Carolina State University, Raleigh, North Carolina 27695, USA
% }%
% current address
\altaffiliation{Current address: Computational Mechanics and Materials Department, Idaho National Laboratory, Idaho Falls, Idaho 83415, USA}
\email{sourabhbhagwan.kadambi@inl.gov}

% \author{Fadi Abdeljawad}
% \affiliation{%
% Department of Mechanical Engineering, Department of Materials Science and Engineering, Clemson University, Clemson, South Carolina 29634, USA
% } 

\author{Srikanth Patala}%
 \email{srikanth.patala@gmail.com}
\affiliation{%
Department of Materials Science and Engineering, North Carolina State University, Raleigh, North Carolina 27695, USA \hspace{0.1cm}
}%

% \cortext[cor1]{Corresponding author}
% \author[NCSU]{Srikanth Patala\corref{cor1}}
% \ead{spatala@ncsu.edu}

\begin{abstract}
Microalloying elements tend to segregate to the matrix-precipitate phase boundaries to reduce the interfacial energy. The segregation mechanism is emerging as a novel design strategy for developing precipitation-hardened alloys with significantly improved coarsening resistance for high temperature applications. In this paper, we report a nanoscopic diffuse-interface thermodynamic model that describes multicomponent segregation behavior in two-phase substitutional alloys. Following classical approaches for grain boundaries, we employ the regular solution thermodynamics to establish segregation isotherms. We show that the model recovers the Guttmann multicomponent isotherm describing local interfacial concentrations, and the generalized Gibbs adsorption isotherm that governs the total solute excess and interfacial energy. A variety of multicomponent segregation behaviors are demonstrated for a model two-phase quaternary alloy. The nature of interfacial parameters and the resulting analytic solutions make the model amenable for parameterization and comparison with atomistic calculations and experimental characterizations.

\begin{description}
\item[Keywords] Solute Segregation, Phase Boundary, Phase-field Model, Gibbs Adsorption,\\ Multicomponent Thermodynamics
\end{description}
\end{abstract}

% \pacs{Valid PACS appear here}% PACS, the Physics and Astronomy
\maketitle

%\tableofcontents

% \begin{bibunit}[apsrev4-1]
\section{Introduction} \label{sec:intro}

Development of structural alloys with high-temperature stability is crucial for automotive, aerospace and nuclear industries. While superior mechanical properties are conventionally realized by two-phase microstructures--via a high density of finely-sized secondary precipitates in the matrix--their properties deteriorate at elevated temperatures due to coarsening. This limits the operating temperatures of, for example,  $\theta^\prime$-Al$_2$Cu-strengthened Al-Cu alloys to $~250^\circ$C \citep{roy2017comparative}, Mg$_2$Sn-strengthened Mg-Sn alloys to $~170^\circ$C \citep{mendis2006enhanced} and the Al$_3$Sc-strengthened Al-Sc alloys to $~300^\circ$C \citep{seidman2002precipitation}. Therefore, alloy design approaches to enhance the coarsening resistance of two-phase microstructures are crucial.
Ni and Co-based alloys/superalloys, which inherently possess high-temperature strength for applications in gas turbine engines, also require further improvements in thermal stability for realizing energy-efficient operation \citep{pollock2006nickel,bauer2012creep}. More examples of thermal stability requirements can be found in emerging multi-principle-element alloys containing multiple phases \citep{zhao2018thermal}. 
% {Therefore, novel alloy design strategies are sought to enhance the thermal stability of precipitation micro-structures.}
 
 \begin{figure*}[!htp]
\captionsetup[subfigure]{justification=centering} % {labelformat=empty}
\centering
% \hspace{0.5cm}
\begin{subfigure}{0.7\textwidth}
  \centering
  \includegraphics[width=1\linewidth]{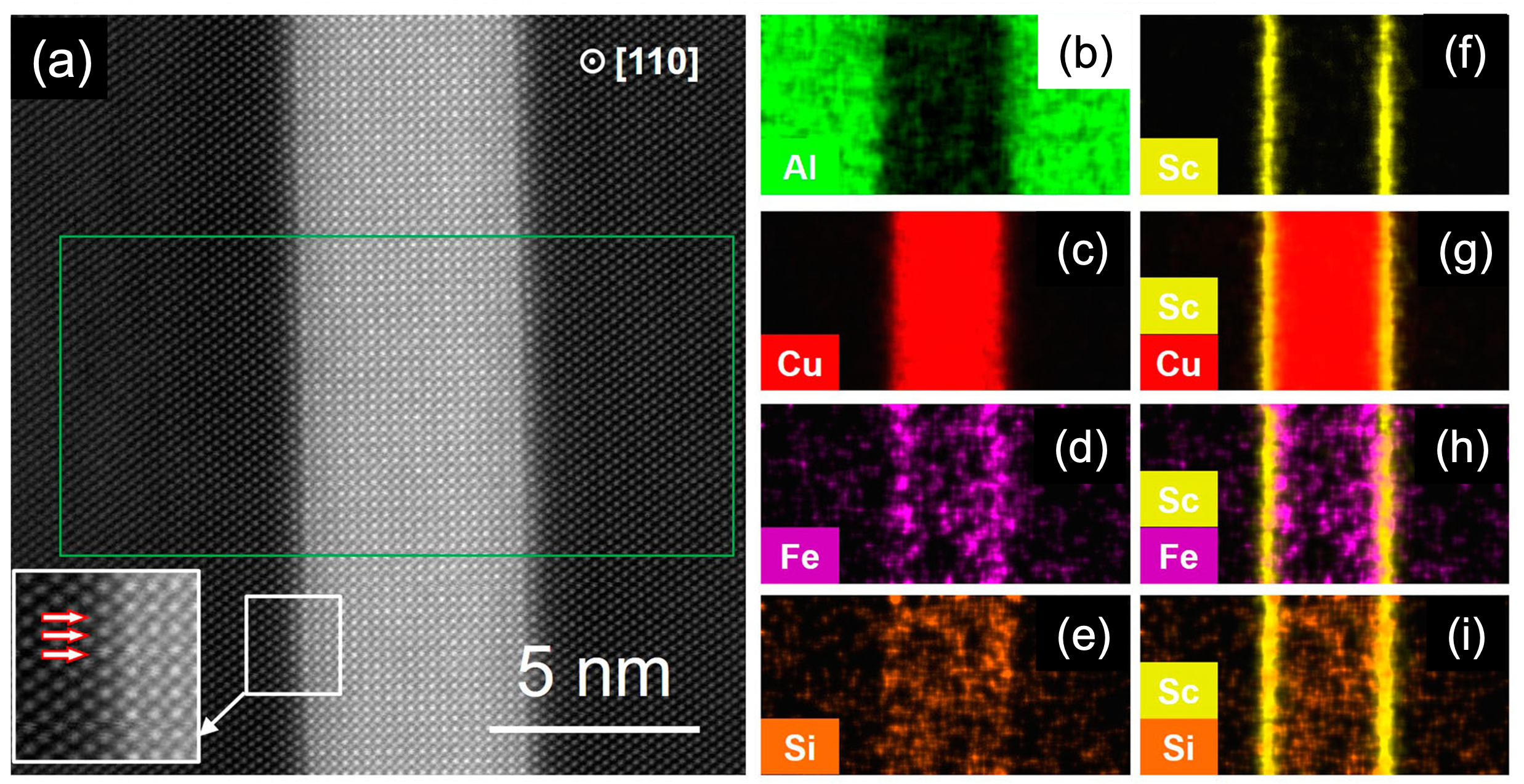}
%   \caption*{}
\end{subfigure}
\caption{Atomc-scale details of a representative $\theta'-$Al$_2$Cu/matrix interface in an (Al-Cu)-Sc-Fe-Si alloy at 300$^\circ$C. (a) HAADF-STEM image. (b-i) EDS maps of elemental distribution demonstrating simultaneous segregation of Sc and Fe at the interface. This alloy was found to be thermodynamically and kinetically stable relative to the base Al-Cu and (Al-Cu)-Sc alloys. (Figure is reproduced with permission from \citep{gao2020segregation}.)}
\label{fig:exp_seg}
\end{figure*}

Approaches to enhance thermal stability of multiphase microstructures can be identifying from the classical theory of precipitate coarsening.
The rate of coarsening primarily depends on two factors: the excess interfacial energy $\gamma$ of the matrix-precipitate interphase boundary (IB); and the diffusion kinetics of elements across the microstructure \citep{ratke2013growth}. Any modification to the alloy composition is likely to alter both the energetic and kinetic factors.
% Thereby, we c the coarsening kinetics can be controlled by modifying the alloy composition: adding selectively chosen microalloying elements. 
% Thereby, we can infer two main approaches to control coarsening.
The kinetics-based approach typically involves the addition of a slow-diffusing solute \citep{clouet2006complex}, which limits the diffusion controlled kinetics, but is generally effective only at low homologous temperatures. On the other hand, the thermodynamics-based approach involves the addition of solutes that cause a reduction in $\gamma$ by preferentially segregating to the IB \citep{marquis2005coarsening}. This approach can be considered to be more effective as it fundamentally alters the interfacial chemistry and energetics.
% \textcolor{red}{Recent literature suggests that the thermodynamic segregation mechanism is emerging as a novel consideration to design stable precipitate microstructures \textcolor{red}{Ref.}}

Thermodynamic segregation is classically described by the generalized Gibbs adsorption isotherm \citep{cahn1979interfacial}, which describes the variation in $\gamma$ resulting from changes to the solute addition within the bulk phase. For an IB \citep{kaplan2013review} in a multicomponent, two-phase alloy at equilibrium, $d\gamma = - \sum_\xi \Gamma^{(1,2)}_\xi d\mu_\xi $ ($\xi\neq1,2$), where $\mu_\xi$ is the solute chemical potential. $\Gamma^{(1,2)}_\xi$ is the relative solute excess which quantifies IB segregation of a solute $\xi = 3,4,\ldots$ with reference to the binary base components $(1,2)$. The isotherm reveals that a solute $\xi$, which preferentially segregates (i.e. $\Gamma^{(1,2)}_\xi>0$) to the IB, will reduce $\gamma$. Furthermore, multiple solutes can co-segregate to the IB to cause a reduction in $\gamma$. Therefore, the thermodynamic segregation mechanism offers a number of compositional degrees of freedom in multicomponent alloys to tune the energetics of IBs in binary base alloys \citep{frolov2015phases}.

% {Moreover, the solute components $\xi$ allow independent compositional degrees of freedom in tuning the chemistry and energetics of the IB \citep{frolov2015phases}. Therefore, alloys designed using the thermodynamic IB solute segregation mechanism will possess enhanced coarsening resistance due to lower $\gamma$.}
% ( via its chemical potential )

A number of experimental studies have demonstrated the practical feasibility and potential of the IB segregation mechanism. Precipitation hardening alloys microalloyed with IB-segregating solutes have exhibited enhanced age-hardening response, coarsening resistance, creep resistance and elevated-temperature retention of mechanical properties.
Methods have been developed to derive a quantitative measure of the classical IB solute excess $\Gamma^{(1,2)}_\xi$ from composition profiles obtained using atom probe tomography (APT) \citep{marquis2006composition,amouyal2008segregation,biswas2011precipitates}. To complement experimental observations, first-principles calculations have been used to validate the occurrence of segregation and evaluate the segregation energies for various solutes to different IB types and lattice sites \citep{shin2017solute,samolyuk2020eqsegdft}.
Examples of experimentally observed solutes segregated at matrix-precipitate IBs in various alloys are: Ag, Sn, Si, Sc segregation at Al/$\theta^\prime$-Al$_2$Cu IB in Al-Cu; Mg at Al/Al$_3$Sc in Al-Sc; Zn at Mg/Mg$_2$Sn in Mg-Sn; W at $\gamma$/$\gamma^\prime$ IB in Ni-based superalloy; Ni and Mn at $\alpha$-Fe/Cu-rich precipitate.

Engineering alloys often contain multiple minority elements--either added by design or arising as impurities from manufacturing process--wherein solutes have been found to co-segregate at the IB. In many such alloys, the combined presence of solutes is found to have an effect distinct from that of individual solutes.
In (Al-Cu)-Mg-Ag, Ag and Mg co-segregate at Al-matrix/$\Omega$-Al$_2$Cu IB \citep{hutchinson2001origin}: this has been found to lower $\gamma$, via formation of strong Mg-Ag chemical bond at the IB, and stabilize the precipitate. However, individual additions of Mg and Ag appears to be ineffective as thermodynamic segregants \citep{sun2009first, rosalie2012silver, kang2014determination}.
In (Al-Cu)-Si-Mg-Zn, Si and Mg co-segregate at Al/$\theta^{\prime}$-Al$_2$Cu IB \citep{biswas2010simultaneous}. However, Si and Mg have a negligible effect individually.
In (Al-Cu)-Mn-Zr-Si, enhanced co-segregation of Mn and Zr at Al-matrix/$\theta^{\prime}$-Al$_2$Cu IB was recently found to increase the stability of $\theta^{\prime}$ to $~350^{\circ}$C \citep{shyam2019elevated}. However, individual additions of Zr or Mn stabilized $\theta^{\prime}$ to less than $300^{\circ}$C or $250^{\circ}$C, respectively. Recently, in (Al-Cu)-Sc-Fe-Si, co-segregation of Sc and Fe to Al-matrix/$\theta^{\prime}$ was realized, and was found to provide an unprecedented creep resistance at 300$^\circ$C \citep{gao2020segregation} (see Fig. \ref{fig:exp_seg}). The choice of solutes in \citep{gao2020segregation} was informed via first principles DFT calculations of segregation energies of various single and di-solutes. Another known example is the co-segregation of Re and Ru at $\gamma$/$\gamma^\prime$ in Ni-based superalloy \citep{wang2008alloying,wu2020unveiling}.
Some of these studies also demonstrate the potential for computationally-informed, segregation-driven design of precipitation alloys.

In the aforementioned examples, distinct interatomic interactions within the bulk and the IB bonding regions are involved.
% equilibrium IB segregation and its effect on microstructural stability. 
While the thermodynamics of bulk phases are well understood, and that of GB phases or complexions have significantly advanced \citep{kaplan2013review}, thermodynamics of IB is lacking beyond the classical models \citep{blum2020integral}. The classical Gibbs adsorption isotherm in its differential form cannot be readily integrated to obtain analytic governing equations for absolute changes in IB energy $\Delta \gamma$ as a function of measurable solute concentrations in the matrix phase. One needs to first determine the exact functional dependence of the chemical potentials $\mu_\xi$ and solute excesses $\Gamma^{(1,2)}_\xi$ on the solute concentration. So far, such estimations of $\Delta \gamma$ from APT-based composition profiles and calculations of $\Gamma^{(1,2)}_\xi$ have involved limiting assumptions of ideal or dilute solution behavior \citep{marquis2006composition,amouyal2008segregation,biswas2011precipitates}. A formal description of the distinct thermodynamic solution behavior of the IB phase, accounting for distinct atomic interactions at the IB, is of critical importance to advance our understanding of segregation thermodynamics.

In this paper, we present a diffuse-interface framework for describing multicomponent segregation thermodynamics in two-phase alloys. The paper is outlined as follows. In Sec.\ref{sec:method}, we propose the general phase-field model and then present the equilibrium solutions for a one-dimensional system with planar IB. Assuming regular solution behavior, we derive governing expressions for $\gamma$ and $\Gamma^{(1,2)}_\xi$ and demonstrate their validity with the Gibbs adsorption isotherm. Detailed analytic derivations are presented in the Appendix~\ref{app:profile}\textendash\ref{app:GA}. In Sec.\ref{sec:parametric}, we apply the model to study multicomponent segregation behavior in hypothetical quaternary alloys. Using multicomponent grain boundary segregation as a reference, we present a preliminary classification of the segregation behavior in the quaternary two-phase alloy.
In Sec.\ref{sec:conclusions}, we summarize and provide future directions for parameterization and comparison with experiments.

\section{Phase-field Model} \label{sec:method}

\subsection{Microstructure Formulation} \label{subsec:model}

We invoke the concept of an interfacial phase or complexion for the interphase boundary (IB). That is, a compositionally-homogeneous layer of IB phase $i$ is governed by a fundamental thermodynamic equation distinct from that of the adjacent matrix $m$ and precipitate $p$ phases. This description follows from the classical treatment of solute segregation to free surfaces and grain boundaries (GB) \citep{cahn1979interfacial,mcleangrain,sutton1995interfaces,guttmann1979interfacial,frolov2015phases}, and the non-classical treatment of diffuse segregation to GBs in phase-field models \citep{cha2002phase,kim2008GBsegregation,abdeljawad2015stabilization,abdeljawad2017grain,kim2016GBsegregation}.

A non-conserved phase field $\phi(\boldsymbol{x})$ variable is used to represent the phases: $m$ at $\phi=0$, $i$ at $\phi=0.5$, and $p$ at $\phi=1$. Any infinitesimal volume around the spatial location $\boldsymbol{x}$ is defined as a hypothetical thermodynamic mixture of $m$, $i$ and $p$ phases. The alloy is composed of the primary components $1$ and $2$, which act as solvents to from the two-phase system, and additional components $3,4,\ldots,\mathcal{N}$ which act as solutes. The mole fractions corresponding to a phase are represented by the phase concentration variables as $c_{\theta m}(\boldsymbol{x})$, $c_{\theta i}(\boldsymbol{x})$ and $c_{\theta p}(\boldsymbol{x})$; where, $\theta=1,2,3,\ldots, \mathcal{N}$. 
% Assuming substitutional solution and identical molar volumes of the components across the alloy, the local phase concentrations satisfy $c_{1 \psi}(\boldsymbol{x}) = 1-\sum_{\theta=2:\mathcal{N}} c_{\theta \psi}(\boldsymbol{x})$ ($\psi = m, i, p$). 

We introduce interpolating functions $M(\phi)$, $I(\phi)$ and $P(\phi)$ for each of the phases $m$, $i$ and $p$. These functions serve as local phase fractions to define effective properties at $\boldsymbol{x}$ as a mixture of the phase properties.
The effective concentrations $c_\theta(\boldsymbol{x})$ are defined by the mixture rule as
\begin{flalign} \label{eq:conc}
    c_\theta = M(\phi)c_{\theta m} + I(\phi)c_{\theta i} + P(\phi)c_{\theta p}. &&
\end{flalign}
The local effective free energy density $f(\boldsymbol{c},\phi)$ is defined using the local phase-specific free energy densities $f^\psi(c_{2 \psi},c_{3 \psi},\ldots,c_{\mathcal{N} \psi})$ as
\begin{flalign} \label{eq:f_local}
    f(\boldsymbol{c},\phi) = M(\phi)f^m(\boldsymbol{c}_m) + I(\phi) f^i(\boldsymbol{c}_m) + P(\phi)f^p(\boldsymbol{c}_p), && \hspace{0.5cm}
\end{flalign}
{where $\boldsymbol{c}=\{c_2,c_3,\ldots,c_{\mathcal{N}}\}$.} Under the assumption of substitutional solution and identical molar volumes of the components across the alloy, the effective concentrations satisfy $c_{1}(\boldsymbol{x}) = 1-\sum_{\theta=2:\mathcal{N}} c_{\theta}(\boldsymbol{x})$. 

The phase concentrations $c_{\theta \psi}(\boldsymbol{x})$ at any given point are constrained by the condition of equal diffusion potential between the phases as
\begin{flalign} \label{eq:equal_diff_pot}
    \frac{\partial f^m}{\partial c_{\theta m}} = \frac{\partial f^i}{\partial c_{\theta i}} = \frac{\partial f^p}{\partial c_{\theta p}} \equiv {\mu}_{\theta 1},  &&
\end{flalign} 
where ${\mu}_{\theta 1} \equiv \mu_\theta - \mu_1$ ($\theta = 2:\mathcal{N}$) are the diffusion potentials. The above conditions follow from the KKS (Kim-Kim-Suzuki) phase-field formulation---originally developed for two-phase alloy solidification \citep{kim1999phase}, later applied for GB segregation in two-grain alloy with GB phase \citep{cha2002phase,kim2008GBsegregation,kim2016GBsegregation}, and recently developed for IB segregation in two-phase alloy with IB phase \citep{kadambi2020phase,kadambi2020acta,kadambi2020phd}.

\begin{figure}[!htp]
    \centering
    \includegraphics[width=0.4\textwidth,trim=4 4 4 4,clip]{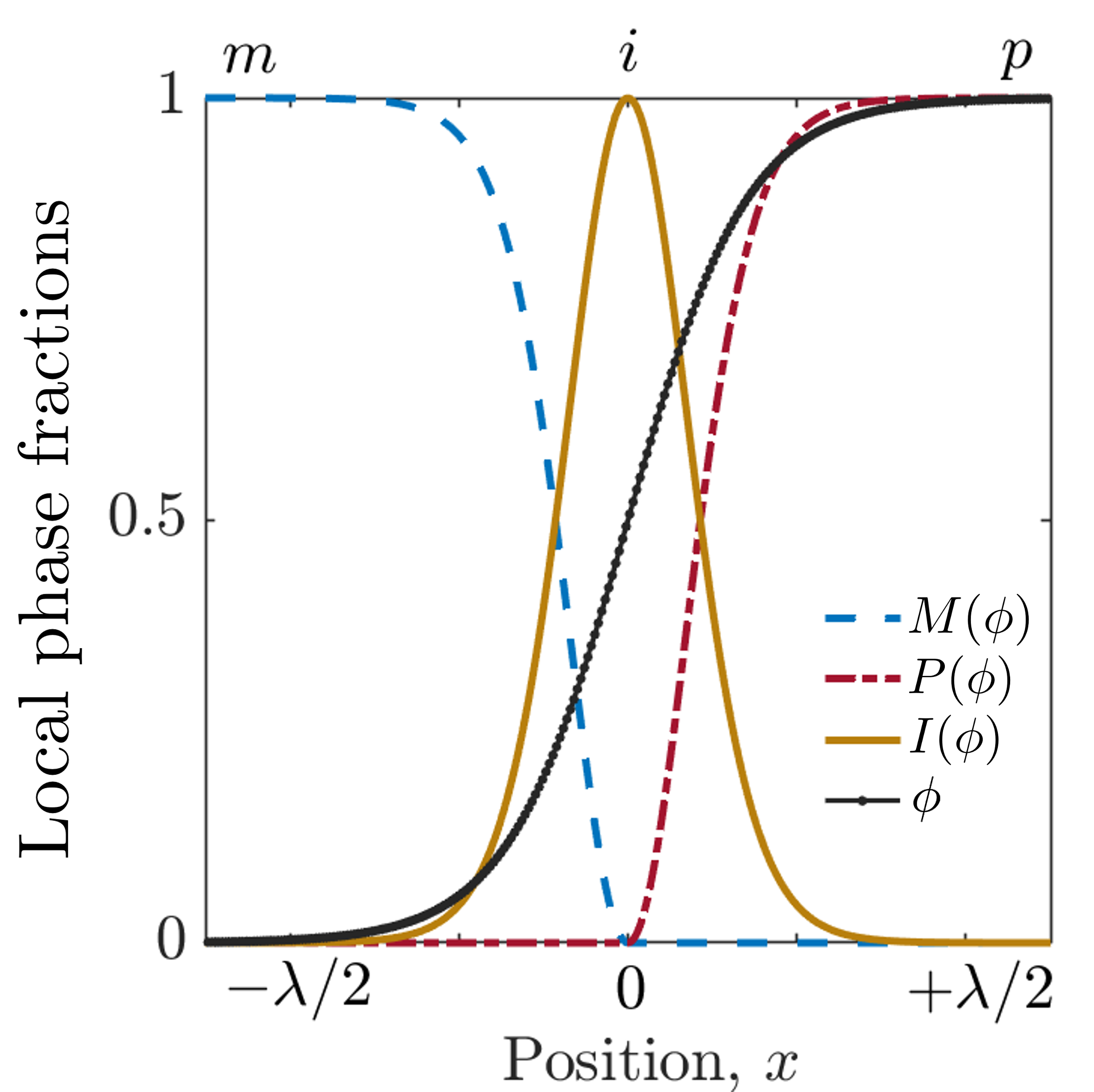}
    \caption{Phase field $\phi({x})$ of width $\lambda$ representing the matrix phase $m$ at $\phi=0$, the interfacial phase $i$ at $\phi=0.5$, and the precipitate phase $p$ $\phi=1$ in a one-dimensional system. Interpolating functions or phase fractions $M(\phi)$, $I(\phi)$ and $P(\phi)$ associated with the $m$, $i$ and $p$ phases, respectively.}
    \label{fig:interp}
\end{figure}

{The functional forms of the interpolating functions are chosen to satisfy the definition of exclusive phases as $M(\phi=0)=1$, $I(\phi=0.5)=1$ and $P(\phi=1)=1$, and the phase-mixture rule $M(\phi)+P(\phi)+I(\phi)=1$. 
For the sake of continuity in the derivatives, $M(\phi)$ and $P(\phi)$ $\in C^1[0,1]$ are chosen to be differentiable, piecewise functions. A convenient choice of interpolating functions constructed from the double-well form $I(\phi) = 16\phi^2(1-\phi)^2$ is shown in Fig.~\ref{fig:interp}. Note that for $\phi \in [0,0.5)$, $M(\phi) = 1 - I(\phi)$ and $P(\phi)=0$; for $\phi \in (0.5, 1]$, $M(\phi)=0$ and $P(\phi) = 1 - I(\phi)$.}

Given the local microstructure and its properties thus far, the overall free energy of the domain of volume $V$ can be defined by the functional,
\begin{flalign} \label{eq:F_functional}
    \mathcal{F}
    = \int_V
    \left[ f(\boldsymbol{c},\phi) + \frac{\varepsilon^2}{2} \left|\nabla \phi
    \right|^2 \right]
\mathrm d V, &&
\end{flalign} 
where the second term in the integrand is the gradient energy density due to $\phi$, with $\varepsilon^2$ representing the gradient energy coefficient.

Following classical linear irreversible thermodynamics, the concentration fields evolve temporally under mass conservation as
\begin{flalign} \label{eq:cahn-hilliard}
    \frac{\partial c_\theta}{\partial t} = \nabla \cdot \left[\sum_{\xi=2:\mathcal{N}} \mathcal{M}_{\theta \xi} \nabla \left(\frac{\delta \mathcal{F}}{\delta c_\xi} \right) \right]; \hspace{0.5 cm} && 
\end{flalign}
where $\theta = 2:\mathcal{N}$. The variational derivatives take the simplified definition as $\delta \mathcal{F}/ \delta c_\theta = \partial f/ \partial c_\theta = {\mu}_{\theta 1}$. $\mathbf{\mathcal{M}}$ is the diffusion mobility matrix with components 
$\mathcal{M}_{\theta \xi} = \sum_{j=1:\mathcal{N}}(\delta_{k\xi j}-c_\xi)(\delta_{j\theta}-c_\theta)c_j \mathcal{M}_j$. $\mathcal{M}_{\theta}$ is the atomic mobility of component $\theta$ ($=1:\mathcal{N}$); $\mathcal{M}_{\theta}$ can be expressed in terms of the phase-dependent atomic mobilities as $\mathcal{M}_{\theta}=M(\phi)\mathcal{M}_{\theta}^m + I(\phi)\mathcal{M}_{\theta}^i+ P(\phi)\mathcal{M}_{\theta}^p$. 

Following Allen-Cahn dynamics, the non-conserved field $\phi$ evolves temporally as
\begin{flalign} \label{eq:allen-cahn}
    \frac{\partial \phi}{\partial t} = - \mathcal{L} \frac{\delta \mathcal{F}}{\delta \phi}, &&
\end{flalign}
where $\mathcal{L}$ is the kinetic mobility parameter for the interface.

\subsection{Equilibrium Planar Interface} \label{subsec:1D}

We now consider a one-dimensional, planar IB at stationary equilibrium. The microstructure fields become invariant with time and will be denoted by $\phi_e(x)$ and $\boldsymbol{c}^e(x)$. Eqs.~\ref{eq:cahn-hilliard} and \ref{eq:allen-cahn} reduce to
\begin{flalign} \label{eq:1D_eq_conc}
    \frac{\delta \mathcal{F}}{\delta c_\theta} &= \frac{\partial f}{\partial c_\theta^e} \equiv {\mu}_{\theta 1}^e (\text{const.}), \\ \label{eq:1D_eq_phi}
    \frac{\delta \mathcal{F}}{\delta \phi} &= \frac{\partial f}{\partial \phi_e} - \varepsilon^2 \frac{d^2 \phi_e}{dx^2} = 0,     &&
\end{flalign}
where $\mu_{\theta 1}^e$ are the equilibrium diffusion potentials. $\mu_{\theta 1}^e$ are constant across $x$ as required by the exchange of atoms between substitutional sites \citep{balluffi2005kinetics}. The conditions (Eq.~\ref{eq:equal_diff_pot}) for the equality of diffusion potentials at a point now read
\begin{flalign} \label{eq:equilibrium_diff_pot}
     \frac{\partial f^m(x)}{\partial c_{\theta m}} = \frac{\partial f^i(x)}{\partial c_{\theta i}} = \frac{\partial f^p(x)}{\partial c_{\theta p}} \equiv {\mu}_{\theta 1}^e (\text{const.}).  &&
\end{flalign} 
The above relations imply that the phase-concentrations become spatially constant, i.e. $c_{\theta \psi}(x)=c_{\theta \psi}^e$. Therefore, the effective concentration fields (Eq.~\ref{eq:conc}) are given by
\begin{flalign} \label{eq:conc_profile} \nonumber
    c_\theta^e(x) =& M\left(\phi_e(x)\right)\,c_{\theta m}^e + I\left(\phi_e(x)\right)\,c_{\theta i}^e \\ 
    &+ P\left(\phi_e(x)\right)\,c_{\theta p}^e.  &&
\end{flalign}

Multiplying Eq.~\ref{eq:1D_eq_phi} with $d\phi_e/dx$ and integrating piecewise with respect to $x$ from $-\infty$ to $0$ and $0$ to $+\infty$ yields the following equilibrium relations between $m$, $i$ and $p$ (see Appendix~\ref{app:profile} for derivation). Here, the limits of integration $-\infty$, $0$, and $+\infty$ refer to the far-field matrix, the exclusive IB and the far-field precipitate phases, respectively. 
\begin{flalign} \label{eq:parallel_tangent_We} \nonumber
    & f^i_e - f^m_e - \sum (c_{\theta i}^e-c_{\theta m}^e){\mu}_{\theta 1}^e \\
    &= f^i_e - f^p_e - \sum (c_{\theta i}^e-c_{\theta p}^e){\mu}_{\theta 1}^e \equiv W_e, &&
\end{flalign}
and
\begin{flalign} \label{eq:common_tangent}
    f^p_e - f^m_e = \sum (c_{\theta p}^e - c_{\theta m}^e){\mu}_{\theta 1}^e.  &&
\end{flalign}
Eq.~\ref{eq:equilibrium_diff_pot} and ~\ref{eq:common_tangent} constitute the well-known common tangent hyperplane condition for equilibrium between the bulk phase free energy hypersurfaces $f^m$ and $f^p$. Eq.~\ref{eq:parallel_tangent_We} and Eq.~\ref{eq:equilibrium_diff_pot} constitute the parallel tangent hyperplane conditions for the equilibrium of the IB phase $f^i$ with respect to the bulk phases. For known convex functions of $f^m$, $f^i$ and $f^p$, the above conditions uniquely determine the equilibrium phase concentrations $c^e_{\theta \psi}$. $W_e$ represents the vertical distance between the two parallel tangent hyperplanes of $i$ and $m/p$. Physically, $W_e$ describes the free energy for the formation of a unit volume of equilibrium IB phase from the equilibrium matrix phase (or equivalently from the equilibrium precipitate phase) \citep{hillert1975lectures}.

Substituting Eqs.~\ref{eq:f_local} and \ref{eq:parallel_tangent_We} in Eq.~\ref{eq:1D_eq_phi}, we obtain the expression for the stationary phase-field profile $\phi_e(x)$ as (see Appendix~\ref{app:profile} for derivation)
\begin{flalign} 
    \varepsilon^2\frac{d^2\phi_e}{dx^2} = W_e\frac{dI(\phi_e)}{d\phi_e},  &&
\end{flalign}
which can be integrated to obtain 
\begin{flalign} \label{eq:eq_phase-field_final}
  \varepsilon^2 \left(\frac{d\phi_e}{dx}\right)^2 = 2W_eI(\phi_e). &&
\end{flalign}
The above expression depends only on the functional form of $I = 16\phi^2(1-\phi)^2$. $W_e$ depends on the equilibrium phase concentrations, which are constants across the microstructure. Therefore, $W_e$ is a constant and the above expression can be readily integrated to obtain the well-known closed-form solution
\begin{flalign} \label{eq:hyperbolic_tangent}
    \phi_e(x) = \frac{1}{2}\left[1+\tanh\left(\frac{2\sqrt{2W_e}}{\varepsilon}x\right)\right].  &&
\end{flalign}
The width $\lambda$ of the diffuse IB (defined by the bounds $\phi_e = 0.1$ and $\phi_e=0.9$) is given by
{\begin{flalign} \label{eq:width}
    \lambda = \int_{0.1}^{0.9} \frac{dx}{d\phi_e} d\phi_e \approx \frac{1.1\varepsilon}{\sqrt{2W_e}}.  &&
\end{flalign}}

\subsection{Gibbs excess Properties} \label{subsec:GA}

We can now obtain expressions for the classical excess quantities from the equilibrium diffuse IB. The excess IB energy per unit IB area, $\gamma$, is the excess grand potential evaluated as (see Appendix~\ref{app:IB_energy} for derivation)
\begin{flalign} \label{eq:gamma_int}
    \gamma = \int_{-l}^{l} 2W_eI(\phi_e(x)) dx \approx \frac{2\varepsilon \sqrt{2W_e}}{3}, &&
\end{flalign}
where $\pm l$ are locations within the bulk phases far-field of the interface. The integrand $2W_eI(\phi_e(x)) \equiv \Omega(x)$ describes the energy density across the system in excess of the thermodynamic mixture of the equilibrium bulk phases. The integrand vanishes within the bulk phases since $I\left(\phi_e(\pm l)\right) \rightarrow 0$, and therefore, the integral converges. This property makes $\gamma$ an invariant thermodynamic quantity that is independent of the specific choice of the layer thickness \citep{mcfadden2002gibbs}. 

The extensive solute excess quantities are evaluated by adapting Cahn's layer approach \citep{cahn1979interfacial} to the phase-field model (see Appendix~\ref{app:GA} for derivations). The generalized Gibbs adsorption equation relating the change in $\gamma$ to the independent variations of intensive quantities, temperature $T$ and $\mu^e_\theta$, is obtained as \citep{cahn1979interfacial,frolov2015phases}
\begin{flalign}
    d\gamma = -\Gamma^{(1,2)}_S dT - \sum_{\xi = 3:\mathcal{N}} \Gamma^{(1,2)}_\xi d\mu^e_\xi , &&
\end{flalign}
where $\Gamma^{(1,2)}_S$ is the relative entropy excess and $\Gamma^{(1,2)}_\xi$ is the relative solute excess for $\xi$, defined per unit interface area. The chemical potentials of the base components $1$ and $2$ have been eliminated with the aid of equilibrium conditions between the two bulk phases, and thus, $\Gamma^{(1,2)}_{1/2}=0$.
The excess quantity $\Gamma^{(1,2)}_{S/\xi}$ describes the difference in the amount of entropy/component-$\theta$ between the chosen layer---containing the diffuse interface---and that of a hypothetical system. The hypothetical system is constructed from the two homogeneous bulk phases in such a proportion that the total amount of $1$ and $2$ is equal to that in the original layer \citep{cahn1979interfacial,frolov2015phases}.
Analytic expressions for these quantities were realized from the equilibrium phase and concentration fields as
\begin{flalign} \label{eq:solute_excess_analytic} \nonumber
   \Gamma^{(1,2)}_{\theta} =& C^{xs}_{\theta} - \frac{(c^e_{1m} c^e_{\theta p} - c^e_{\theta m} c^e_{1p})}{(c^e_{1m} c^e_{2p} - c^e_{2m} c^e_{1p})} C^{xs}_2 \\
   &- \frac{(c^e_{\theta m} c^e_{2p} - c^e_{2m} c^e_{\theta p})}{(c^e_{1m} c^e_{2p} - c^e_{2m} c^e_{1p})} C^{xs}_1 \hspace{0.3 cm}  && 
\end{flalign}
where $\theta = 1:\mathcal{N}$, $c^e_{1\psi}=1-\sum_{k = 2:\mathcal{N}} c^e_{k \psi}$, and
{\begin{flalign} \label{eq:Gibbs_exs}
    C^{xs}_\xi = \frac{\varepsilon}{3\sqrt{2W_e}} \left(2c^e_{\theta i} - c^e_{\theta m} - c^e_{\theta p} \right). &&
\end{flalign}}
$C^{xs}_\xi$ is the Gibbs solute excess evaluated from the effective concentration fields using the Gibbs dividing surface at $x=0$. Since the equilibrium concentrations are generally unequal in the two bulk phases, $C^{xs}_\xi$ depends on the specific location of the dividing surface within the diffuse IB. Note that $\Gamma^{(1,2)}_\xi$ is independent of the position of the dividing surface, and is an invariant thermodynamic quantity like $\gamma$. A similar expression is obtained for the $\Gamma^{(1,2)}_S$ (Eq.~\ref{aeq:entropy_excess_analytic}).

The Gibbs adsorption isotherm can be written in terms of the derivatives of the matrix-phase concentrations $c^e_{\theta m}$ as
\begin{flalign} \label{eq:GA_conc_form}
    \frac{\partial \gamma}{\partial c^e_{\xi m}} = - \sum_{\theta = 3}^{\mathcal{N}} \Gamma^{(1,2)}_\theta \frac{\partial \mu^e_\theta}{\partial c^e_{\xi m}}; \hspace{1cm} (\xi=3:\mathcal{N}), &&
\end{flalign}
where $\mu^e_\theta = f^m_e + \sum_{k = 2:\mathcal{N}} \left(\delta_{\theta k} - c^e_{k m} \right) \mu^e_{k 1}$ ($\delta_{\theta k}$ is the Kronecker delta) \citep{lupis1983chemical}.

\subsection{Multicomponent Thermodynamics} \label{sec:soln_therm}

We assume regular solution behavior for the bulk and IB phases. 
The free energy density-composition dependence characteristic to $\psi$ is given by
\begin{flalign} \label{eq:RS_phase_FE} \nonumber
    f^{\psi} =& \,\hspace{0.1cm} \sum_{\theta = 1}^{\mathcal{N}} G_\theta^\psi c_{\theta \psi} + \sum_{\theta = 1}^{\mathcal{N}} \sum_{\xi \neq \theta} L^\psi_{\theta \xi} c_{\theta \psi} c_{\xi \psi} \\
    &+ \frac{RT}{v_m} \left(\sum_{\theta = 1}^{\mathcal{N}} c_{\theta \psi} \ln{c_{\theta \psi}} \right). &&
\end{flalign}
Here, the energetics of $\theta$ in all the phases $\psi$ are defined from the same pure component reference state $\{\pi \theta\}$ \citep{dehoff2006thermodynamics}.
The reference states can have crystal structures distinct from that of the solid-solution $\psi$. 
$G^\psi_\theta \{\pi \theta \}$ accounts for the pure component energy of $\theta$ in $\psi$ with reference to ${\pi \theta}$. For mixing (second term), the structure of each component must be identical to that of the final solid-solution $\psi$. $L^\psi_{\theta \xi} \{\psi \theta,\psi \xi\}$ accounts for the non-ideal interaction energy between distinct components $\theta$ and $\xi$ in $\psi$, with reference to the pure components in $\psi$. 
Therefore, the parameters $G^\psi_\theta$ and $L^\psi_{\theta \xi}$ account for the distinct chemical and structural energetics of the bulk and interfacial phases. The final term in the equation is the ideal configurational entropy for mixing and $v_m$ is the molar volume.

Using Eq.~\ref{eq:RS_phase_FE}, the equilibrium compositions in the bulk phases are established by Eqs.~\ref{eq:equilibrium_diff_pot} and ~\ref{eq:common_tangent}. These constitute $\mathcal{N}$ equations in $2\left(\mathcal{N}-1 \right)$ bulk-phase concentrations and allow $\mathcal{N}-2$ compositional degrees of freedom. We choose the matrix-phase solute concentrations $c^e_{3m}, \ldots , c^e_{\mathcal{N}m}$ as the control variables, while the rest are to be uniquely determined.

The equilibrium composition in the IB phase $\psi = i$ is then established using Eq.~\ref{eq:RS_phase_FE} in $\partial f^i/\partial c^e_{\theta i} = \partial f^m/\partial c^e_{\theta m}$ of Eq.~\ref{eq:equilibrium_diff_pot}. These constitute $\mathcal{N}-1$ segregation equations relating $\mathcal{N}-1$ unknowns $(c^e_{2i}, \ldots ,c^e_{\mathcal{N}i})$ to the known $(c^e_{2m}, \ldots , c^e_{\mathcal{N} m})$. They can be written as
\begin{flalign} \label{eq:seg_isotherm_RS}
    \frac{c^e_{\theta i}}{c^e_{1i}} = \frac{c^e_{\theta m}}{c^e_{1m}} \exp \left(\frac{\Delta E^{\theta 1}_{\text{seg}}}{RT/v_m} \right); \hspace{0.3cm} (\theta = 2:\mathcal{N}), &&
\end{flalign}
where
\begin{flalign} \label{eq:seg_isotherm2}
    \Delta &E^{\theta 1}_{\text{seg}} = (G^i_1-G^m_1-G^i_\theta+G^m_\theta) \\ \nonumber
    &+\left[L^m_{1\theta} -L^i_{1\theta} + 2(L^i_{1\theta}c^e_{\theta i} - L^m_{1\theta}c^e_{\theta m})\right] \\ \nonumber
    &+ \sum_{\xi \neq \theta} \left[(L^i_{1\xi}+L^i_{1\theta}-L^i_{\xi \theta})c^e_{\xi i} - (L^m_{1\xi}+L^m_{1\theta}-L^m_{\xi \theta})c^e_{\xi m}\right]. &&
\end{flalign}
$\Delta E^{\theta 1}_{\text{seg}}$ describes the segregation energy resulting form the exchange of $\theta$ in $m$ with component $1$ in $i$. The above coupled and non-linear equations must be solved simultaneously to determine the equilibrium IB phase composition.

The segregation equations obtained in our model are equivalent to the generalized multicomponent grain boundary segregation isotherms of Guttmann \citep{guttmann1975equilibrium}.
For an ideal alloy ($L=0$), the equations reduce to the decoupled multicomponent versions of Langmuir-McLean \citep{mcleangrain,lejcek2010grain}. For a non-ideal ($L\neq0$) alloy, the second term in Eq.~\ref{eq:seg_isotherm2} is identical to that of the Fowler-Guggenheim isotherm \citep{guggenheim1985thermodynamics,lejcek2010grain} representing the interaction between $\theta$ and $1$.
The third and following terms are the cross terms relating segregation of $\theta$ with segregation of $\xi$. For IB segregation, these interactions become important for a quaternary (or higher-order) alloy as shown in the next section. 

% The $\mathcal{N}-2$ compositional degrees of freedom following from the Gibbs interface phase rule \citep{frolov2015phases} will allow independent tuning of the IB's solute concentrations: $c^e_{3i},c^e_{4i},\ldots,c^e_{\mathcal{N} i}$. 
\section{Parametric Study} \label{sec:parametric}

\subsection{Non-dimensionalization}
In this section, we apply the phase-field model to explore segregation behavior in quaternary alloys. 
The model is non-dimensionalized by setting $x=l_o\tilde{x}$, $t = t_o \tilde{t}$, $T = T_o \tilde{T}$ and $f^\psi = \left(RT_o/v_m\right)\tilde{f}^\psi$ ($\psi = m,i,p$). $l_o$, $t_o$ and $T_o$ are the characteristic length, time and temperature, respectively, and $RT_o/v_m$ is the characteristic energy. The dimensionless quantities are denoted by the tilde symbol. The free energy density for a quaternary alloy now takes the form, $\tilde{f}^\psi = \tilde{G}^\psi_1 c_{1 \psi} + \tilde{G}^\psi_2 c_{2 \psi} + \tilde{G}^\psi_3 c_{3 \psi} + \tilde{G}^\psi_4 c_{4 \psi} + \tilde{L}^\psi_{12} c_{1\psi} c_{2\psi} + \tilde{L}^\psi_{13} c_{1\psi} c_{3\psi} + \tilde{L}^\psi_{23} c_{2\psi} c_{3\psi} + \tilde{L}^\psi_{14} c_{1\psi} c_{4\psi} + \tilde{L}^\psi_{24} c_{2\psi} c_{4\psi} + \tilde{L}^\psi_{34} c_{3\psi} c_{4\psi} + \tilde{T} \left(c_{1\psi}\ln{c_{1\psi}} + c_{2\psi}\ln{c_{2\psi}} + c_{3\psi}\ln{c_{3\psi}} + c_{4\psi}\ln{c_{4\psi}} \right)$.  The equilibrium equations in Sec.~\ref{subsec:model} are replaced with the dimensionless parameters: $\tilde{\varepsilon}^2=\varepsilon^2/\left[l_o^2\left(RT_o/v_m\right)\right]$ and $\tilde{\mu}^e = \mu^e/\left(RT_o/v_m\right)$. 
% Finally, the kinetic parameters are modified as $\tilde{\mathcal{M}}=\left(\mathcal{M} R T_o t_o\right)/\left(l_o^2 v_m \right)$ and $\tilde{\mathcal{L}} = \left(\mathcal{L} RT_o t_o \right)/v_m$. 

For the parametric study of the model, we assume $(\tilde{\varepsilon}^2,\tilde{T})=(1,1)$.
The dimensionless results can be dimensionalized as $\gamma = \left(RT_ol_o/v_m\right)\tilde{\gamma}$ and $\Gamma_\xi^{(1,2)}=(l_o/v_m)\tilde{\Gamma}_\xi^{(1,2)}$. For example, setting the characteristic scales to be $l_o = 1$ nm, $T_o = 300$ K and $v_m = 7\times 10^{-6}$ m$^3$/mol yields $\gamma = 0.36 \tilde{\gamma}$ J/m$^2$ and $\Gamma_\xi^{(1,2)}=\left(1.43\times10^{-4}\right)\tilde{\Gamma}_\xi^{(1,2)}$ mol/m$^2$.

\begin{table}[!htp]
\centering
\caption{Default ideal solution parameters of the binary and ternary two-phase alloys.} \label{table:base_params}
\includegraphics[width=0.35\textwidth]{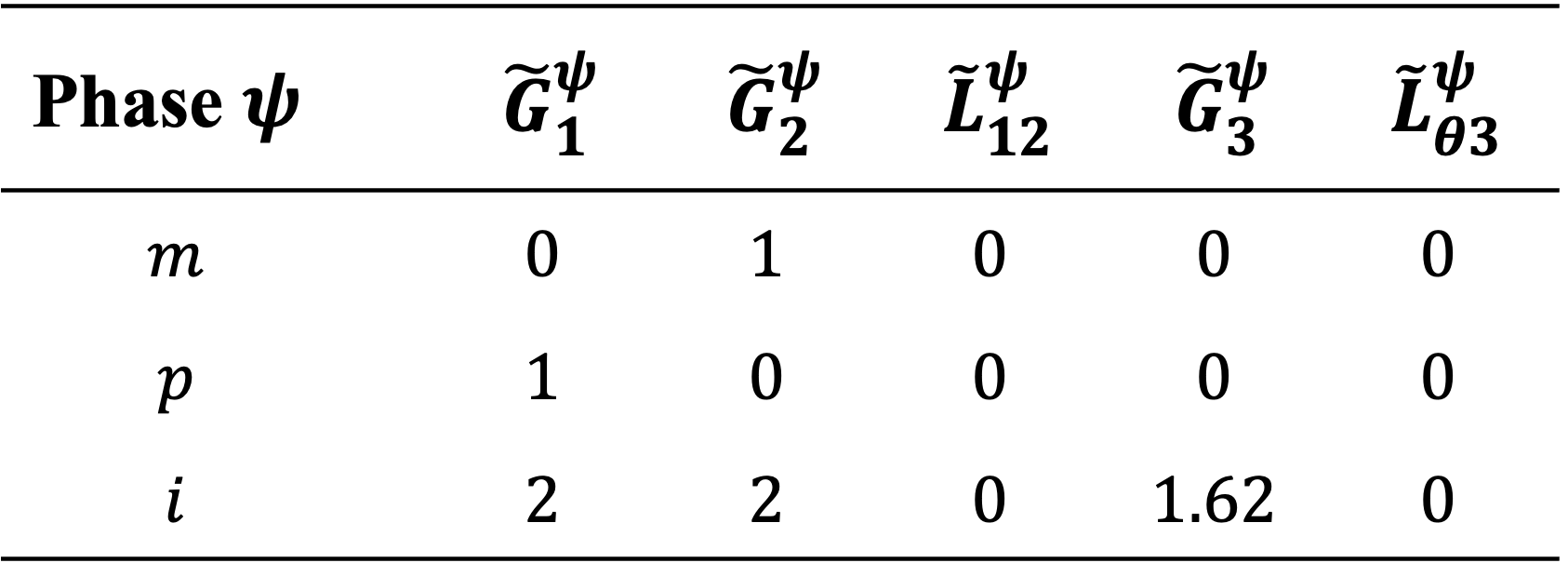}
\end{table}

\subsection{Binary and Ternary References}

\begin{figure*}[!htp]
\captionsetup[subfigure]{justification=centering}
\centering
\begin{subfigure}{0.3\textwidth}
  \centering
  \includegraphics[width=1\linewidth,trim=4 4 4 4,clip]{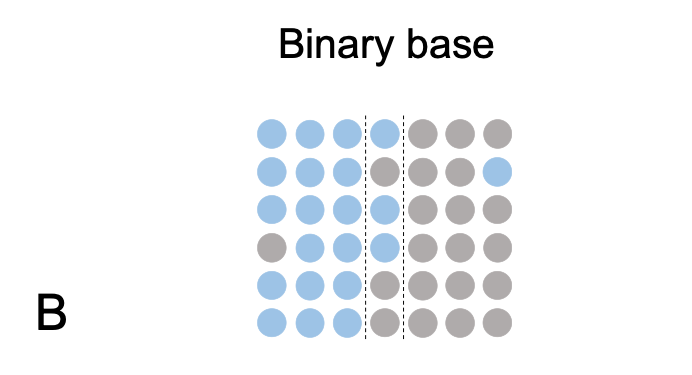}
%   \caption*{}
\end{subfigure}
\begin{subfigure}{0.3\textwidth}
  \centering
  \includegraphics[width=1\linewidth,trim=4 4 4 4,clip]{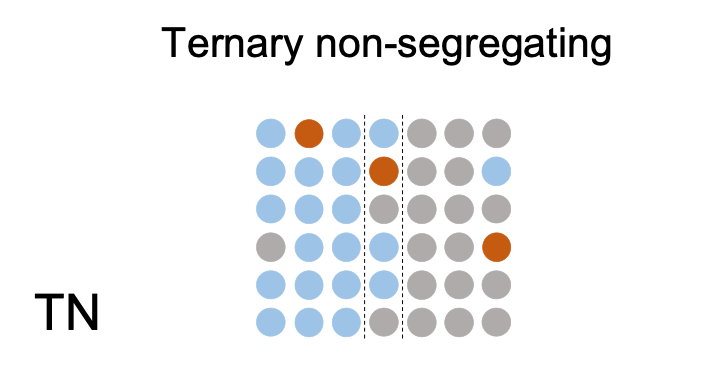}
%   \caption*{}
\end{subfigure}
\begin{subfigure}{0.3\textwidth}
  \centering
  \includegraphics[width=1\linewidth,trim=4 4 4 4,clip]{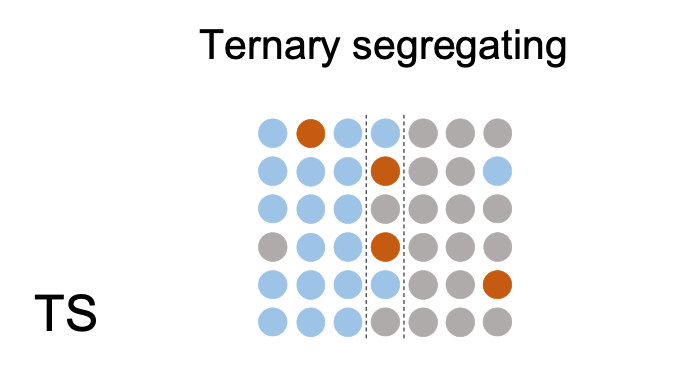}
%   \caption*{}
\end{subfigure}
\begin{subfigure}{0.29\textwidth}
  \centering
    \includegraphics[width=1\linewidth,trim=4 4 4 4,clip]{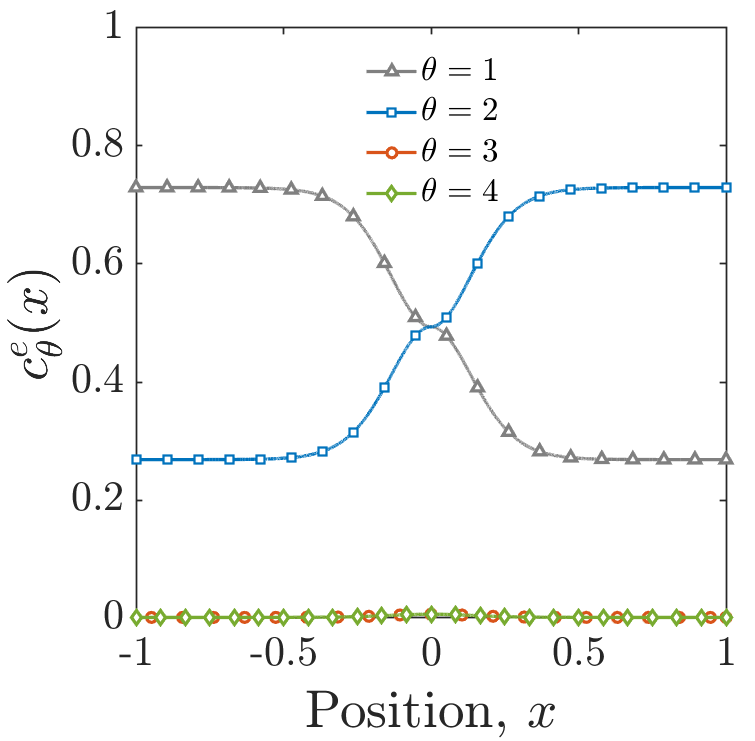}
%   \caption*{}
\end{subfigure}
\begin{subfigure}{0.3\textwidth}
  \centering
  \includegraphics[width=1\linewidth,trim=4 4 4 4,clip]{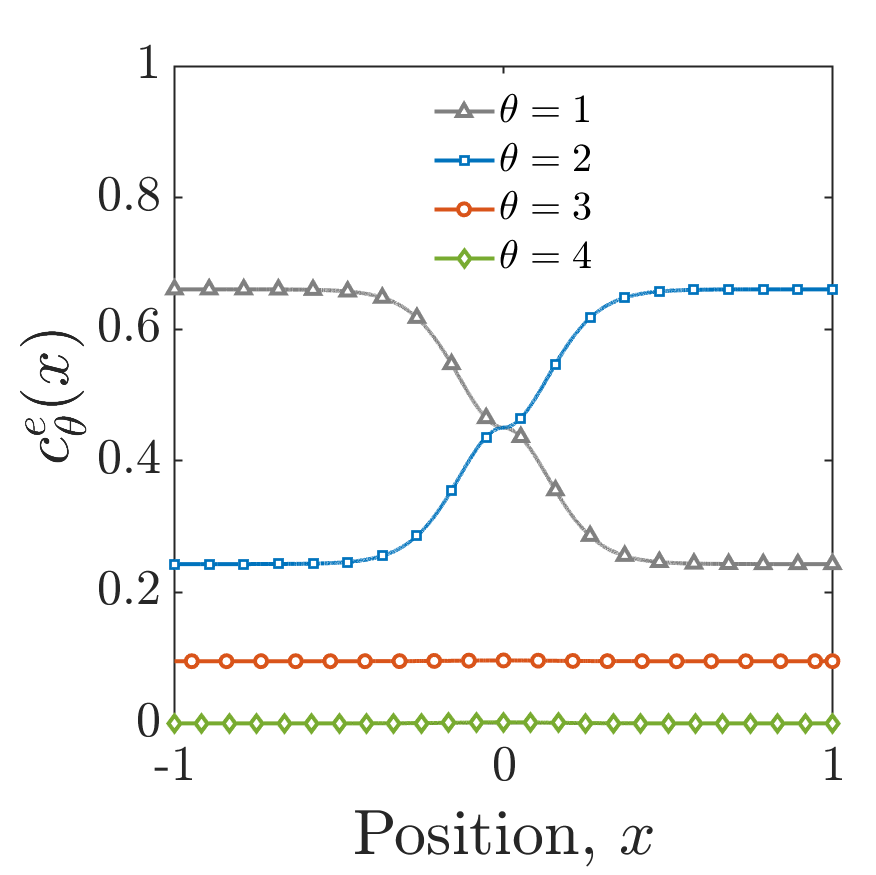}
%   \caption*{}
\end{subfigure}
\begin{subfigure}{0.3\textwidth}
  \centering
  \includegraphics[width=1\linewidth,trim=4 4 4 4,clip]{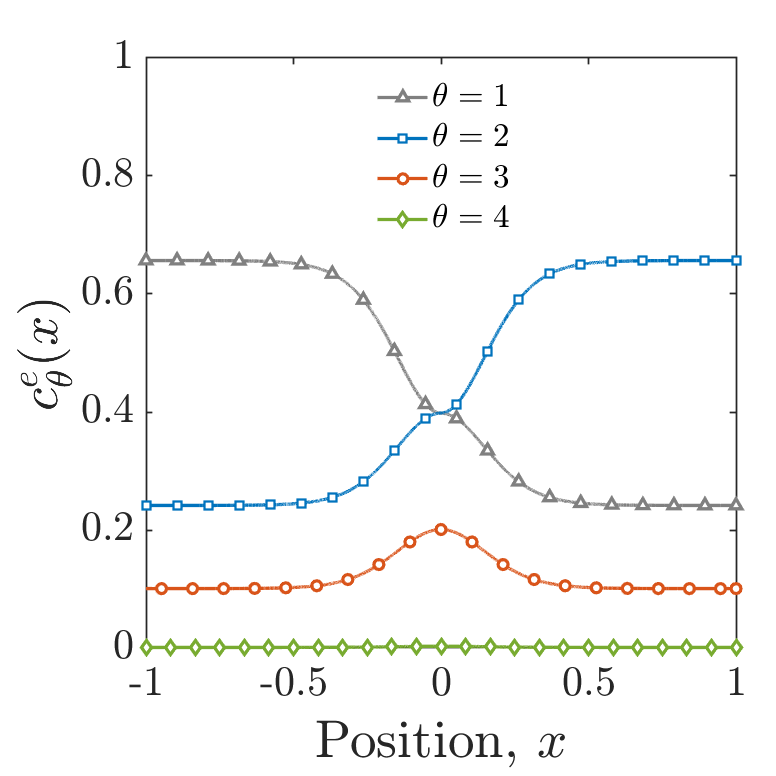}
%   \caption*{}
\end{subfigure}
\caption{Top: Representative schematics of elemental distribution in the two bulk phases and interfacial phase modeled as a random substitutional alloy. Bottom: Equilibrium concentration profiles across a one-dimensional, planar two-phase interface obtained from the phase-field model parameterized using Table~\ref{table:base_params}. (B) Binary ($1$-$2$) alloy. (TN) Ternary ($1$-$2$)-$3$ alloy with a non-segregating solute. (TS) Ternary ($1$-$2$)-$3$ alloy with a segregating solute. Color legend: grey is atom $1$, blue is $2$, red is $3$, green is $4$. }
\label{fig:base_profiles}
\end{figure*}

We choose an ideal binary base alloy ($1$-$2$) that forms a two-phase coexistence region in the bulk phase diagram (see Table~\ref{table:base_params} and Ref.~\citep{kadambi2020phase}). 
The alloy microstructure and the equilibrium concentration profiles are shown in Fig.~\ref{fig:base_profiles} and represented by the schematic with the label {B}.
The matrix phase $m$ is rich in $1$ ($c^e_{2m}\approx0.27$) while the precipitate phase $p$ is rich in $2$ ($c^e_{2p}\approx0.73$). The IB phase $i$ has been chosen to have an equilibrium concentration as the average of $m$ and $p$ phases ($c^e_{2m}\approx0.5$). 
The phase-field calculations yield a relative excess entropy of $\tilde{\Gamma}^{(1,2)}_S=0$ and an excess IB energy of $\tilde{\gamma}_B=1.2$ units. 
% The IB energy can be controlled by adding solute components as demonstrated for a ternary system in \citep{kadambi3539253interphase}. 

Next we consider the addition of solutes $3$ and $4$ individually to form the ternaries ($1$-$2$)-$\xi$ ($\xi=3/4$). Two types of solute behaviors are considered as references--ternary non-segregating ({TN}) and ternary segregating ({TS}). The alloy microstructures of TN and TS are shown via representative schematics and equilibrium concentration profiles in Fig.~\ref{fig:base_profiles}. {TN} alloys are realized for $L^i_{\theta \xi}$ = $L^{m/p}_{\theta \xi}$ ($\theta=1,2$), i.e. when the regular solution interactions are identical or ideal across all phases in the alloy. For these cases, $\tilde{G}^i_\theta \approx 1.62$ units was derived from Eq.~\ref{eq:seg_isotherm_RS} such that $c^e_{\xi i}=c^e_{\xi {m/p}}$ is satisfied.
Calculations of excess quantities yields $\tilde{\Gamma}^{(1,2)}_\xi=0$ and $\tilde{\gamma}=\tilde{\gamma}_B$.
TS alloys, on the other hand, are realized for $L^i_{\theta \xi} < L^{m/p}_{\theta \xi}$, i.e. when the solute $\xi$ interacts more favorably with $\theta$ in $i$. These can further be noted for the ternary parameters $L^i_{\theta \xi}$ and $L^{m/p}_{\theta \xi}$ listed in Table~\ref{table:quaternary_params}. Calculation of excess quantities yield a positive segregation $\tilde{\Gamma}^{(1,2)}_\xi>0$ and a reduced IB energy $\tilde{\gamma}<\tilde{\gamma}_\text{B}$.
For simplicity, we omit segregation cases arising from variations in the pure energies $\tilde{G}^\psi_\xi$. We also simplify the parameter space by considering that the solutes interact identically within the bulk phases $m$ and $p$, and with solvent component $1$ and $2$. This condition is captured in Table~\ref{table:quaternary_params} by $L^{m/p}_{\theta 3}$ and $L^{m/p}_{\theta 4}$, and lead to vanishing cross terms between the solute and solvent in Eq.~\ref{eq:seg_isotherm2}, $\tilde{L}^\psi_{12}+\tilde{L}^\psi_{1\xi}-\tilde{L}^\psi_{2\xi}=0$. 

For the sake of completeness, we note here that the variation of excess quantities for the ternaries can be read from the limits $c^e_{\xi m}\rightarrow0$ of $\tilde{\Gamma}^{(1,2)}_\xi$ and $\tilde{\gamma}$ contour plots presented for quaternary alloys in Fig.~\ref{fig:quaternary_exs_plots}. As the matrix-phase concentration $c^e_{\xi m}$ is increased, TN alloys (in {I}--{III}) exhibit constant values of $\tilde{\Gamma}^{(1,2)}_\xi = 0$ and $\tilde{\gamma}=\tilde{\gamma}_\text{B}$. TS alloys (in {V}--{VIII}) exhibit a positive and increasing $\tilde{\Gamma}^{(1,2)}_\xi$, with a correspondingly decreasing $\tilde{\gamma}$ from that of $\tilde{\gamma}_\text{B}$. The model parameters used for the reference alloys here were inspired by our earlier works on phase boundary segregation in binary \citep{kadambi2020phase} and ternary \citep{kadambi2020acta} systems. \citet{kadambi2020acta} rigorously demonstrated the validity of the current phase-field approach with the Gibbs adsorption isotherm $d\tilde{\gamma} = -\tilde{\Gamma}^{(1,2)}_\xi d\mu^e_\xi$. 
% exhibits an invariant value along the ternaries, which is equal to that of the binary base alloy: in accordance with the generalized Gibbs adsorption isotherm,
% $\tilde{\gamma}(\tilde{\Gamma}^{(1,2)}_3=0,\tilde{\Gamma}^{(1,2)}_4=0)=\tilde{\gamma}_{\text{B}}$.

\begin{table*}[!htp]
\centering
\caption{Parametric cases for various solute segregation behaviors: regular solution parameters $L$ for solutes $3$ and $4$ characteristic to the bulk ($m/p$) and interfacial ($i$) phases ($\theta=1,2$).} 
\label{table:quaternary_params}
\includegraphics[width=0.8\textwidth]{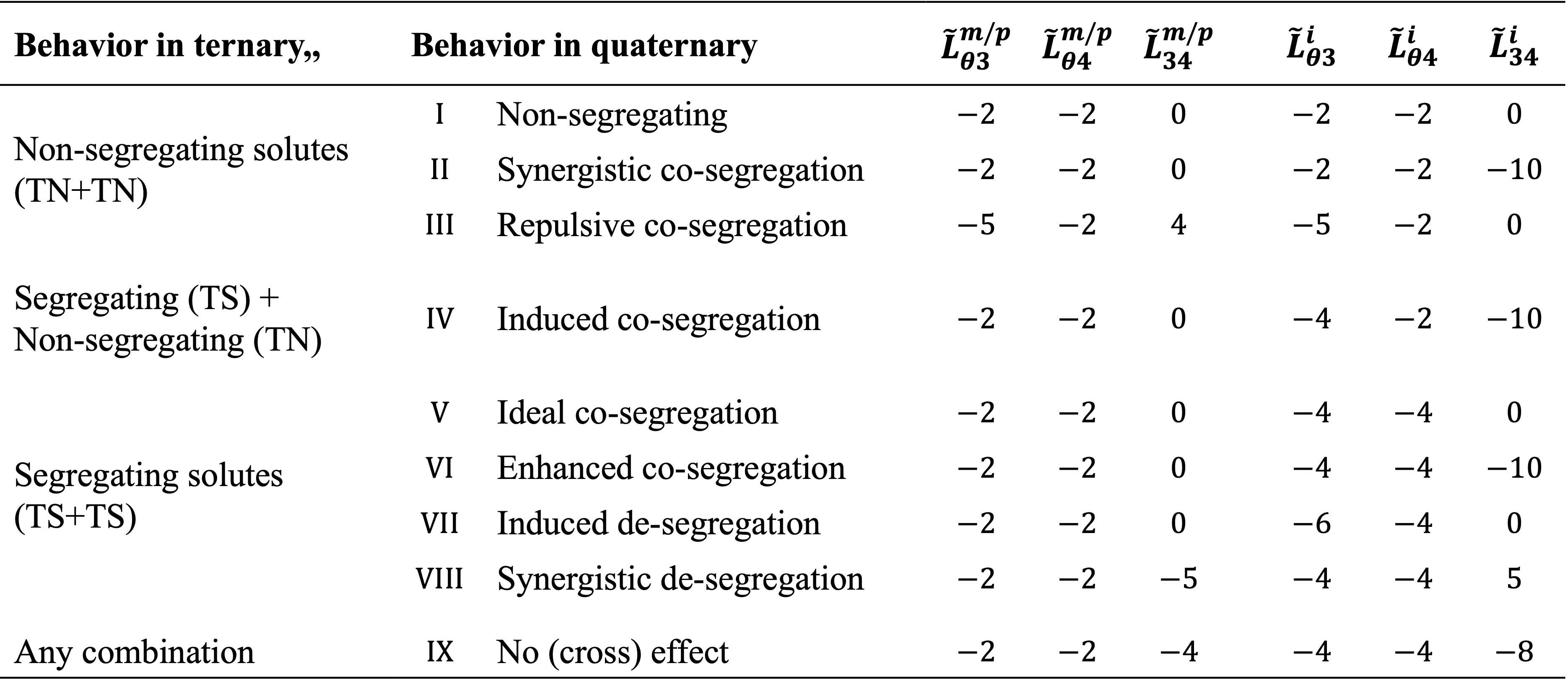}
\end{table*}

\subsection{Segregation in the Quaternary Alloy}

In the case studies presented below, we examine the effect of combined solute addition in the quaternary ($1$-$2$)-$3$-$4$ alloy. The interaction between solutes governed by $\tilde{L}^{m/p}_{34}$ and $\tilde{L}^{i}_{34}$ forms the basis for understanding the segregation behavior in ($1$-$2$)-$3$-$4$ relative to ($1$-$2$)-$3$ (TN or TS) and ($1$-$2$)-$4$ (TN or TS). The different cases examined here are labeled I through IX in Table~\ref{table:quaternary_params}. 

\subsubsection{Non-segregating} \label{subsec:base}

In case I, both the solutes are of non-segregating type in the ternaries (TN+TN). In the quaternary, they are mutually ideal/non-interacting in all the phases $\tilde{L}^{i/m/p}_{34}=0$ (Table~\ref{table:quaternary_params}). 
The concentration profiles (Fig.~\ref{fig:quaternary_profiles}\,I) demonstrate the non-segregating nature of $3$ and $4$ in the quaternary. The quantitative excess contour plots (Fig.~\ref{fig:quaternary_exs_plots}\,I) show a non-altered zero solute excess and IB energy energy, identical to the binary and ternary reference alloys. Alternatively, the solutes can possess a non-ideal interaction that is uniform across the interface and bulk, $\tilde{L}^{i}_{34}=\tilde{L}^{m/p}_{34}$. Even in these scenarios, the solutes are non-segregating in the quaternary and the excess quantities are not altered over the binaries and ternaries. 
Overall, the individual solutes do not possess the tendency to segregate even on combined addition due to ideal or non-preferential mutual interaction across the alloy.

\subsubsection{Synergistic Co-segregation}

In case II, both the solutes are non-segregating in the ternaries (TN+TN). However, they possess an attractive interaction in $i$ that is stronger than in $m/p$, i.e. $\tilde{L}^i_{34} < \tilde{L}^{m/p}_{34}$. 
The concentration profiles (Fig.~\ref{fig:quaternary_profiles}\,II) for the quaternary at $c^e_{3 m}=c^e_{4 m}=0.1$ demonstrate segregation of $3$ and $4$ at the diffuse IB. The quantitative excess contour plots (Fig.~\ref{fig:quaternary_exs_plots}\,II) show a positive and increasing solute excess on combined solute addition, $\tilde{\Gamma}^{(1,2)}_{3/4}(c^e_{3m},c^e_{4m})>0$. Correspondingly, the IB energy plot shows a decreasing contour $\tilde{\gamma}<\tilde{\gamma}_\text{B}$, with a steep decrease at relative large and equal matrix-phase solute concentrations, $c^e_{3m}$ and $c^e_{4m}$. 
The plot also shows that reduction of IB energy to very small magnitudes $\tilde{\gamma}\rightarrow0$ is possible. 
Overall, individual non-segregating solutes can strongly co-segregate on combined addition due to a favorable mutual interaction at the IB.

\subsubsection{Repulsive (Co-)Segregation} \label{subsec:rep_seg}

In case III, the solutes are non-segregating in their ternaries (TN+TN), with $3$ having a more favorable attractive interaction with the binary base alloy compared to $4$, i.e. $\tilde{L}^{m/p}_{\theta3}<\tilde{L}^{m/p}_{\theta4}$. Additionally, the solutes
have a repulsive mutual interaction in the bulk phases, relative to the IB, $\tilde{L}^{m/p}_{34}>0$ and $\tilde{L}^{m/p}_{34}>\tilde{L}^{i}_{34}$.
The concentration profiles (Fig.~\ref{fig:quaternary_profiles}\,III) for the quaternary at $c^e_{3 m}=c^e_{4 m}=0.2$ demonstrates segregation of $4$ at the diffuse IB, in preference to $3$. The quantitative excess contour plots (Fig.~\ref{fig:quaternary_exs_plots}\,III) show regions of positive solute excess on combined solute addition, with $\tilde{\Gamma}^{(1,2)}_{4}>\tilde{\Gamma}^{(1,2)}_{3}$ in general. The $\tilde{\Gamma}^{(1,2)}_{\xi}$ contour is asymmetric with respect to $(c^e_{3m},c^e_{4m})$ indicating that a lower $c^e_{\xi m}$ is needed to induce segregation of the other solute. In the presence of unequal solute concentrations, e.g. $(c^e_{3m},c^e_{4m}) = (0.2,0.05)$ or $(0.05,0.2)$, the minority solute in the matrix-phase segregates at the IB. The IB energy plot shows a decreasing contour, $\tilde{\gamma}<\tilde{\gamma}_\text{B}$. All the excess quantities for the considered scenario of $\tilde{L}^{i}_{34}=0$ show only a minor variation compared to that of case II; however, for case III scenarios with $\tilde{L}^{i}_{34}<0$, larger variation can be expected.
Overall, individually non-segregating solutes can exhibit segregation on combined addition due to mutual repulsion from the bulk.

\subsubsection{Induced Co-segregation} \label{subsec:ind_coseg}

Case IV is very similar to case II, with the solutes in the quaternary possessing a mutually attractive interaction in $i$ over $m/p$, i.e. $\tilde{L}^i_{34} < \tilde{L}^{m/p}_{34}$. However, in contrast to case II, solute $3$ is a segregating solute in the ternary, whereas solute $4$ is non-segregating (TS+TN).
The concentration profiles (Fig.~\ref{fig:quaternary_profiles}\,IV) for $c^e_{3 m}=c^e_{4 m}=0.1$ demonstrate a greater segregation of $3$ compared to $4$. The excess plots (Fig.~\ref{fig:quaternary_exs_plots}\,IV) show contours very similar to those of case II. 
% and provides a comprehensive variation across the quaternary composition space. 
However, due to $3$ being of TS type, $\tilde{\Gamma}^{(1,2)}_{3}$ demonstrates a positive segregation with $c^e_{3m}$ along the ternary axis $c^e_{4m}\rightarrow 0$. Correspondingly, $\tilde{\gamma}$ demonstrates an asymmetric variation with the two solutes.
In summary, an individually non-segregating solute can be induced to co-segregate in the presence of a segregating solute due to favorable mutual interaction at the IB,.

\subsubsection{Ideal Co-segregation} \label{subsec:co-seg_noint}

In case V, both the solutes are segregating in the ternaries (TS+TS), with $\tilde{L}^{i}_{\theta\xi}<\tilde{L}^{m/p}_{\theta\xi}$. In the quaternary, they are mutually ideal/non-interacting in all the phases $\tilde{L}^{i/m/p}_{34}=0$. (Alternatively, they may posses a non-deal interaction that is uniform across the phases, $\tilde{L}^{i}_{34}=\tilde{L}^{m/p}_{34}<0$). The concentration profiles (Fig.~\ref{fig:quaternary_profiles}{V}) for the quaternary at $c^e_{3 m}=c^e_{4 m}=0.1$ demonstrate co-segregation. We chose both the solutes to be identically segregating in the ternaries (case VII considers solutes that are non-identically segregating). 
The quantitative excess contour plots (Fig.~\ref{fig:quaternary_exs_plots}\,V) show regions of positive solute excess near the ternaries. The IB energy-composition map shows a planar contour with $\tilde{\gamma}$ decreasing from $\tilde{\gamma}_\text{B}$. The iso-$\tilde{\gamma}$ lines demonstrate a rule-of-mixtures behavior of quaternaries with reference to the ternaries. This behavior is represented by the red interconnected dot-star-dot marker in Fig.~\ref{fig:quaternary_exs_plots}\,V. Accordingly, the consolidated concentration profiles of the quaternary ($c^e_{3 m}+c^e_{4 m}$) was found to be identical to that of the ternary profiles evaluated at the same total amount of solute, i.e. $c^e_{\xi m}=0.2$. 
The reduced segregation observed at larger concentrations in both the ternaries and the quaternary is due to the contribution of entropy becoming significant to the overall free energy, relative to the contribution of enthalpy (Eq.~\ref{eq:RS_phase_FE}).

\begin{figure*}[!htp]
\captionsetup[subfigure]{justification=centering} % {labelformat=empty}
\centering
\hspace{0.5cm}
\begin{subfigure}{0.3\textwidth}
  \centering
  \includegraphics[width=1\linewidth,trim=2 2 2 2,clip]{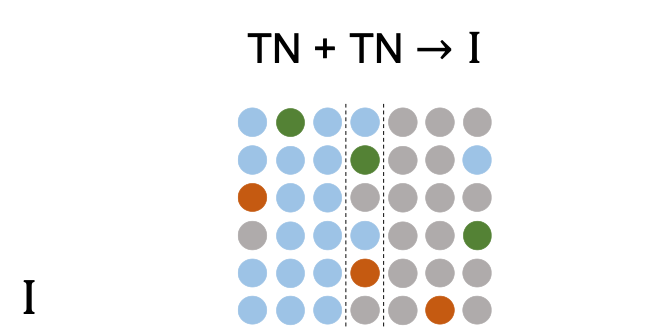}
%   \caption*{}
\end{subfigure}
\begin{subfigure}{0.3\textwidth}
  \centering
  \includegraphics[width=1\linewidth,trim=2 2 2 2,clip]{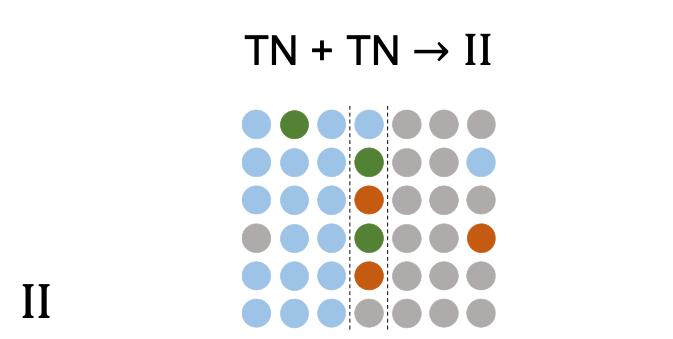}
%   \caption*{}
\end{subfigure}
\begin{subfigure}{0.3\textwidth}
  \centering
  \includegraphics[width=1\linewidth,trim=2 2 2 2,clip]{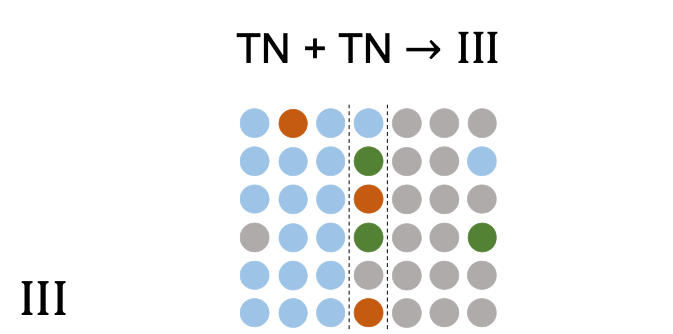}
%   \caption*{}
\end{subfigure}
\vspace{0.25cm}
\begin{subfigure}{0.3\textwidth}
  \centering
  \includegraphics[width=1\linewidth,trim=4 4 4 4,clip]{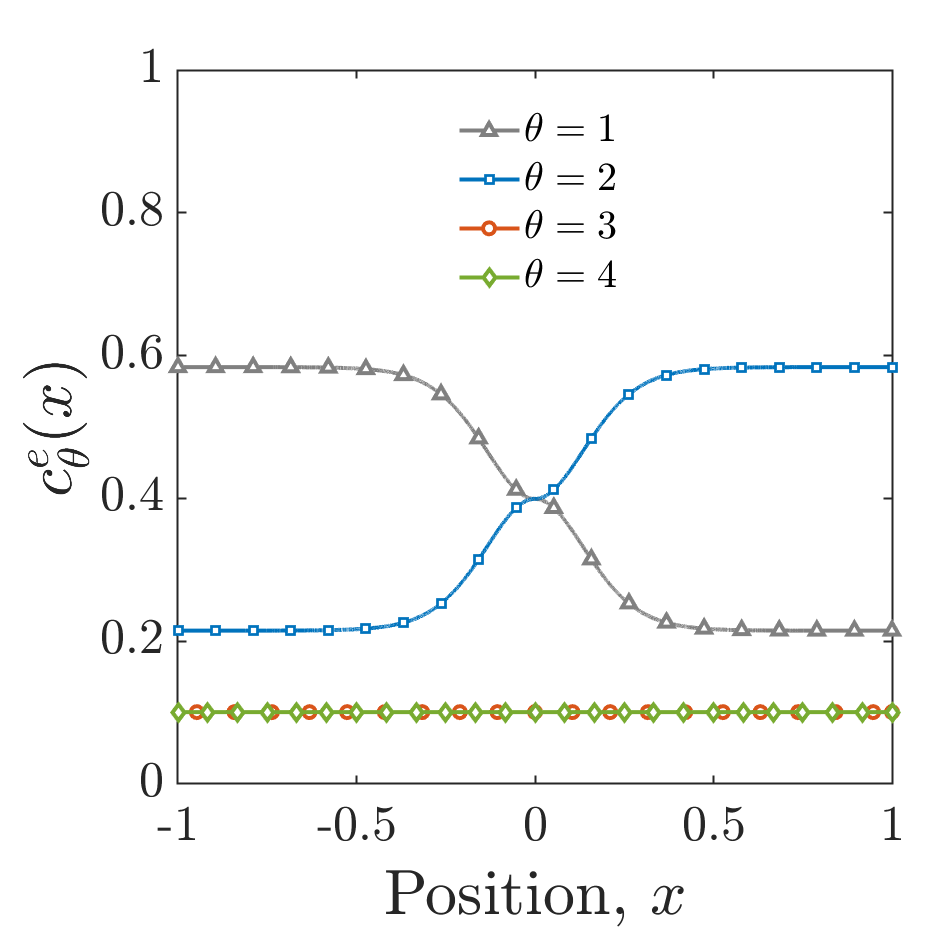}
%   \caption*{}
\end{subfigure}
\begin{subfigure}{0.3\textwidth}
  \centering
  \includegraphics[width=1\linewidth,trim=4 4 4 4,clip]{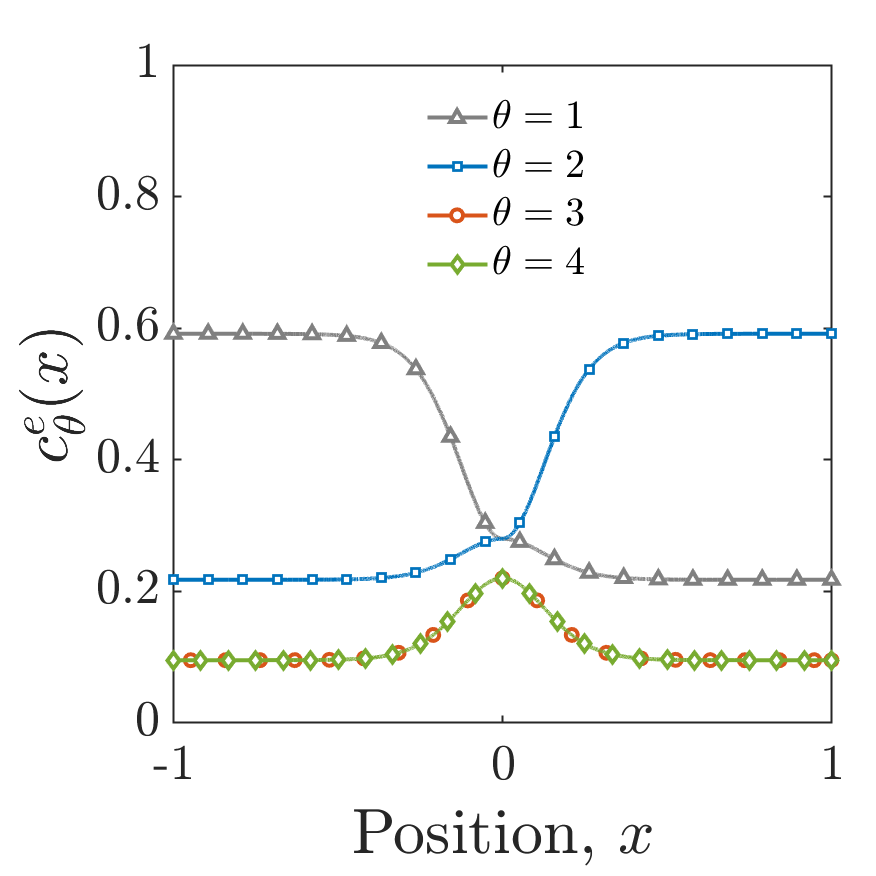}
%   \caption*{}
\end{subfigure}
\begin{subfigure}{0.3\textwidth}
  \centering
  \includegraphics[width=1\linewidth,trim=4 4 4 4,clip]{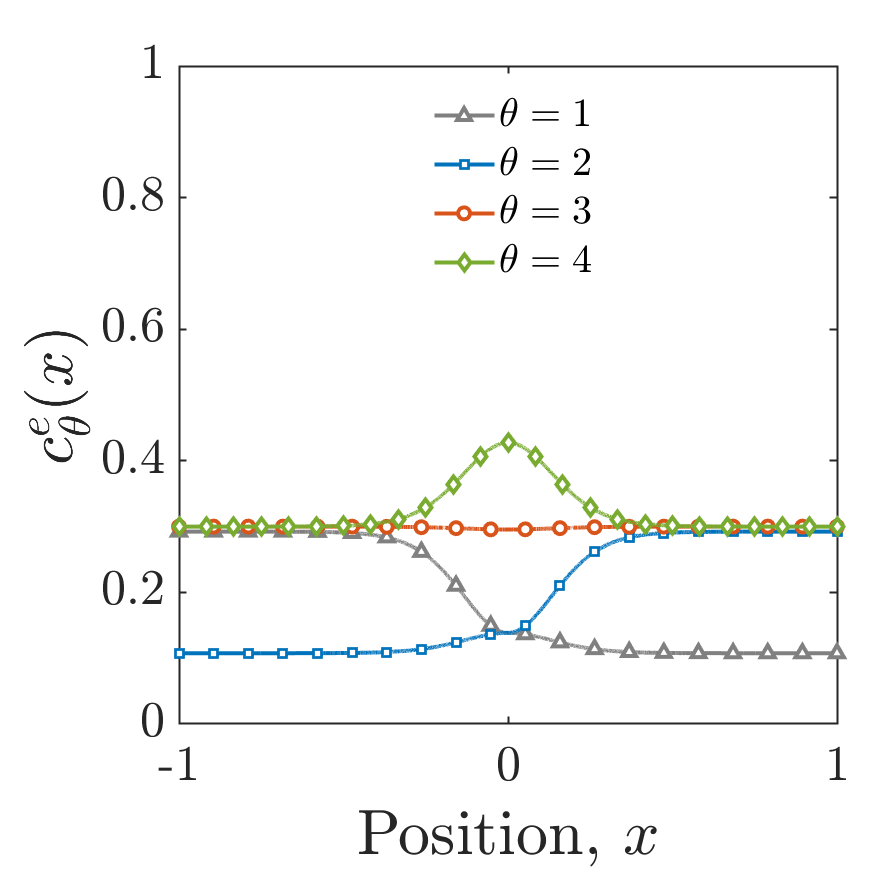}
\end{subfigure}
\begin{subfigure}{0.3\textwidth}
  \centering
  \includegraphics[width=1\linewidth,trim=2 2 2 2,clip]{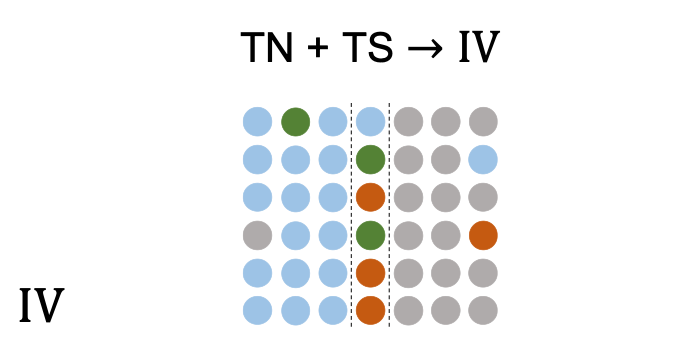}
%   \caption*{}
\end{subfigure}
% \hspace{2cm}
\begin{subfigure}{0.3\textwidth}
  \centering
  \includegraphics[width=1\linewidth,trim=2 2 2 2,clip]{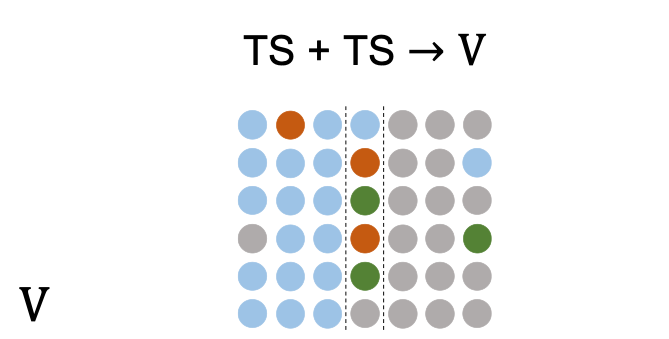}
%   \caption*{}
\end{subfigure}
\begin{subfigure}{0.3\textwidth}
  \centering
  \includegraphics[width=1\linewidth,trim=2 2 2 2,clip]{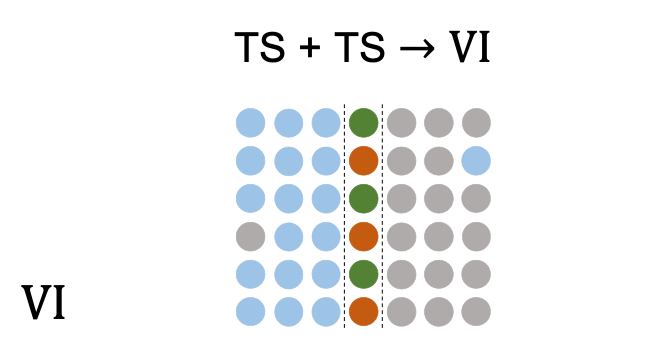}
%   \caption*{}
\end{subfigure}
% \newline
\begin{subfigure}{0.3\textwidth}
  \centering
  \includegraphics[width=1\linewidth,trim=4 4 4 4,clip]{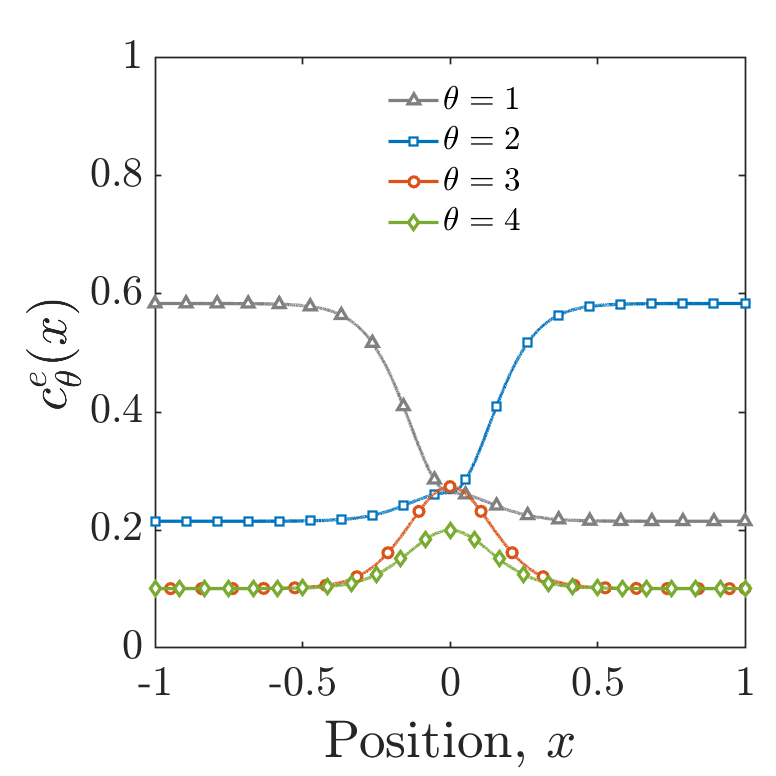}
%   \caption*{}
\end{subfigure}
% \hspace{2cm}
\begin{subfigure}{0.3\textwidth}
  \centering
  \includegraphics[width=1\linewidth,trim=4 4 4 4,clip]{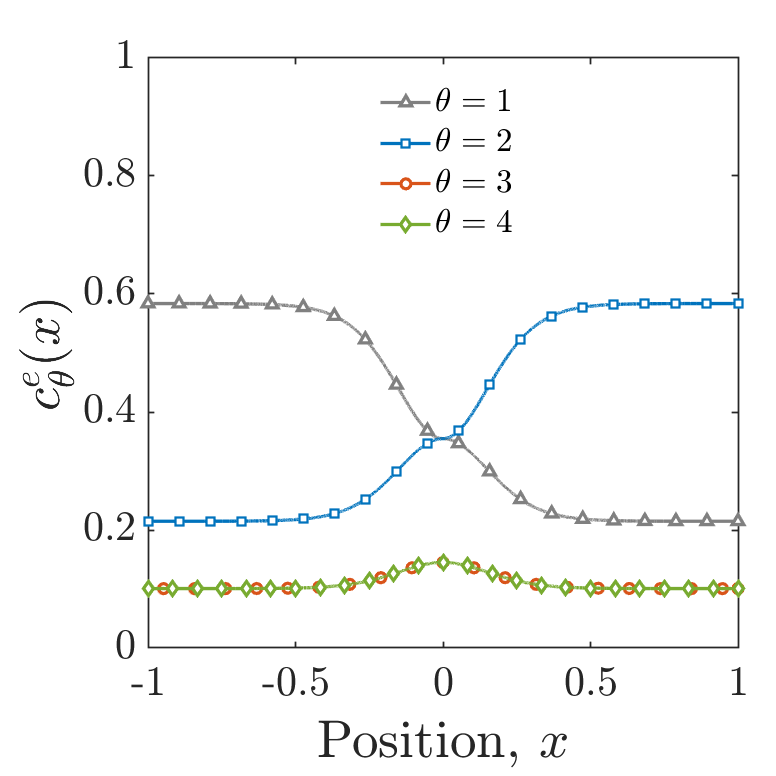}
%   \caption*{}
\end{subfigure}
\begin{subfigure}{0.3\textwidth}
  \centering
  \includegraphics[width=1\linewidth,trim=4 4 4 4,clip]{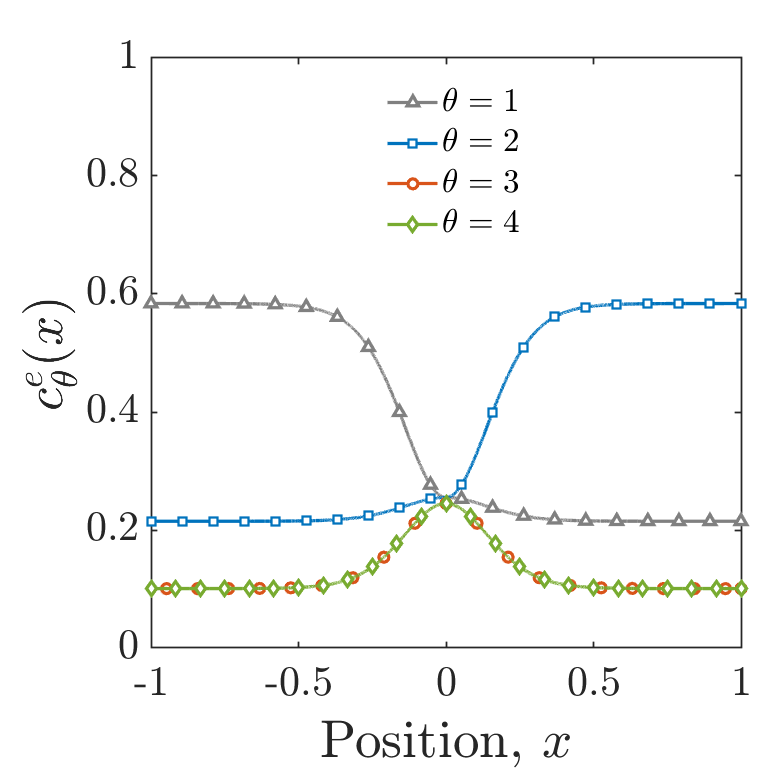}
%   \caption*{}
\end{subfigure}
\vspace{0.5cm}
\rulesep
\end{figure*}

\begin{figure*}[!htp]\ContinuedFloat
\captionsetup[subfigure]{justification=centering}
\centering
\hspace{0.5cm}
\begin{subfigure}{0.3\textwidth}
  \centering
  \includegraphics[width=1\linewidth,trim=2 2 2 2,clip]{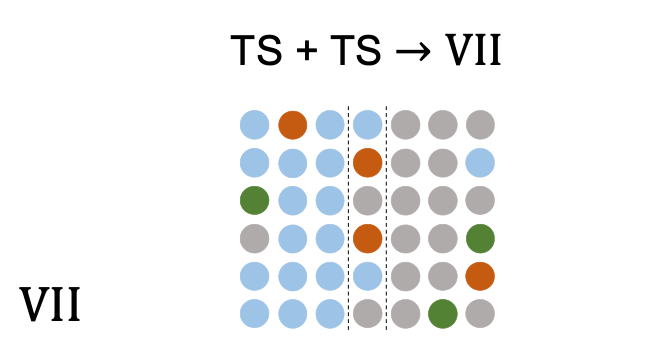}
%   \caption*{}
\end{subfigure}
\begin{subfigure}{0.3\textwidth}
  \centering
  \includegraphics[width=1\linewidth,trim=2 2 2 2,clip]{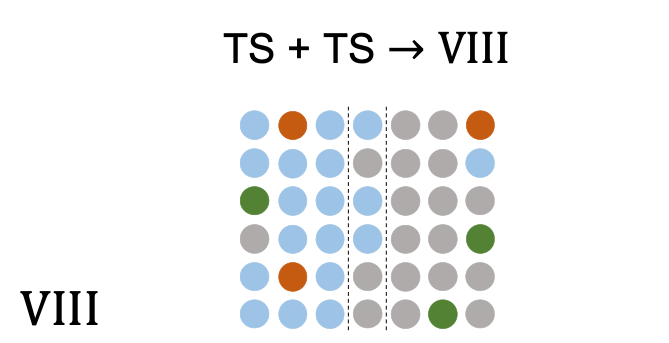}
%   \caption*{}
\end{subfigure}
\begin{subfigure}{0.3\textwidth}
  \centering
  \includegraphics[width=1\linewidth,trim=2 2 2 2,clip]{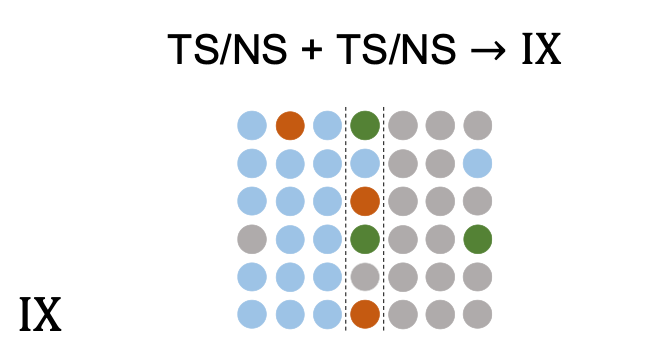}
%   \caption*{}
\end{subfigure}
\begin{subfigure}{0.3\textwidth}
  \centering
  \includegraphics[width=1\linewidth,trim=4 4 4 4,clip]{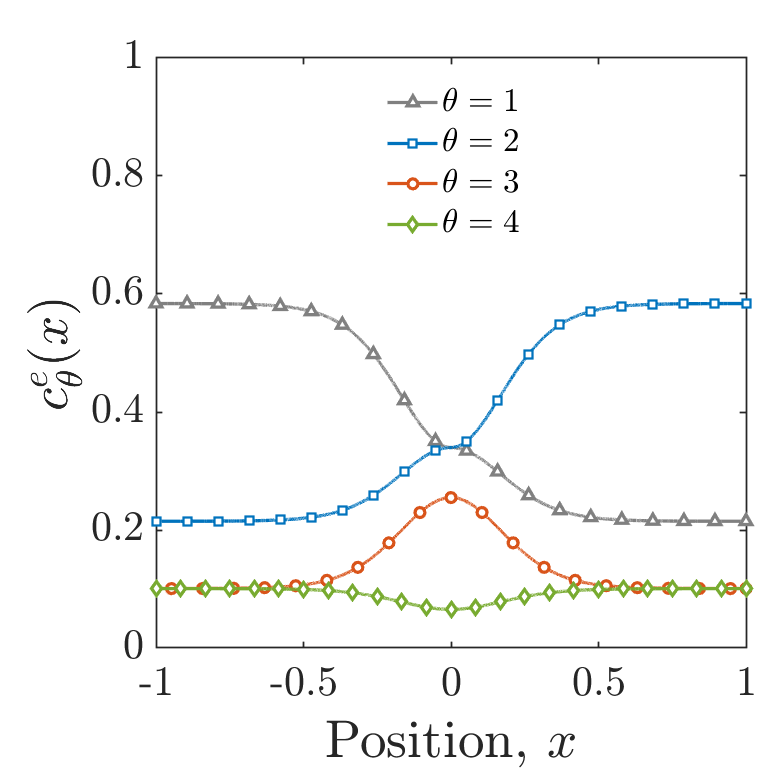}
%   \caption*{}
\end{subfigure}
\begin{subfigure}{0.3\textwidth}
  \centering
  \includegraphics[width=1\linewidth,trim=4 4 4 4,clip]{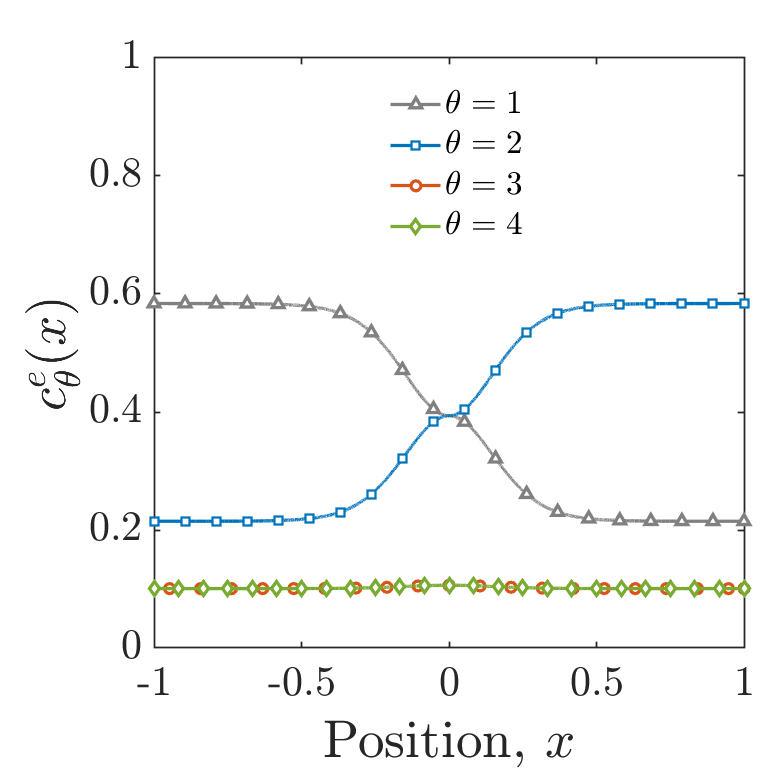}
%   \caption*{}
\end{subfigure}
\begin{subfigure}{0.3\textwidth}
  \centering
  \includegraphics[width=1\linewidth,trim=4 4 4 4,clip]{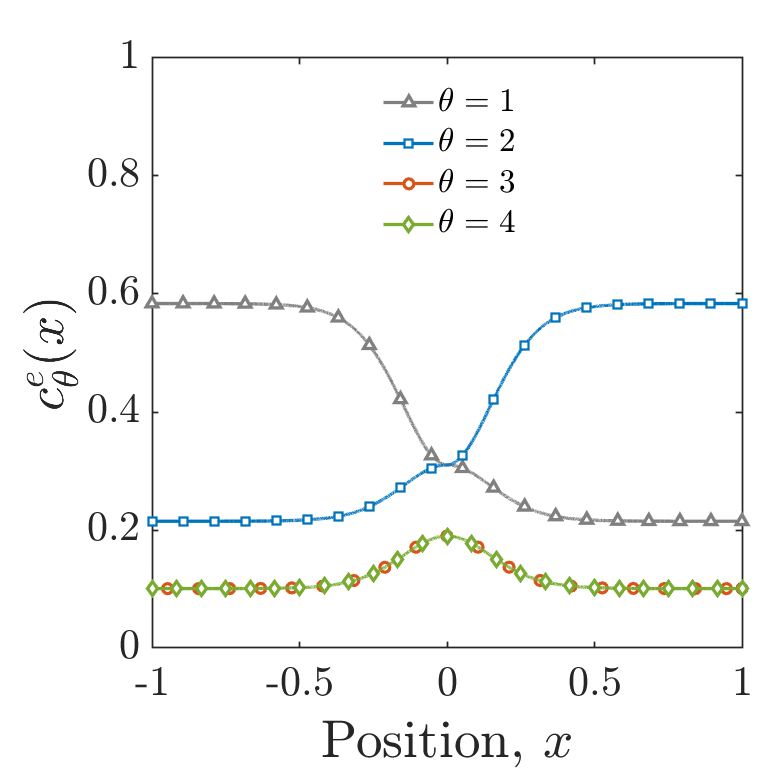}
%   \caption*{}
\end{subfigure}
\caption{Elemental distribution across the two-phase interface in quaternary alloys for parametric cases I-IX listed in Table~\ref{table:quaternary_params}. Top figure in each case: representative schematic of the two bulk phases and the interface layer. Bottom figure in each case: equilibrium concentration profile across a 1D, planar interface obtained from the phase-field model parameterized using Table~\ref{table:quaternary_params}. Color legend: grey is atom $1$, blue is $2$, red is $3$, green is $4$.}
\label{fig:quaternary_profiles}
\end{figure*}

\subsubsection{Enhanced Co-segregation} \label{subsec:coseg_attrint}

Case VI is similar to cases II and IV, with the solutes having a mutually attractive interaction in $i$ over $m/p$. However, both the solutes are segregating in their respective ternaries (TS+TS), as opposed to (TN+TN) or (TN+TS). The concentration profiles (Fig.~\ref{fig:quaternary_profiles}\,IV) for $c^e_{3 m}=c^e_{4 m}=0.1$ demonstrate a greater segregation of the solutes compared to that in the ternaries. The quantitative excess contour plots exhibit contours very similar to cases II and IV. Symmetric contours with respect to solute are observed due to the simplified choice of identically strong segregation of $3$ and $4$.
Overall, individually segregating solutes exhibit enhanced segregation in the quaternary due to the more favorable mutual interaction at the IB.

\subsubsection{Site Competition/Induced Desegregation} \label{subsec:site_comp}

Case VII involves segregating solutes in the ternaries (TS+TS) with $3$ being a stronger segregating species compared to $4$. As in case V, they are chosen to be mutually non-interacting or uniformly interacting in the quaternary. 
In the concentration profiles (Fig.~\ref{fig:quaternary_profiles}\,VII) obtained for $c^e_{3 m}=c^e_{4 m}=0.1$, solute $3$ is enriched at the diffuse IB, while $4$ is depleted. 
The quantitative excess contour plots (Fig.~\ref{fig:quaternary_exs_plots}\,VII) show large compositional regions of $\tilde{\Gamma}^{(1,2)}_{3}>0$ and $\tilde{\Gamma}^{(1,2)}_{4}\leq0$. Starting from the ternary and moving towards the quaternary (red arrow in the figure), solute $4$ desegregates readily even on small addition of solute $3$. The IB energy map shows a greater reduction in $\tilde{\gamma}$ with addition of $3$ than with $4$. Contrary to  cases V and VI, equal additions of $3$ and $4$ is ineffective in reducing $\tilde{\gamma}$.
In summary, a strongly segregating solute is favored to preferentially occupy IB sites over a weakly segregating solute, inducing desegregation of the later in the quaternaries.

\subsubsection{Synergistic Desegregation} \label{subsec:syn_deseg}

In case VIII, both the solutes are identically segregating in the ternaries (TS+TS). However, they have a repulsive mutual interaction at the IB ($\tilde{L}^{i}_{34}>0$) in the quaternary. In addition, they have an attractive interaction in the bulk ($\tilde{L}^{i}_{34}<0$). The concentration profiles (Fig.~\ref{fig:quaternary_profiles}\,VIII) evaluated at $c^e_{3 m}=c^e_{4 m}=0.1$ show negligible segregation of the the solutes at the diffuse IB. The quantitative excess contour plots (Fig.~\ref{fig:quaternary_exs_plots}\,VIII) exhibit a transition from  $\tilde{\Gamma}^{(1,2)}_{\xi}>0$ to $\tilde{\Gamma}^{(1,2)}_{\xi}<0$ on adding the quartary solute ($3$ or $4$) to the ternary alloy with $\xi=4$ or $3$. The IB energy map demonstrates a saddle point (marked by the red star) on combined solute addition $c^e_{3 m}=c^e_{4 m}=0.1$. Beyond the saddle point, combined solute addition leads to an increase $\tilde{\gamma}$. A minimum in $\tilde{\gamma}$ is preferred only along the ternary compositions. If non-identically segregating solutes are chosen for the case study, we would observe the saddle point in $\tilde{\gamma}$ to shift closer to the weakly segregating ternary. Moreover, the segregation region $\tilde{\Gamma}^{(1,2)}_{\xi}>0$ of the strongly segregating solute would be enhanced, while that of the weakly segregating solute would be diminished, thus altering the relative solute concentration at which transition to desegregation is observed. In summary, case VIII demonstrates a synergistic desegregation of both the segregating solutes due to mutually repulsive interaction between the solutes at the IB.

\subsubsection{No Cross Effect on Segregation}

In case IX, the solutes can either be segregating or non-segregating in the ternaries (TS/TN+TS/TN). Segregating solutes are chosen for the purpose of illustration. In all the previous cases, the coupling between solutes $3$ and $4$ in the quaternaries gave rise to concentration profiles that are distinct from the corresponding ternaries at a given $c^e_{\xi m}$. Even for the ideal or non-interacting case V (Fig.~\ref{fig:quaternary_profiles}\,V) with $\tilde{L}^{m/p/i}_{34}=0$, the amount of solute $3$ at the IB was altered from that in its ternary (Fig.~\ref{fig:base_profiles}\,TS), for $c^e_{3m}=0.1$. This coupling arises from the contribution of cross terms $\tilde{L}^\psi_{34} - \tilde{L}^\psi_{\theta3} - \tilde{L}^\psi_{\theta4}$ ($\theta=1,2$) to the segregation isotherm (Eq.~\ref{eq:seg_isotherm_RS}) \citep{guttmann1979interfacial}. For solutes that are non-interacting---mutually and with the base components $1$ and $2$---the aforementioned terms vanish. Moreover, even for mutually interacting solutes, we can find combinations of interaction parameters that lead to $\tilde{L}^\psi_{34} - \tilde{L}^\psi_{\theta3} - \tilde{L}^\psi_{\theta4}=0$ (as in IX of Table~\ref{table:quaternary_params}). The resulting solute concentration profiles are identical to that in their corresponding ternaries for a given $c^e_{\xi m}$.

\begin{figure*}[!htp]
\captionsetup[subfigure]{justification=centering}
\centering
\begin{subfigure}{0.3\textwidth}
  \centering
  \includegraphics[width=1\linewidth,trim=4 4 4 4,clip]{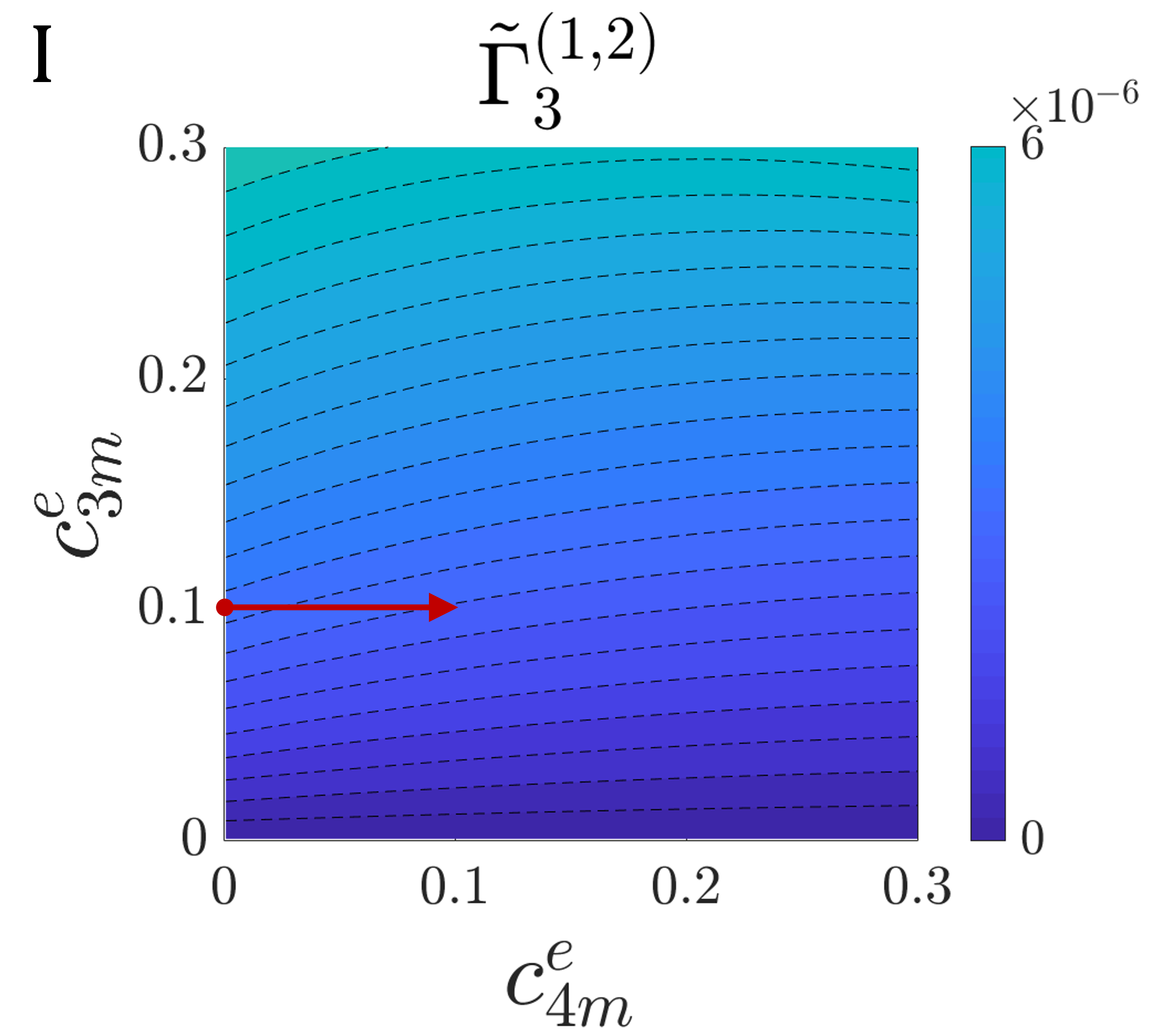}
%   \caption*{}
\end{subfigure}
\begin{subfigure}{0.3\textwidth}
  \centering
  \includegraphics[width=1\linewidth,trim=4 4 4 4,clip]{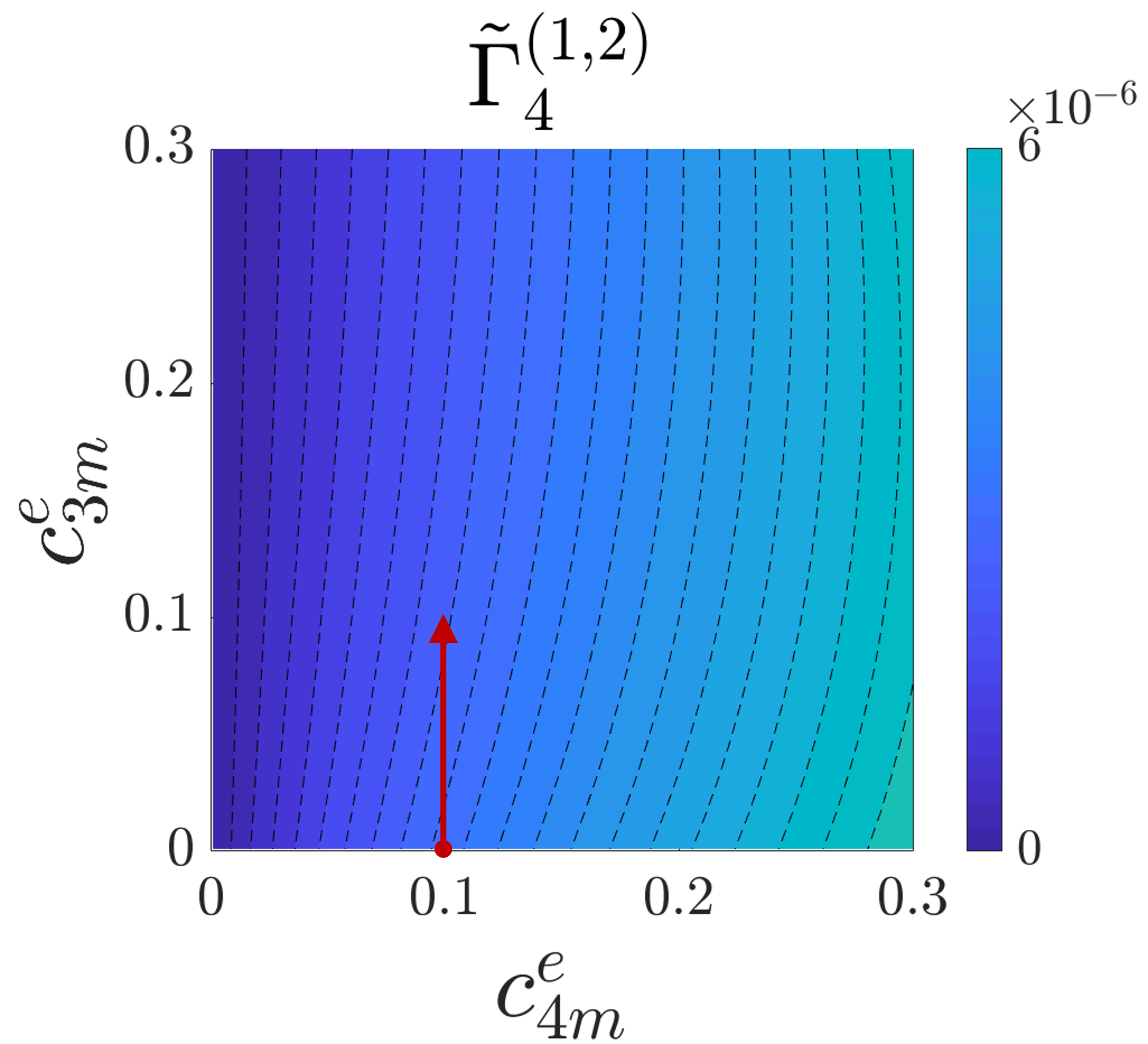}
%   \caption*{}
\end{subfigure}
\begin{subfigure}{0.3\textwidth}
  \centering
  \includegraphics[width=1\linewidth,trim=4 4 4 4,clip]{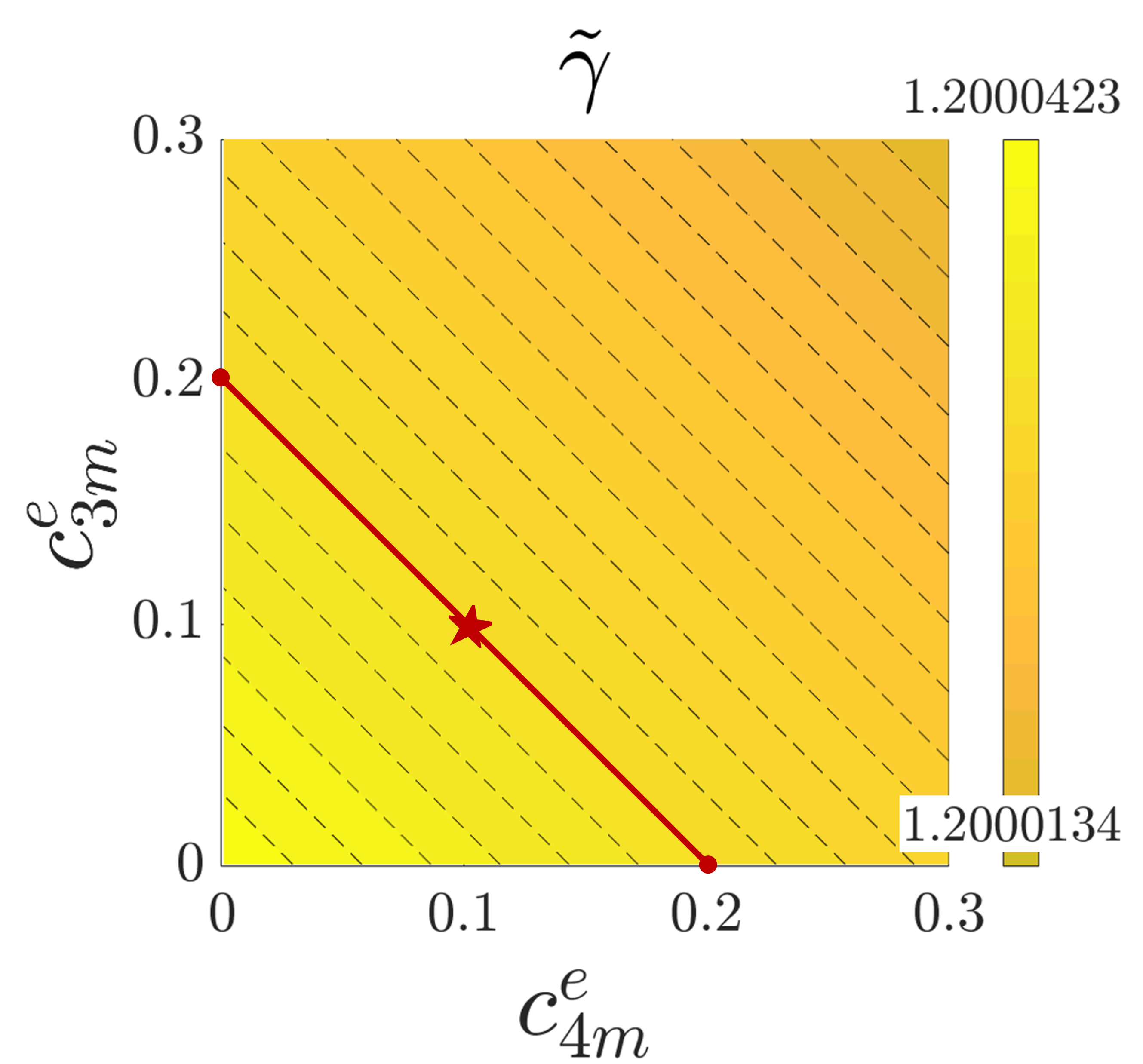}
%   \caption*{}
\end{subfigure}
\vspace{0.25cm}
\begin{subfigure}{0.3\textwidth}
  \centering
  \includegraphics[width=1\linewidth,trim=4 4 4 4,clip]{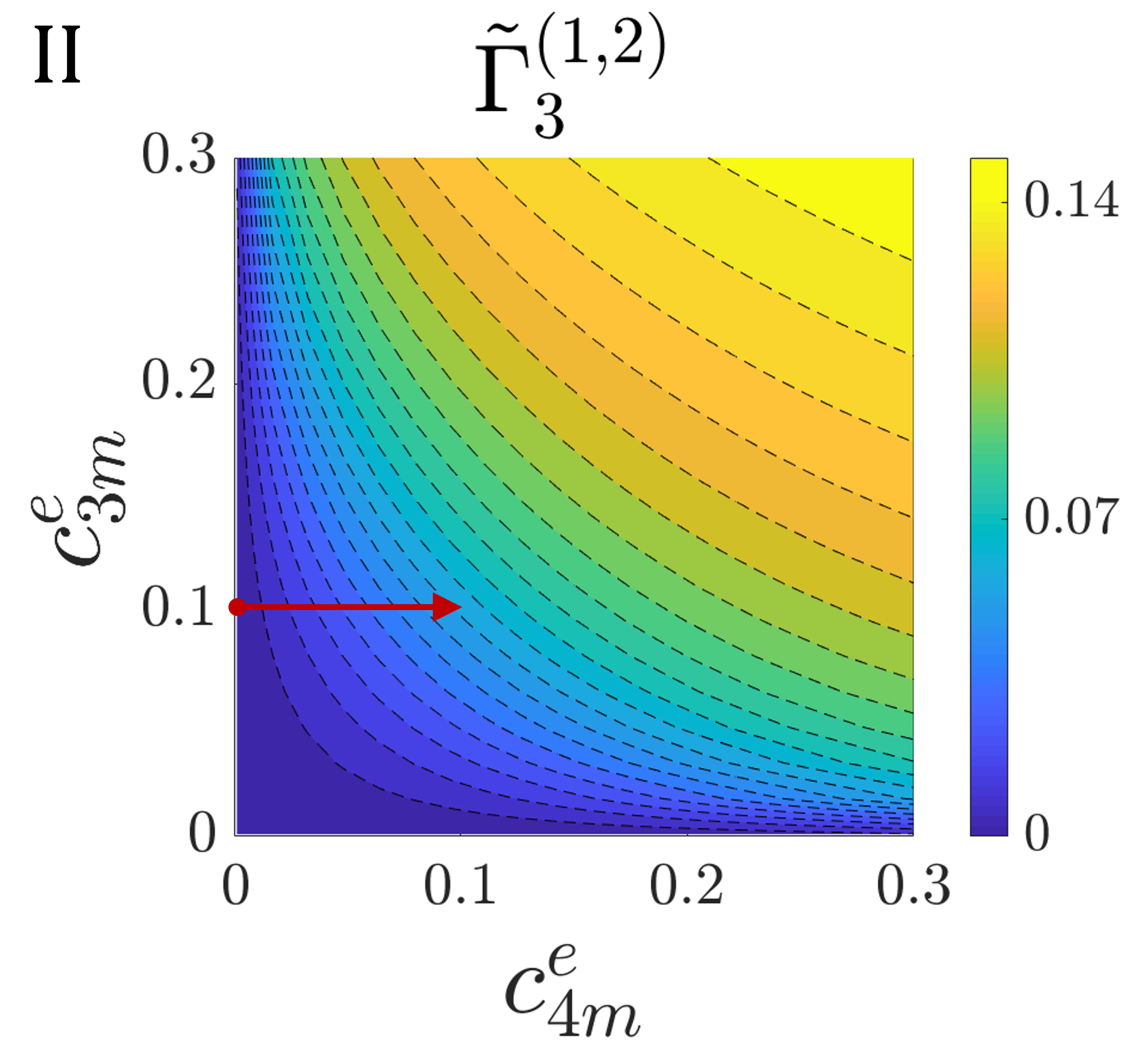}
%   \caption*{}
\end{subfigure}
\begin{subfigure}{0.3\textwidth}
  \centering
  \includegraphics[width=1\linewidth,trim=4 4 4 4,clip]{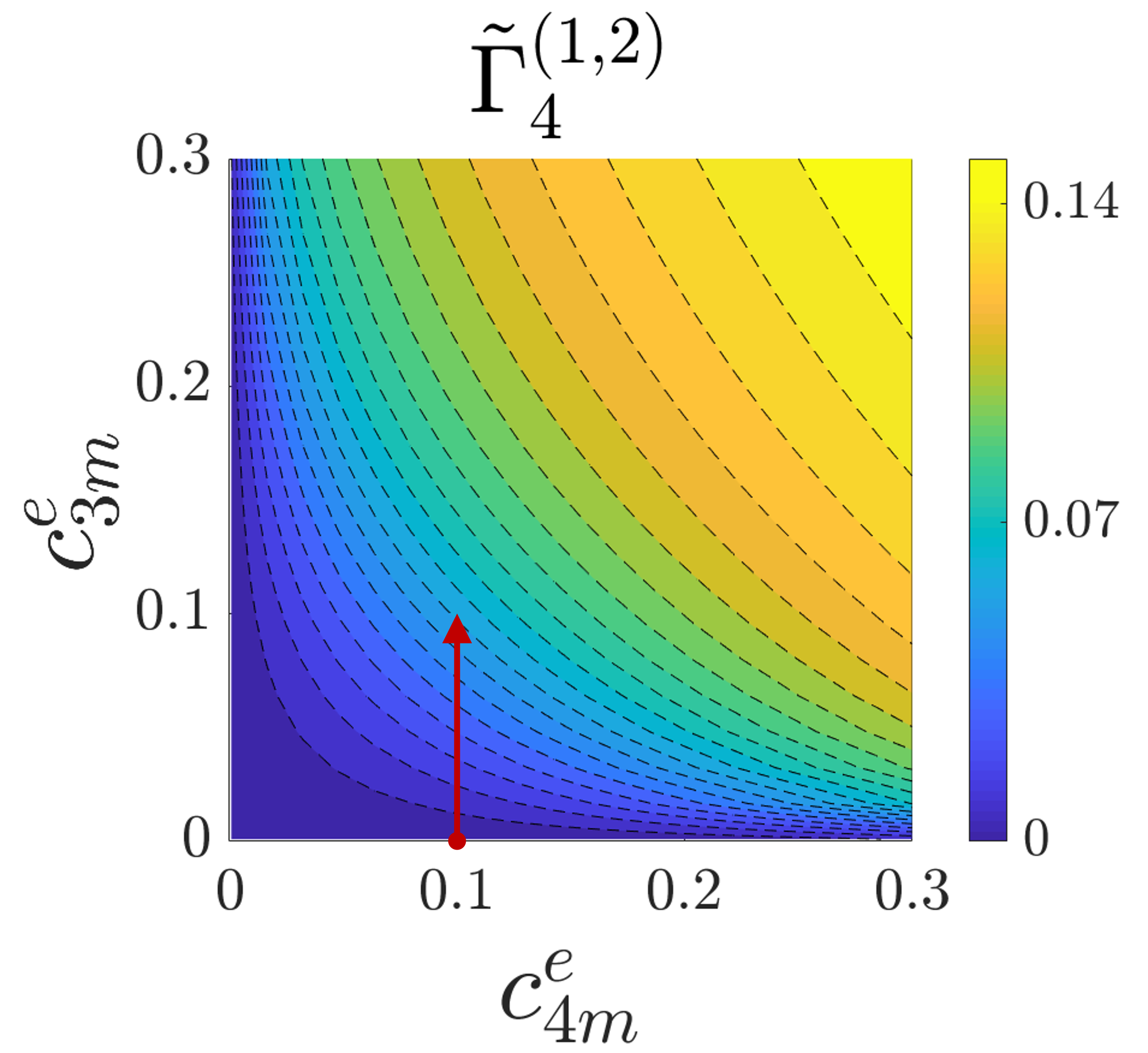}
%   \caption*{}
\end{subfigure}
\begin{subfigure}{0.3\textwidth}
  \centering
  \includegraphics[width=1\linewidth,trim=4 4 4 4,clip]{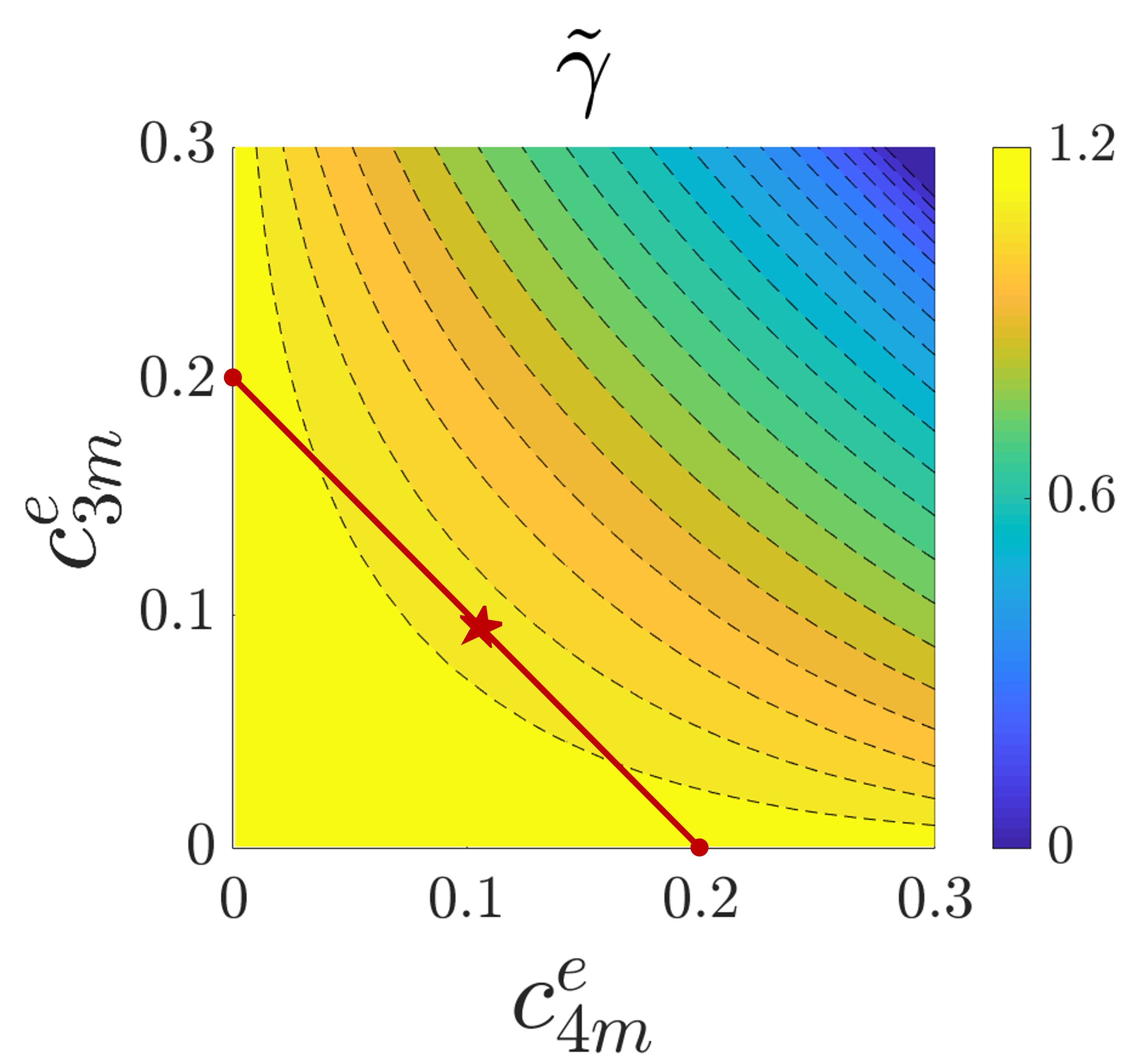}
%   \caption*{}
\end{subfigure}
\vspace{0.25cm}
\begin{subfigure}{0.3\textwidth}
  \centering
  \includegraphics[width=1\linewidth,trim=4 4 4 4,clip]{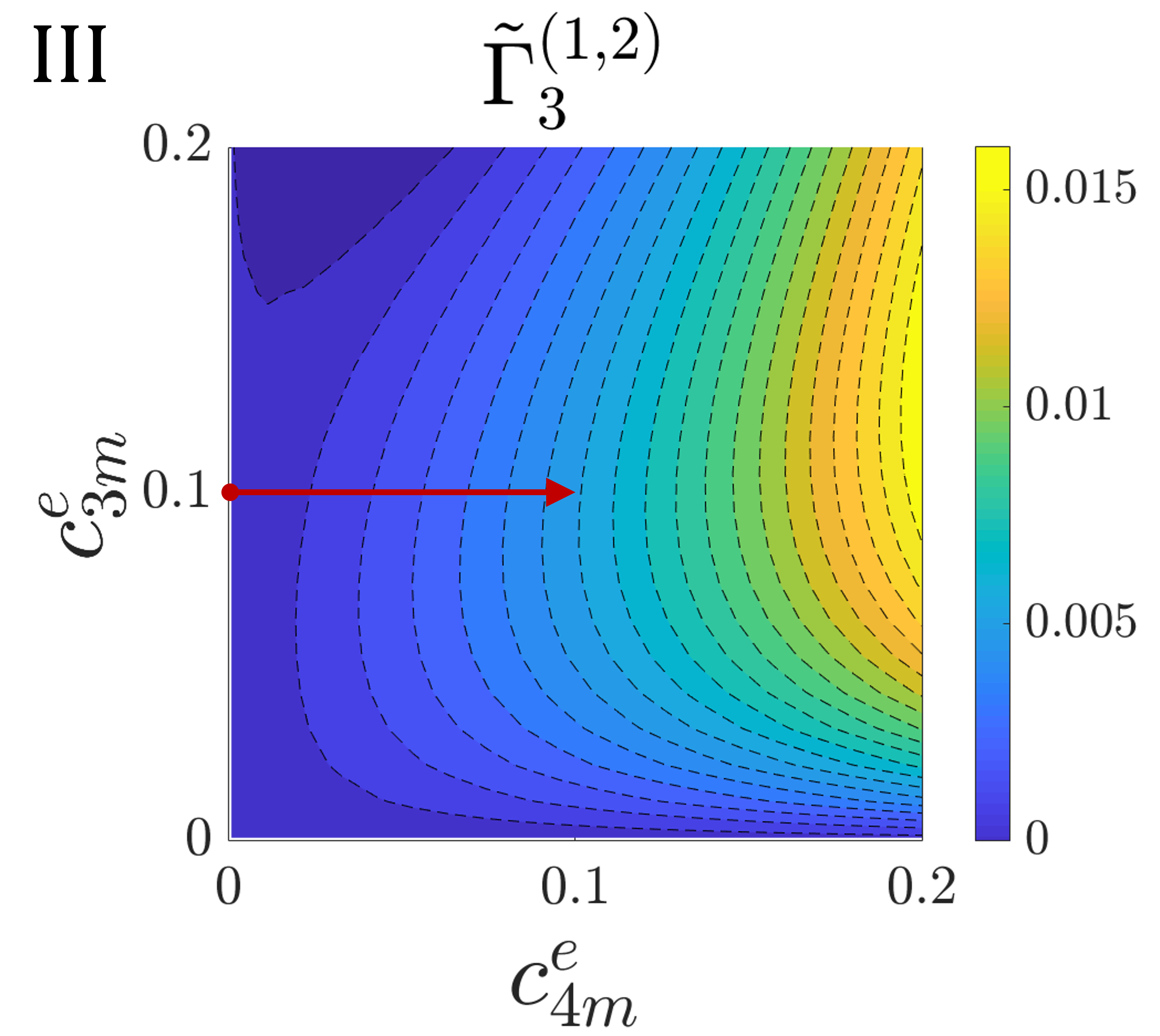}
%   \caption*{}
\end{subfigure}
\begin{subfigure}{0.3\textwidth}
  \centering
  \includegraphics[width=1\linewidth,trim=4 4 4 4,clip]{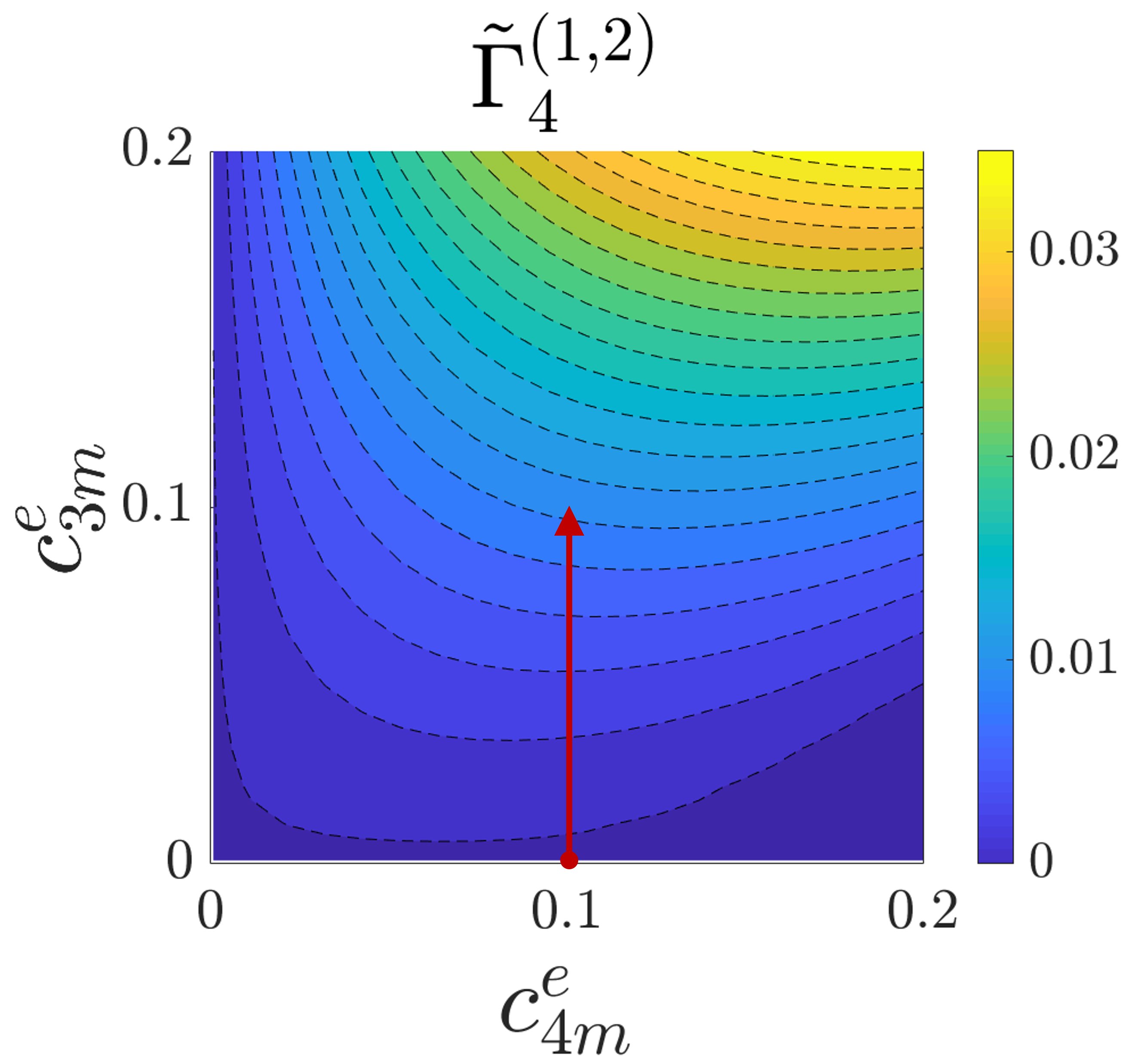}
%   \caption*{}
\end{subfigure}
\begin{subfigure}{0.3\textwidth}
  \centering
  \includegraphics[width=1\linewidth,trim=4 4 4 4,clip]{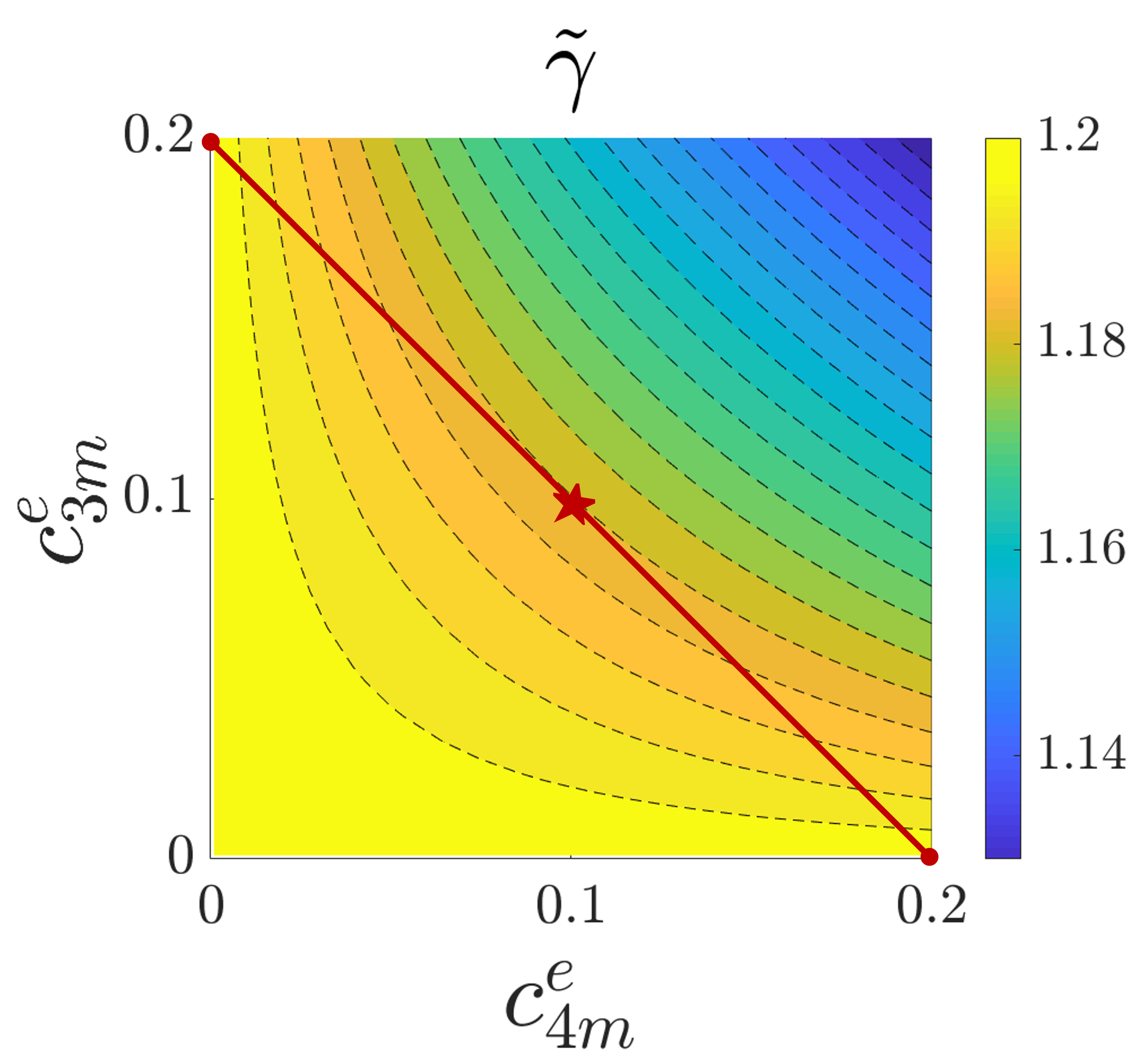}
%   \caption*{}
\end{subfigure}
\vspace{0.25cm}
% \rulesep
\begin{subfigure}{0.3\textwidth}
  \centering
  \includegraphics[width=1\linewidth,trim=4 4 4 4,clip]{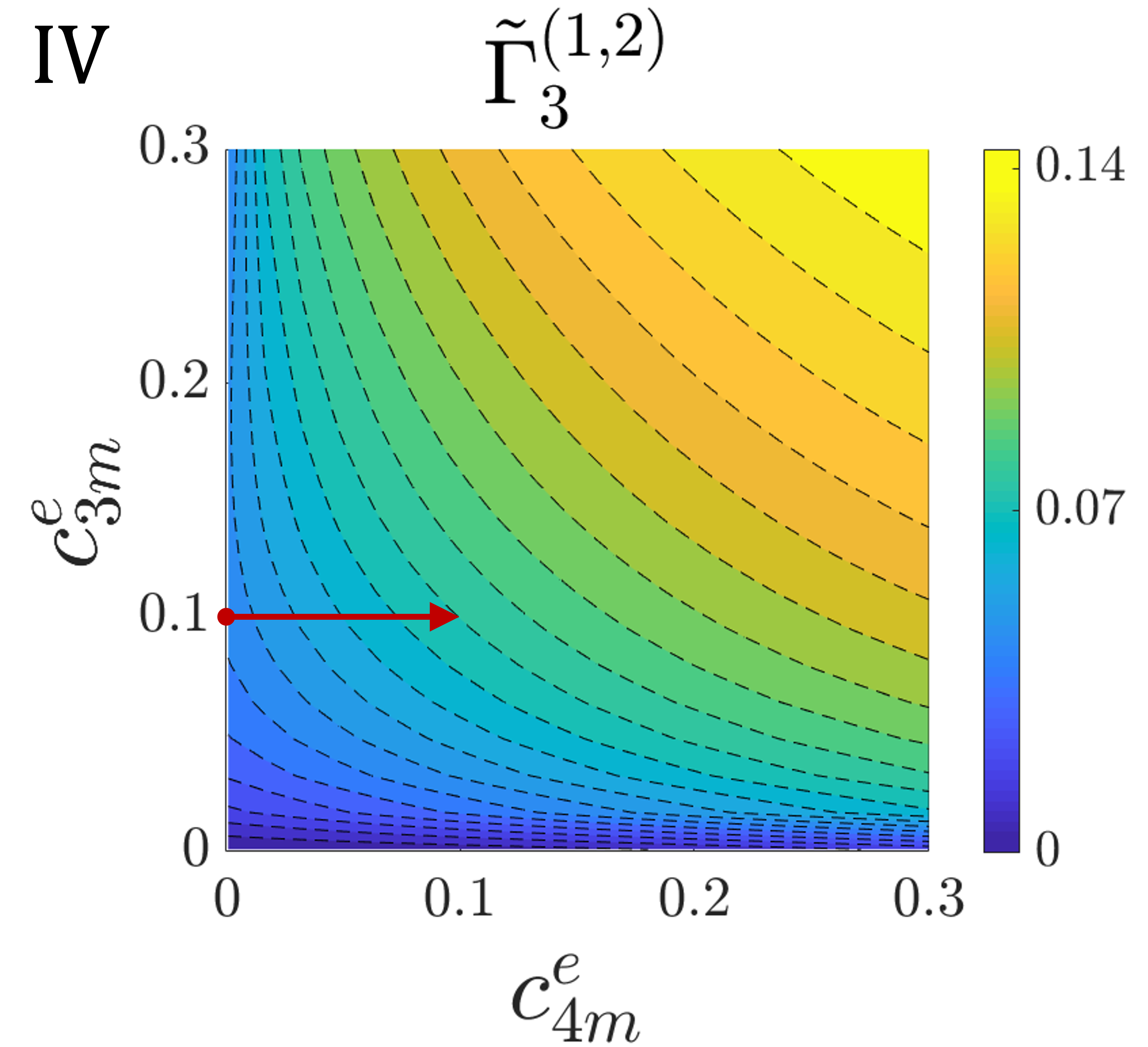}
%   \caption*{}
\end{subfigure}
\begin{subfigure}{0.3\textwidth}
  \centering
  \includegraphics[width=1\linewidth,trim=4 4 4 4,clip]{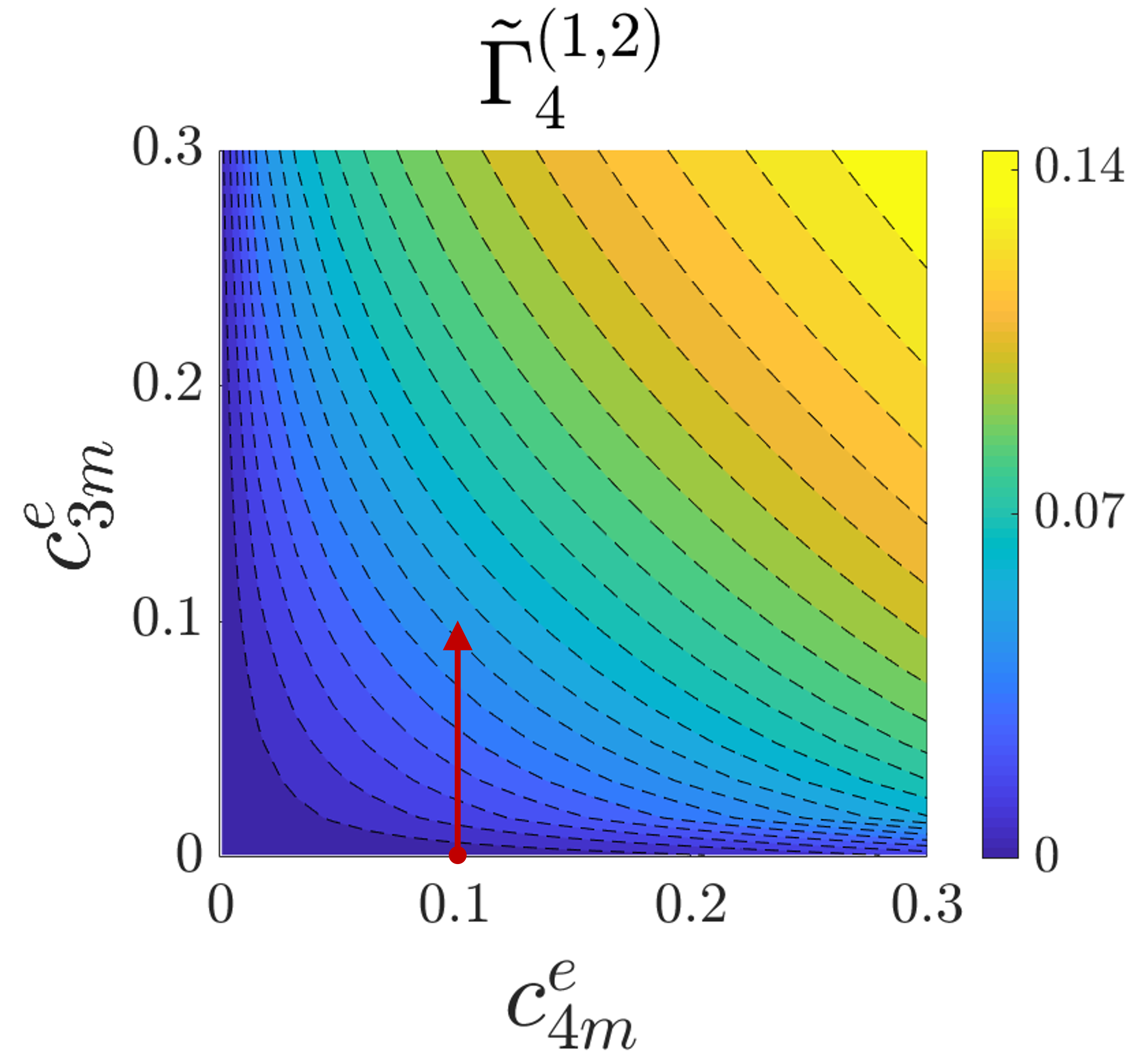}
%   \caption*{}
\end{subfigure}
\begin{subfigure}{0.3\textwidth}
  \centering
  \includegraphics[width=1\linewidth,trim=4 4 4 4,clip]{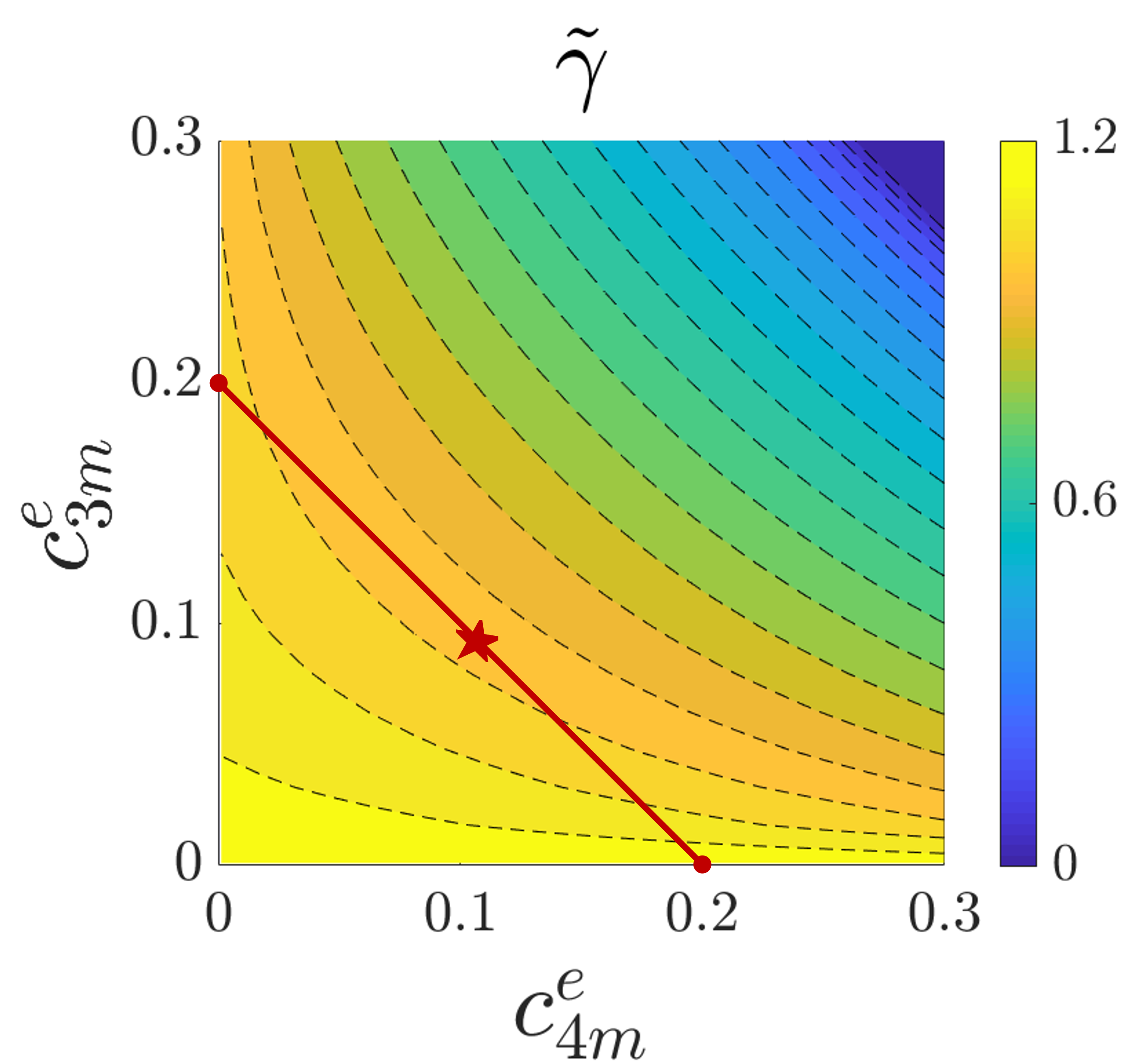}
%   \caption*{}
\end{subfigure}
\vspace{0.5cm}
\rulesep
\end{figure*}

\begin{figure*}[!htp]\ContinuedFloat
% \begin{figure*}[!htp]
\captionsetup[subfigure]{justification=centering}
\centering
\begin{subfigure}{0.3\textwidth}
  \centering
  \includegraphics[width=1\linewidth,trim=4 4 4 4,clip]{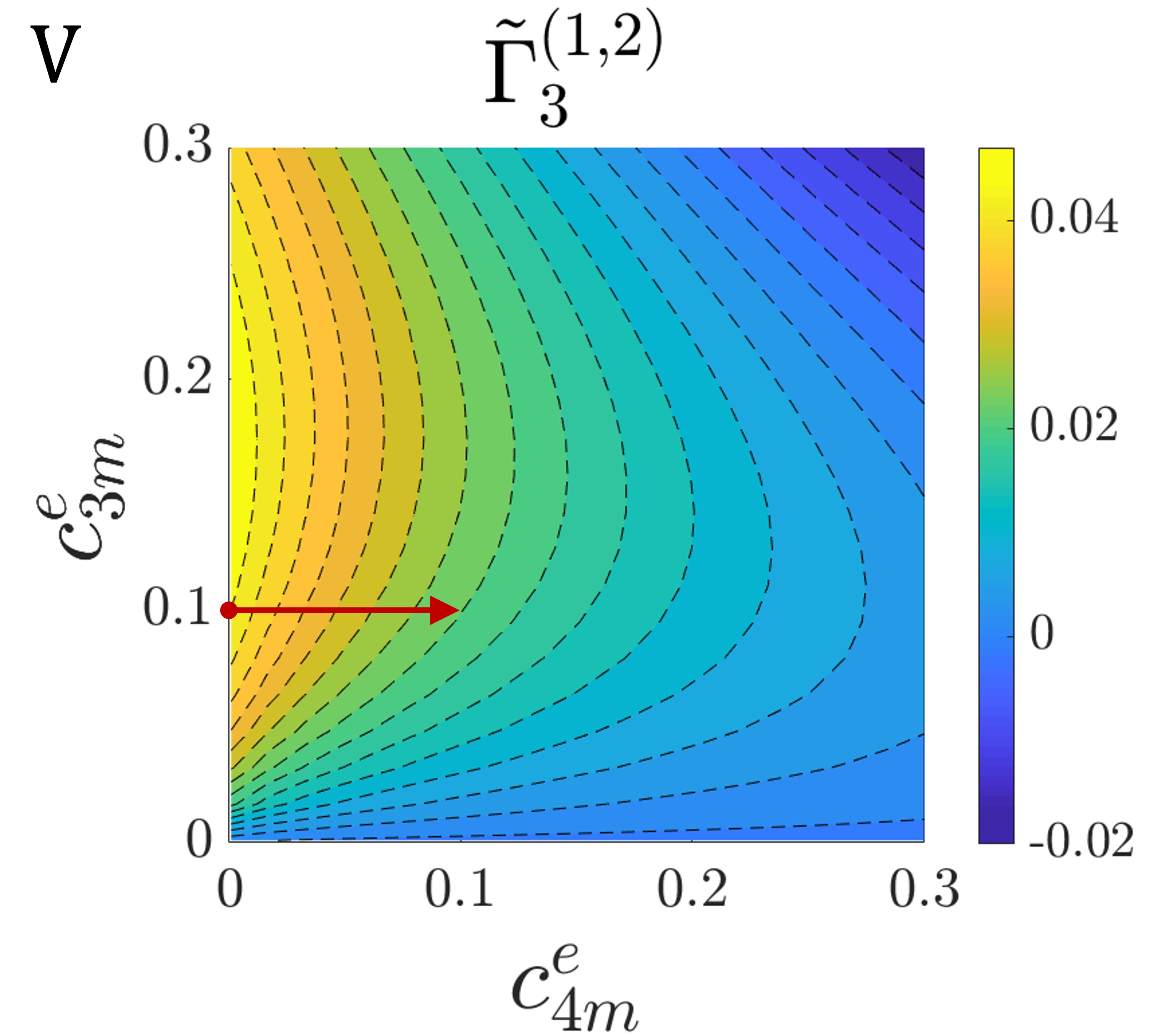}
%   \caption*{}
\end{subfigure}
\begin{subfigure}{0.3\textwidth}
  \centering
  \includegraphics[width=1\linewidth,trim=4 4 4 4,clip]{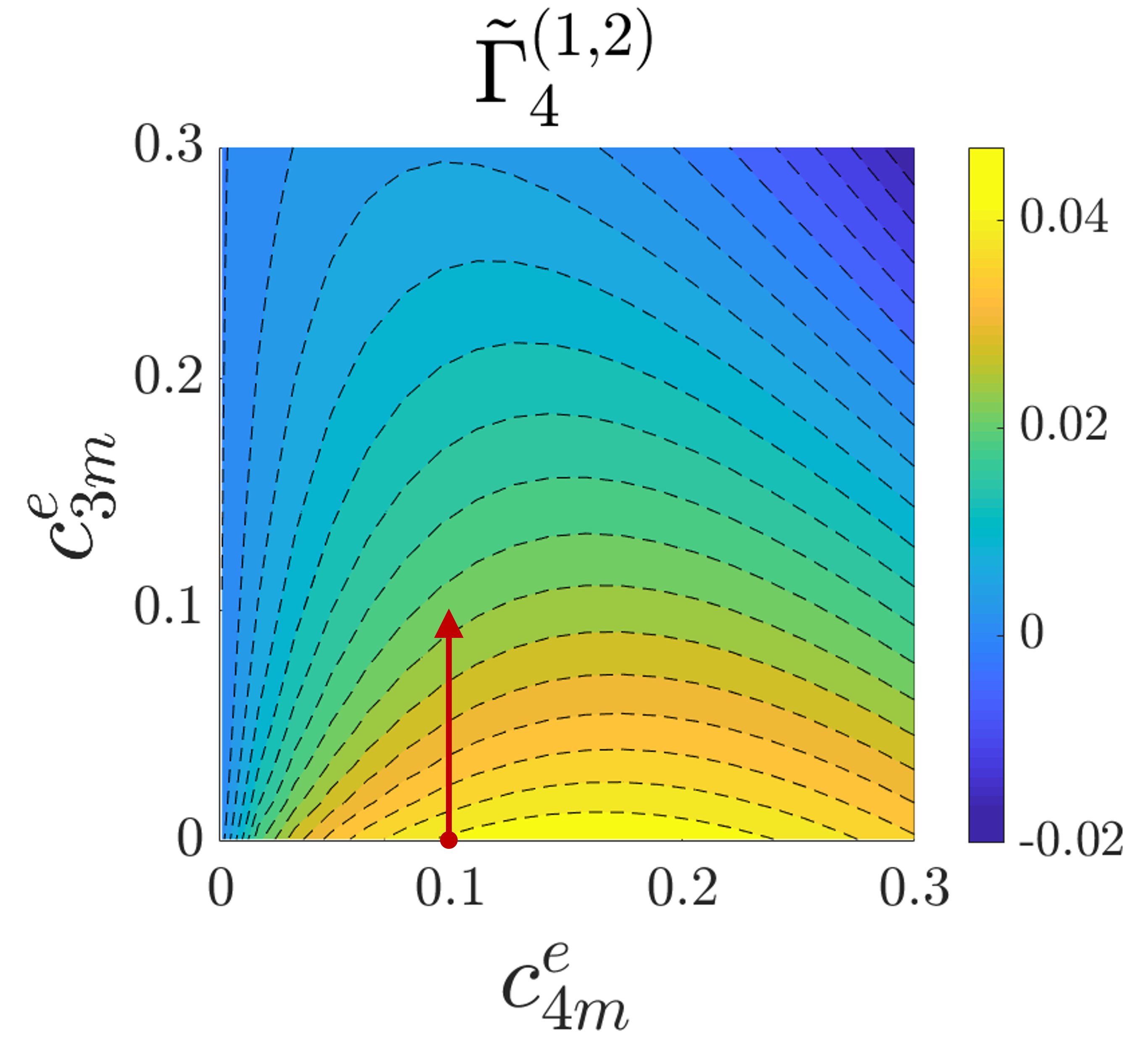}
%   \caption*{}
\end{subfigure}
\begin{subfigure}{0.3\textwidth}
  \centering
  \includegraphics[width=1\linewidth,trim=4 4 4 4,clip]{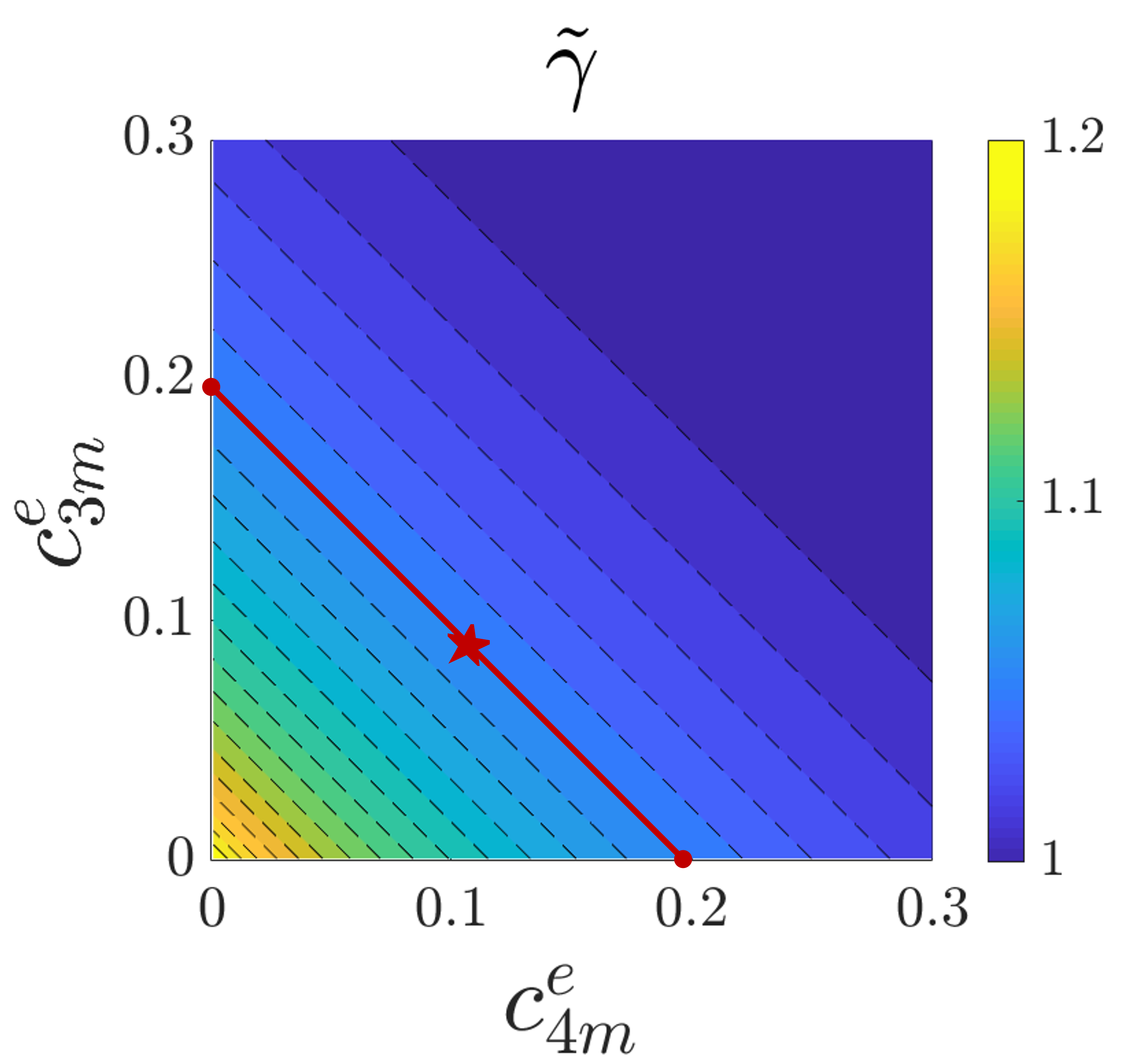}
%   \caption*{}
\end{subfigure}
\vspace{0.25cm}
\begin{subfigure}{0.3\textwidth}
  \centering
  \includegraphics[width=1\linewidth,trim=4 4 4 4,clip]{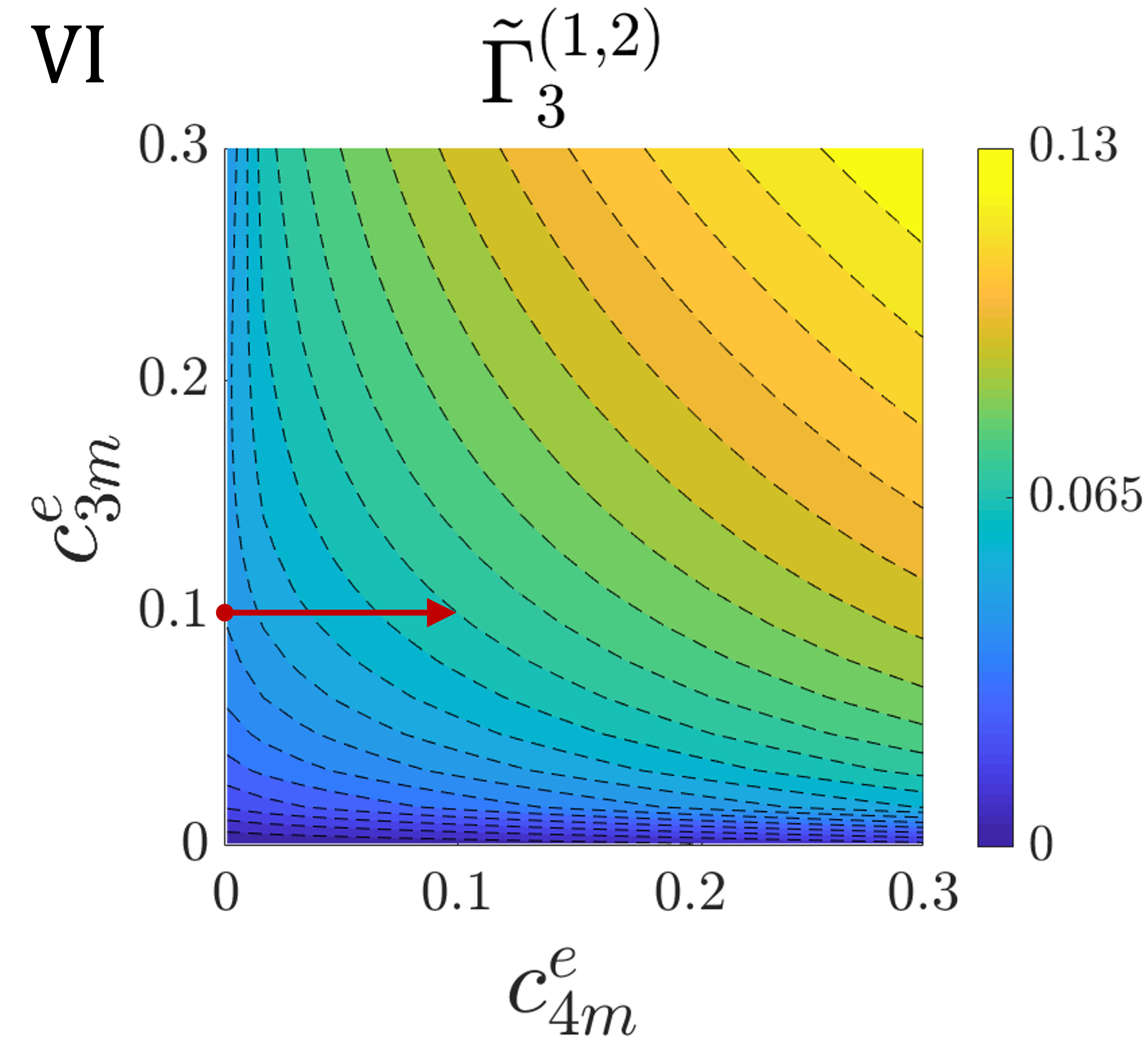}
%   \caption*{}
\end{subfigure}
\begin{subfigure}{0.3\textwidth}
  \centering
  \includegraphics[width=1\linewidth,trim=4 4 4 4,clip]{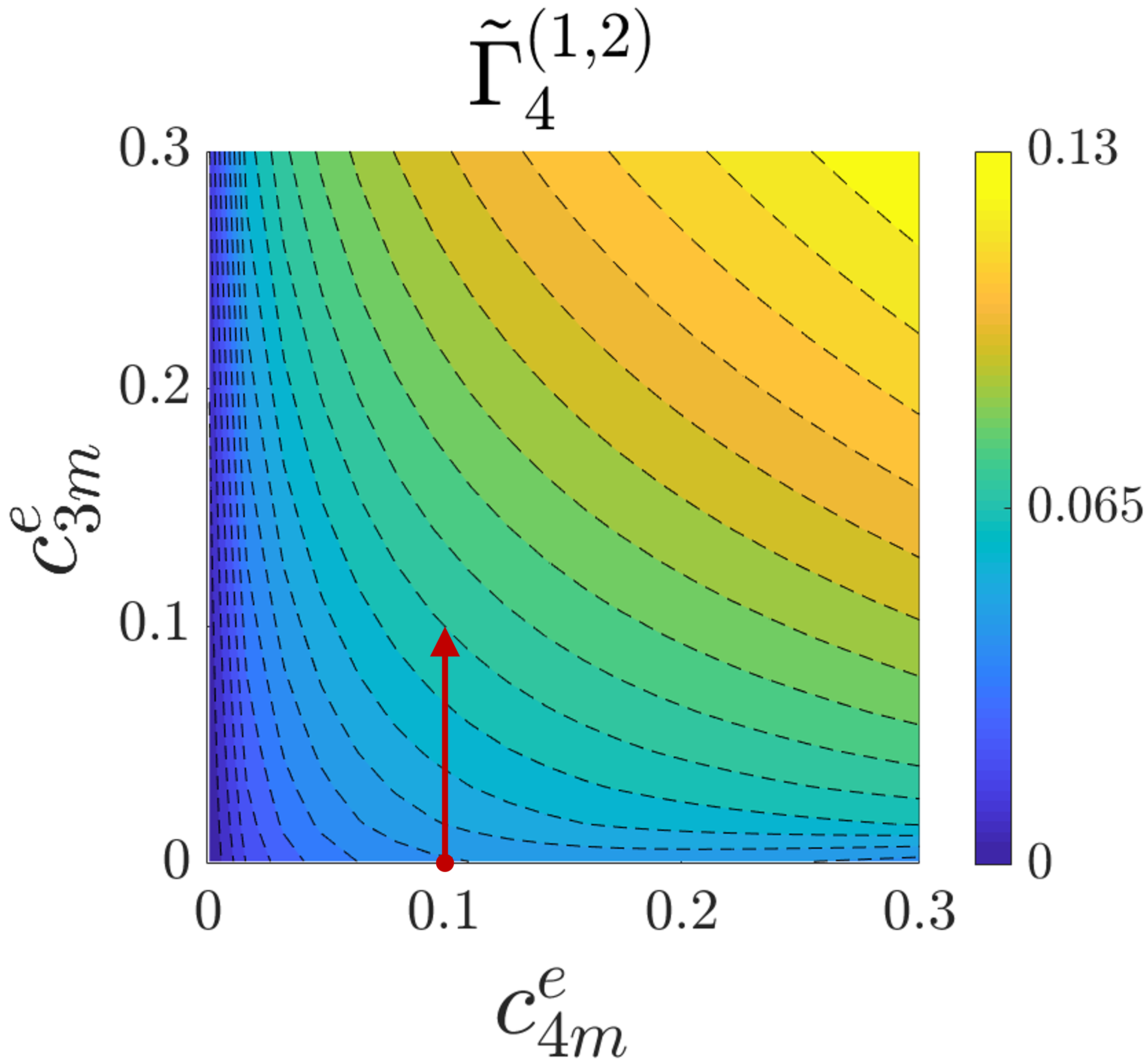}
%   \caption*{}
\end{subfigure}
\begin{subfigure}{0.3\textwidth}
  \centering
  \includegraphics[width=1\linewidth,trim=4 4 4 4,clip]{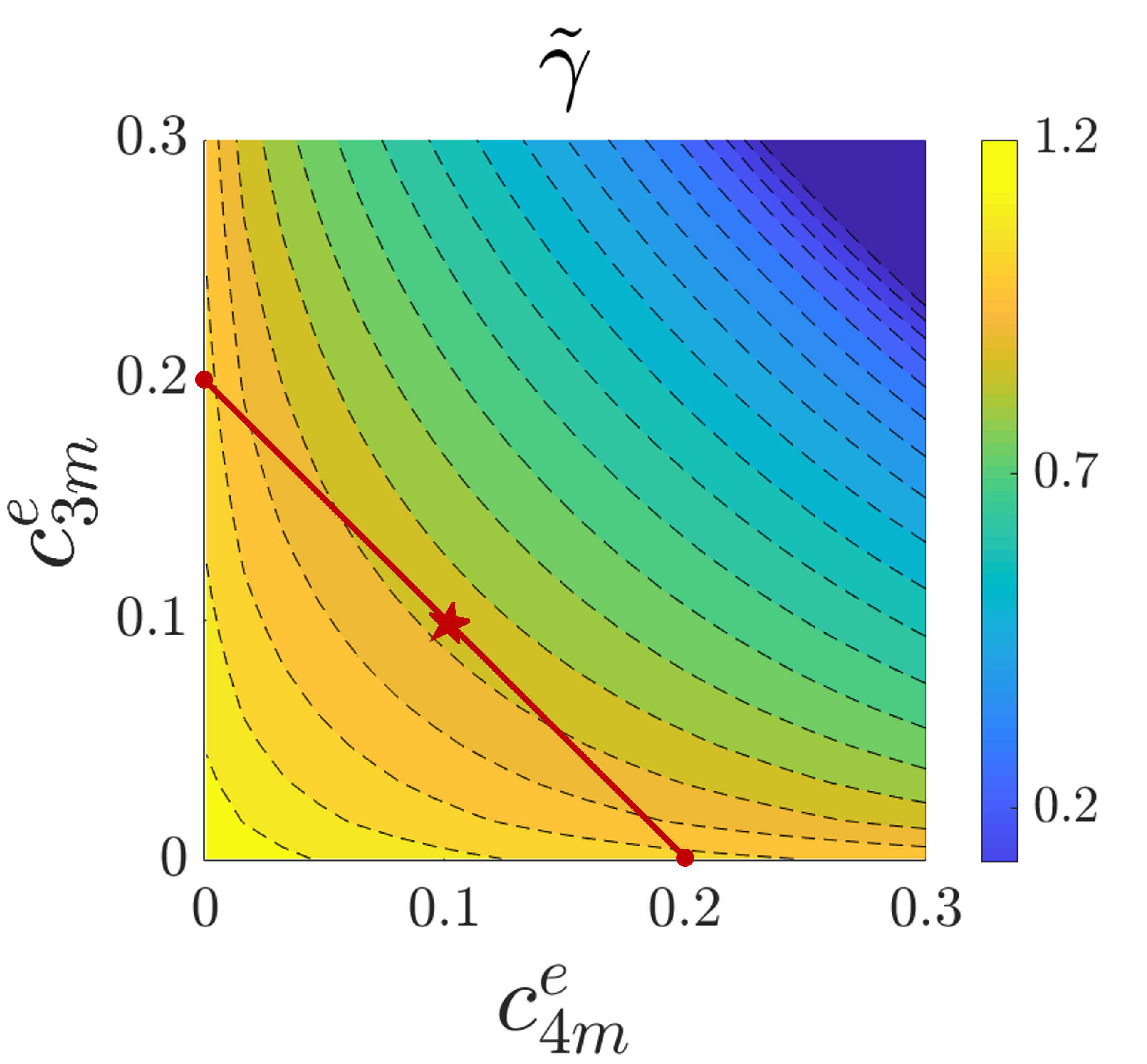}
%   \caption*{}
\end{subfigure}
\vspace{0.25cm}
\begin{subfigure}{0.3\textwidth}
  \centering
  \includegraphics[width=1\linewidth,trim=4 4 4 4,clip]{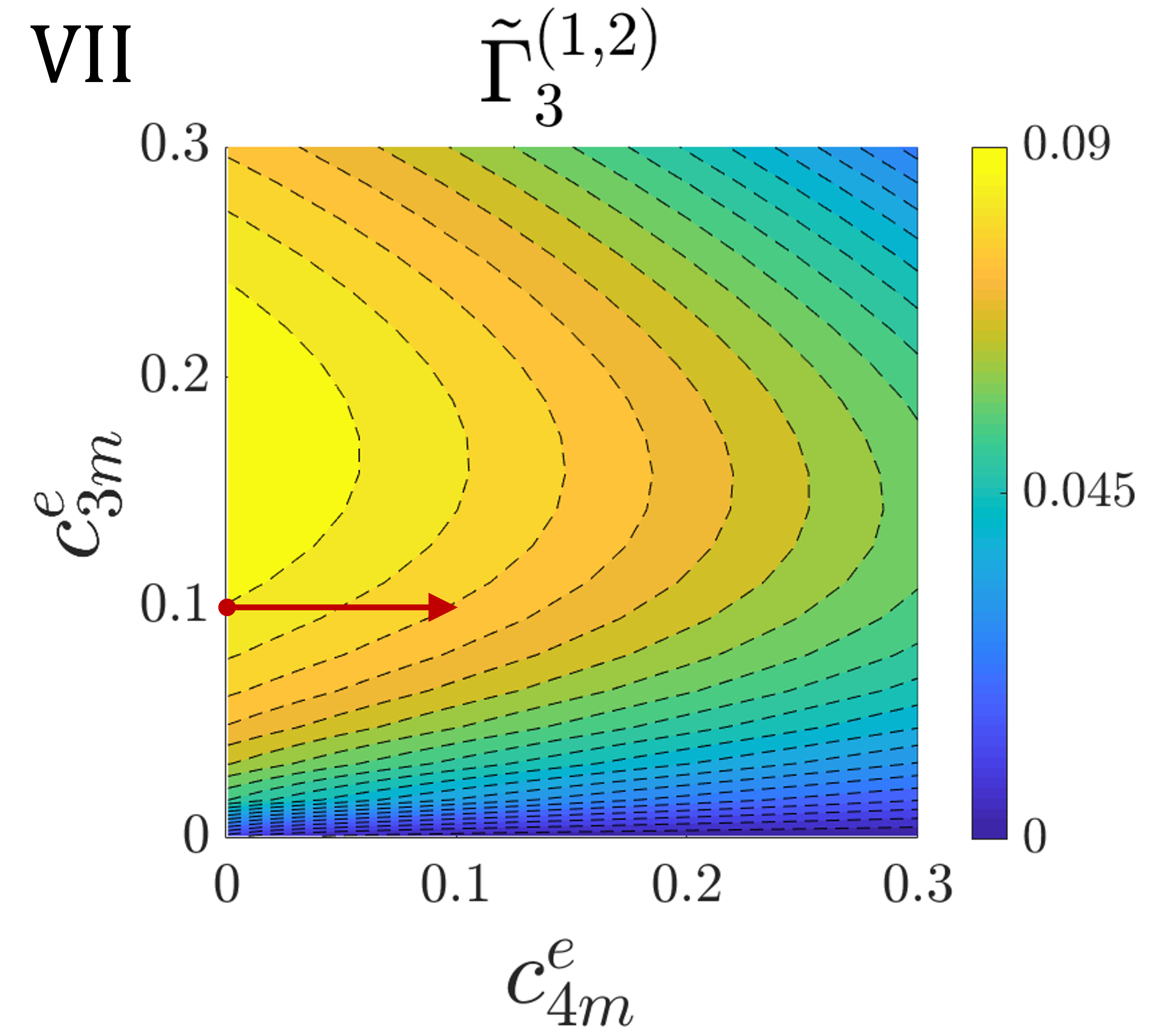}
%   \caption*{}
\end{subfigure}
\begin{subfigure}{0.3\textwidth}
  \centering
  \includegraphics[width=1\linewidth,trim=4 4 4 4,clip]{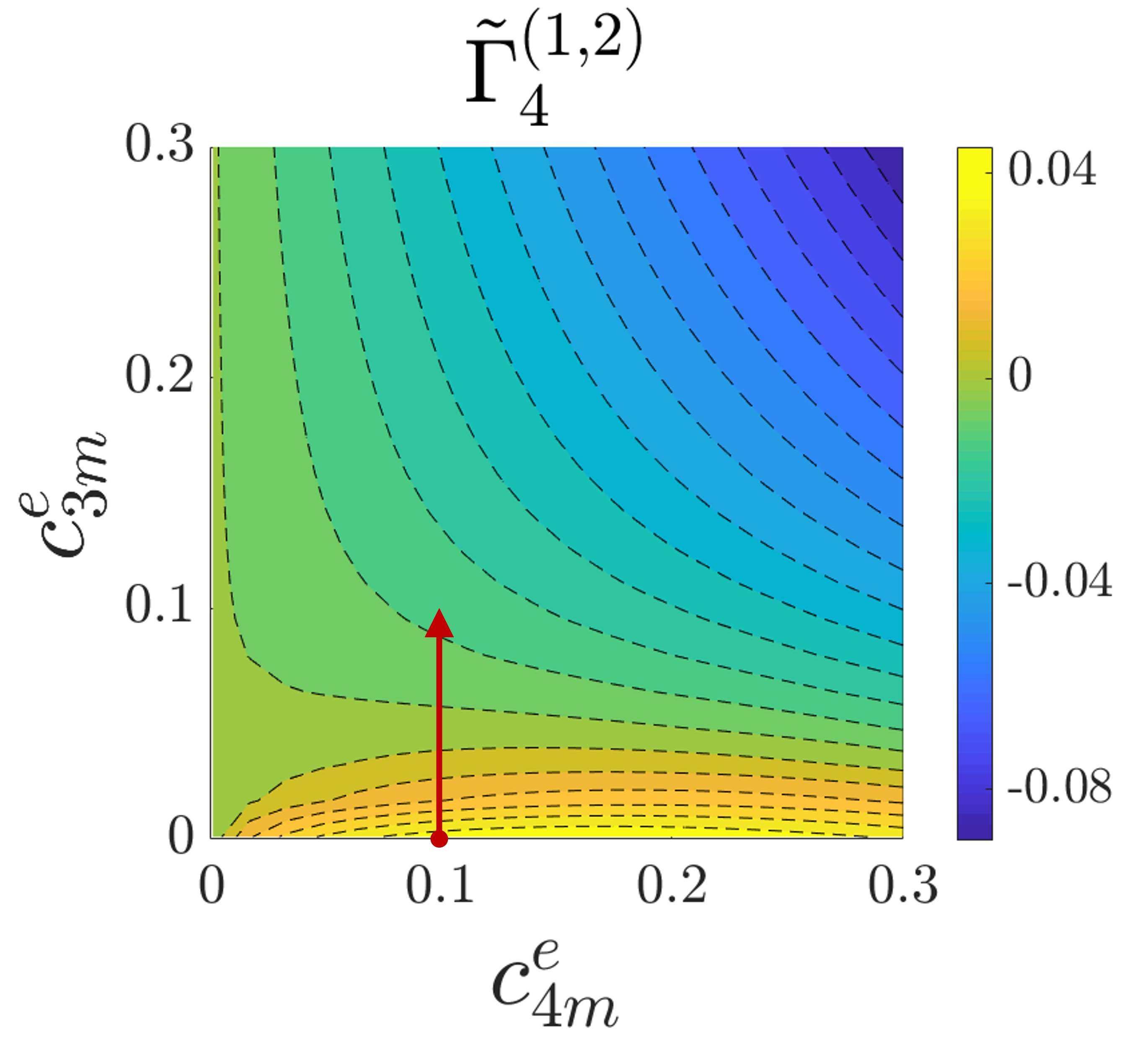}
%   \caption*{}
\end{subfigure}
\begin{subfigure}{0.3\textwidth}
  \centering
  \includegraphics[width=1\linewidth,trim=4 4 4 4,clip]{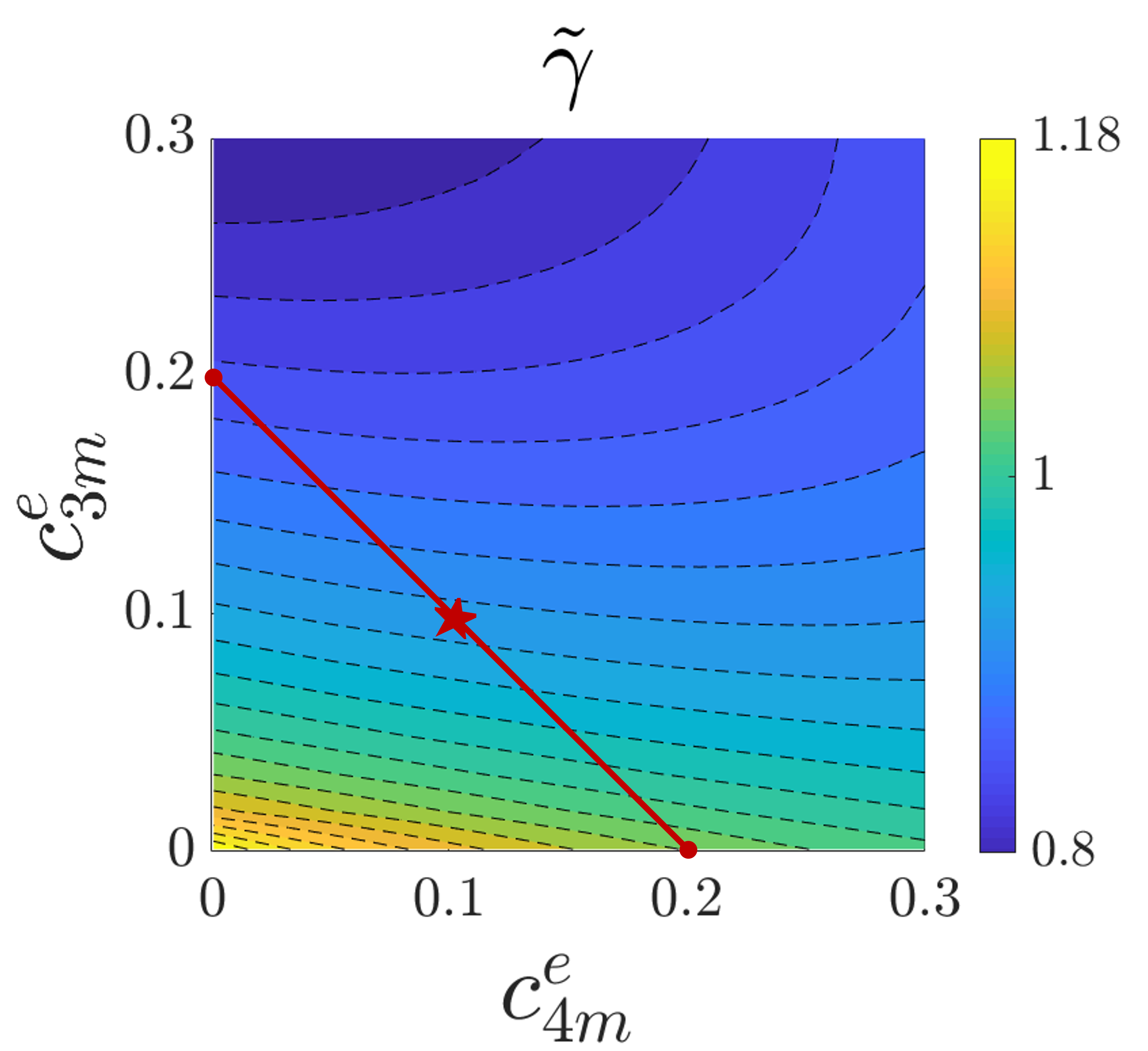}
%   \caption*{}
\end{subfigure}
\vspace{0.25cm}
% \rulesep
\begin{subfigure}{0.3\textwidth}
  \centering
  \includegraphics[width=1\linewidth,trim=4 4 4 4,clip]{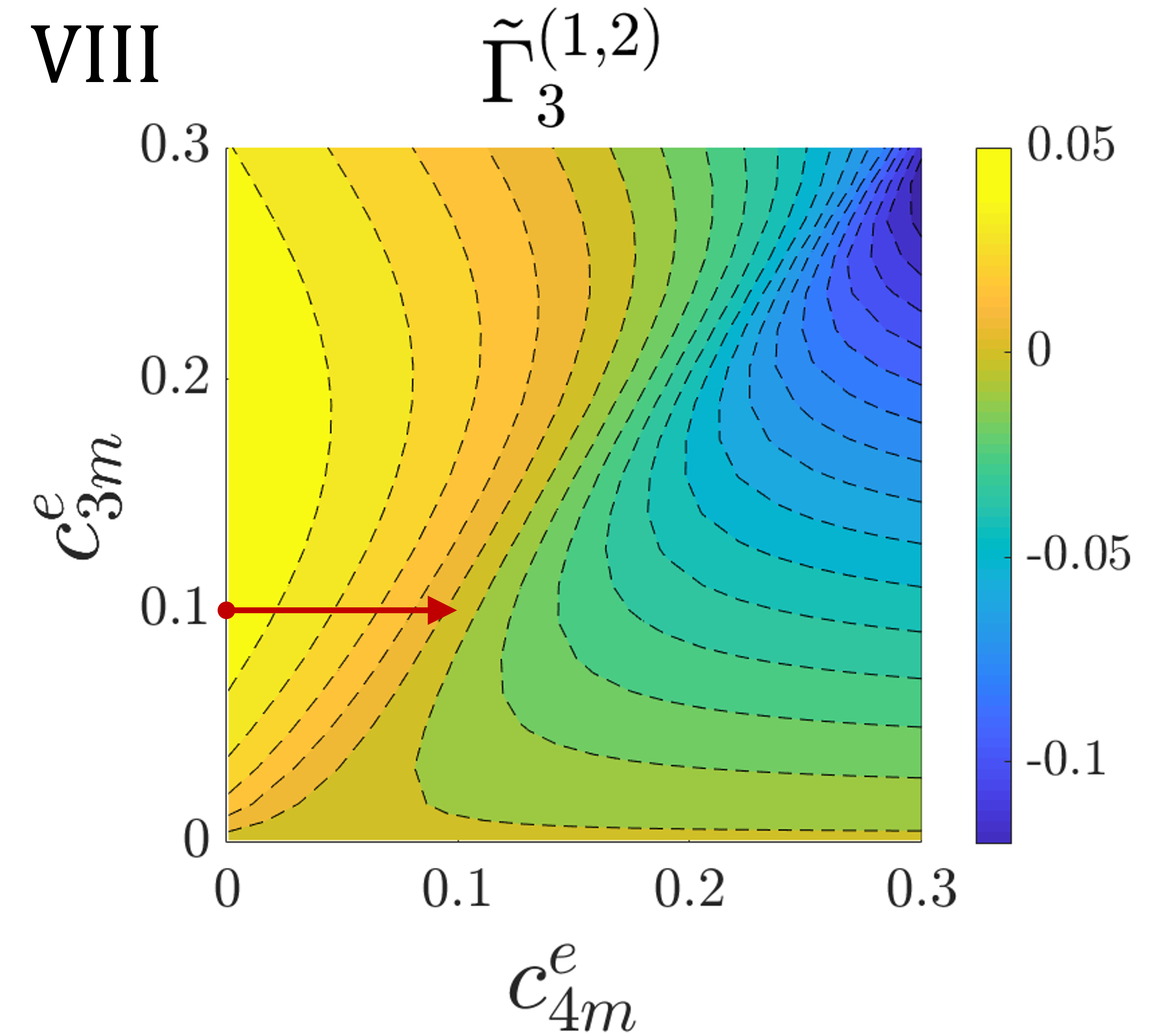}
%   \caption*{}
\end{subfigure}
\begin{subfigure}{0.3\textwidth}
  \centering
  \includegraphics[width=1\linewidth,trim=4 4 4 4,clip]{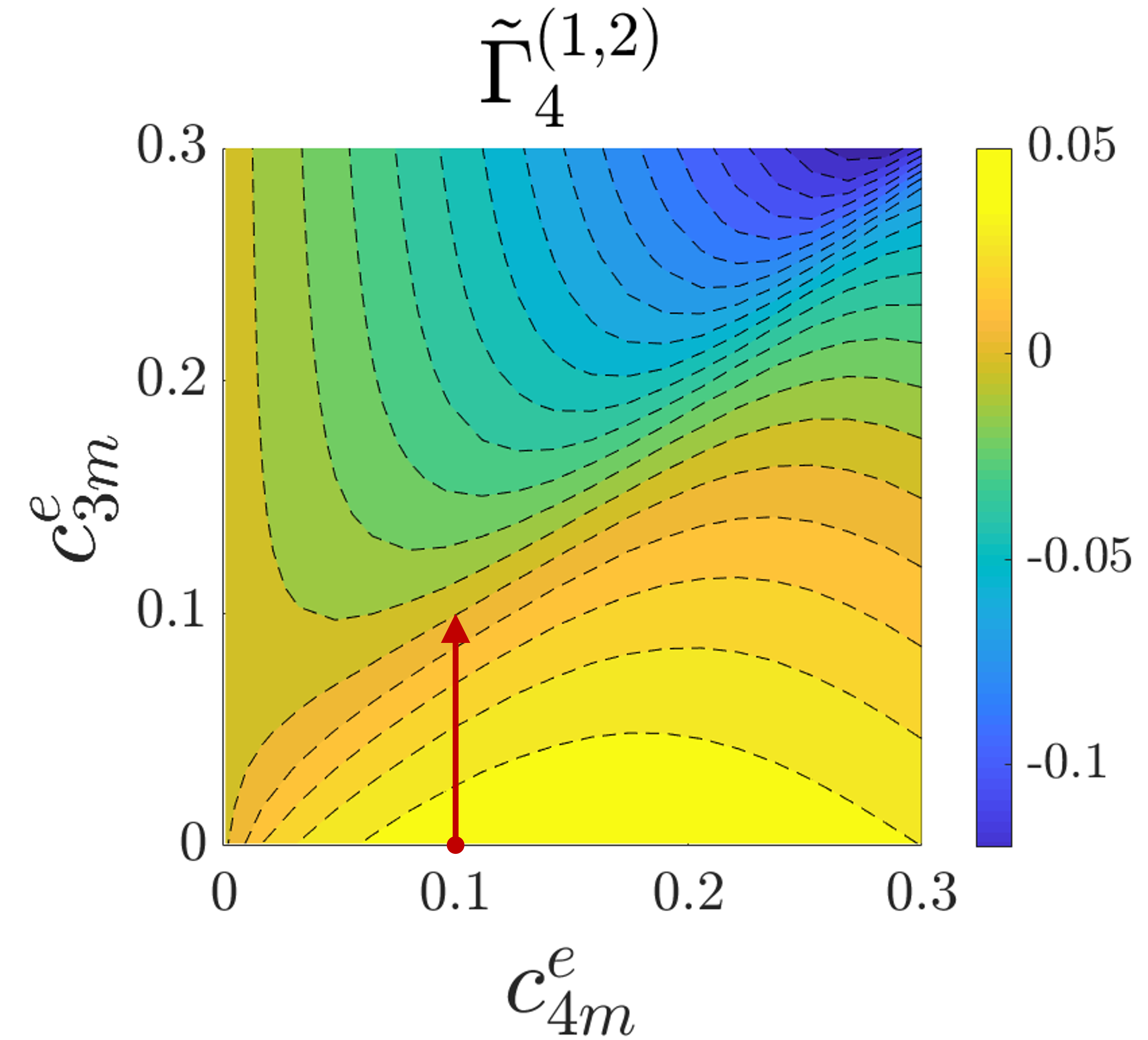}
%   \caption*{}
\end{subfigure}
\begin{subfigure}{0.3\textwidth}
  \centering
  \includegraphics[width=1\linewidth,trim=4 4 4 4,clip]{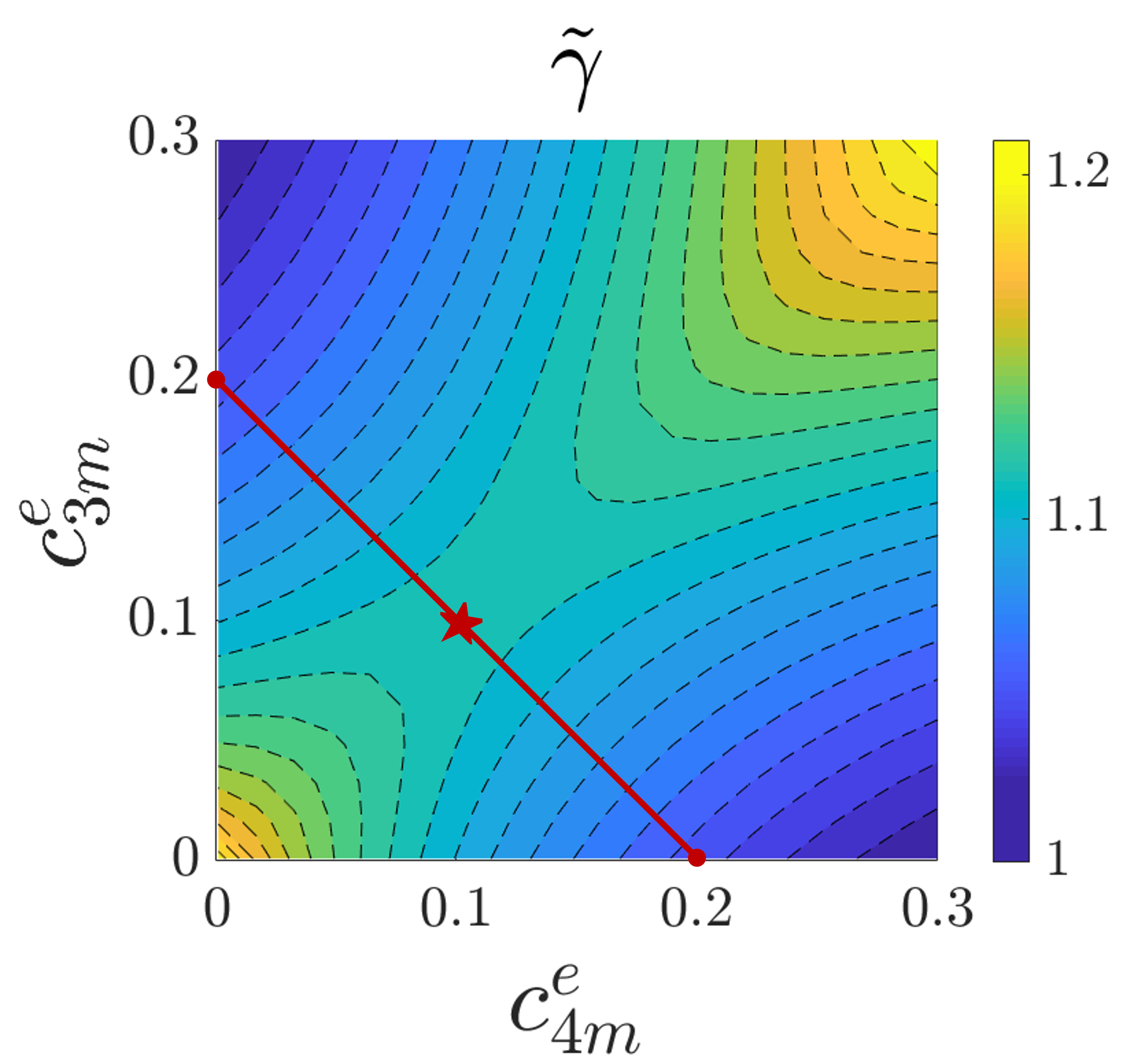}
%   \caption*{}
\end{subfigure}
\caption{Contour plots of relative solute excess $\tilde{\Gamma}^{(1,2)}_{\xi}$ ($\xi=3,4$) and excess boundary energy $\tilde{\gamma}$ for the various parametric cases I-VIII (Table~\ref{table:quaternary_params}). The contours are plotted against variation in the matrix solute concentrations $c^e_{3m}$ and $c^e_{4m}$. The red arrows in $\tilde{\Gamma}^{(1,2)}_\xi$-contour plots suggest a way to read the segregation behavior moving from a ternary to quaternary alloy. The interconnected dot-star-dot marker in $\tilde{\gamma}$-plot suggests a way to read the effect of the quaternary composition (star) relative to the end-ternaries (dots).}
\label{fig:quaternary_exs_plots}
\end{figure*}
\section{Discussion} \label{sec:discussion}

% In the present section, we discuss key features of the diffuse-interface thermodynamics formulation for segregation. We compare the interfacial energy and solute segregation of the interphase boundary (IB) with the more conventional case of the grain boundary (GB).

\subsection{Interfacial Energy}

We can understand the interfacial properties of the diffuse-interface formulation (Sec.~\ref{sec:method}) by first considering the interface to be a GB in a two-component, single-phase alloy. $\phi(x)$ now denotes the variation in crystallographic orientation across the diffuse GB. The bulk-phase free energies, phase concentrations and local phase fractions take the following simplified forms: $f^p\equiv f^m$, $c_{\theta p}\equiv c_{\theta m}$ and $P(\phi) \equiv M(\phi) = 1-I(\phi)$ for $\phi \in [0,1]$. Therefore, the local free energy density (Eq.~\ref{eq:f_local}) becomes $f(c_2,\phi)=f^m(c_{2m})[1-I(\phi)]+f^i(c_{2i})I(\phi)$. This reduced model is equivalent in most aspects to existing phase-field formulations \citep{cha2002phase,abdeljawad2015stabilization,kim2016GBsegregation} for GB segregation. Invoking the regular solution behavior, and imposing the limit of pure component $1$, we get $f(\phi) = G^m_1 + (G^i_1-G^m_1) I(\phi)$. The term $(G^i_1-G^m_1)I(\phi)$ is analogous to the classical, symmetric double-well potential $W_\phi I(\phi)$ used in standard phase-field models for grain growth, alloy solidification and antiphase boundary motion \citep{provatas2011phase}. Since the pure component energies $G^\psi_1 \{\pi1 \}$ are defined from the same pure component reference state $\pi1$ (Sec.~\ref{sec:soln_therm}), $G^i_1-G^m_1>0$ accounts for the excess structural energy of the GB phase over that of the matrix. The GB energy $\gamma$ of this pure metal is given by $\gamma_1 \approx 0.943 \varepsilon \sqrt{G^i_1-G^m_1}$ (Eq.~\ref{eq:gamma_int}). However, for a multicomponent alloy, the equilibrium between the GB phase and the bulk matrix phase is established by the parallel tangent construction, which determines the equilibrium GB phase concentrations $\textbf{c}^e_{i}$ and parallel tangent distance $W_e(\textbf{c}^e_{i},\textbf{c}^e_{m})$. $W_e$ then represents the free energy for formation of a unit volume of the equilibrium GB phase from the equilibrium bulk phase, and therefore, captures not only the the structural energetics but also the compositional dependence \citep{hillert1975lectures}. Therefore, the GB energy for a multicomponent alloy can be obtained (from Eqs.~\ref{eq:common_tangent} and \ref{eq:gamma_int}) by setting $p\equiv m$, giving $\gamma \approx 0.943 \varepsilon \sqrt{W_e(\textbf{c}^e_{i},\textbf{c}^e_{m})}$.

Now consider the interface to be the IB in a two-phase alloy, i.e. $p \not\equiv m$. The local free energy density $f(\textbf{c}^e,\phi_e)$ (Eq.~\ref{eq:f_local}) will take the form of an asymmetric double-well potential as shown below. Reducing the system to the theoretical limit of a pure component $1$, we get $f(\phi) =  G^m_1 + (G^i_1-G^m_1) I(\phi)$ for $\phi \leq 0.5$ and $f(\phi) =  G^p_1 + (G^i_1-G^p_1) I(\phi)$ for $\phi > 0.5$. In conventional phase-field formulations for two structurally-distinct phases, a symmetric double-well potential $W_\phi I(\phi)$ is employed to define the excess structural energy across the diffuse IB. 
In the current formulation, the symmetric part is inherent in the second term $(G^i_1-G^p_1) I(\phi)$ of the piecewise function, while the first terms $G^m_1\neq G^p_1$ results in the asymmetry. 
For a multicomponent alloy, the model additionally captures the compositional dependence: for an equilibrium, planar IB, $f(\textbf{c}^e,\phi_e)=f^m(\textbf{c}^e) + W_e(\textbf{c}^e) I(\phi)$ for $\phi_e \leq 0.5$ and $f(\textbf{c}^e,\phi_e)=f^p(\textbf{c}^e) + W_e(\textbf{c}^e) I(\phi_e)$ for $\phi_e>0.5$; where $W_e(\textbf{c}^e) \equiv f^i(\textbf{c}^e) - f^m(\textbf{c}^e) = f^i(\textbf{c}^e) - f^p(\textbf{c}^e)$. Therefore, analogous to the GB energy discussed in the preceding paragraph, the IB energy is obtained as $\gamma \approx 0.943 \varepsilon \sqrt{W_e(\textbf{c}^e_{i},\textbf{c}^e_{m})}$. Thereby $\gamma$ accounts for the structural and the chemical energetics for the formation of the interfacial phase (via $W_e$) and the associated diffuse regions (via $\varepsilon$). The parametric study (Sec.~\ref{sec:parametric}) showed the possibility of realizing $\gamma \rightarrow 0$ for synergistic, induced and enhanced co-segregation cases of quaternary alloys. However, the gradient energy contribution of the diffuse interface---together with the condition of equi-partitioning of the local excess grand-potential (Eq.~\ref{aeq:phase-field_profile_final})---imposes a theoretical restriction of $\gamma>0$, via $W_e > 0$ in Eq.~\ref{eq:gamma_int}.

% IB energy > 0 (gradient energy) and IB energy = 0 (KKS vs. WBM)
% Conventional non-gradient model for IB segregation: IB phase layer sandwiched by matrix and precipitate bulk phases.
% Equilibrium solutions for composition are obtained under the restrictive condition of equal chemical potentials.
% This is geometrically the common tangent plane condition between the three phases. 
% Therefore, equilibrium corresponds to a vanishing IB energy = 0.  
% The condition of equal diffusion potentials across the system and phases. This reduces to the equal chemical potential condition within the bulk phases (i.e.), but not within the IB phase (here i>m=p). 
% This is due to the positive and finite excess energy (gradient energy coefficient) associated with the gradient in the phase field across the interfacial region. 
% This results in an equilibrium diffuse IB width. 
% For the IB phase, the equilibrium condition is geometrically the parallel tangent plane. This condition is more flexible compared to
% the common tangent plane used in earlier model. Extra dof in concentration.
% The equilibrium solutions are obtained for  gamma > 0.

\subsection{Multicomponent Segregation}

The regular solution assumption (Sec.~\ref{sec:soln_therm}) for the bulk and IB phase thermodynamics allowed a variety of segregation/de-segregation behaviors to be realized. Parameterization was readily possible based on the physical significance afforded by regular solution thermodynamic parameters. 
{Compared to IBs, multicomponent segregation in GBs have been well studied, both experimentally and theoretically \citep{guttmann1979grain, erhart1981equilibrium, guttmann1982thermodynamics, lejcek2010grain, lejvcek2013effect, xing2018solute,xing2019stability}. 
IB segregation in quaternary, two-phase alloys can be found to be similar in many aspects to GB segregation in ternary, single-phase alloys.
% For GBs, the binary-component alloy accommodates a compositional degree of freedom in tailoring the chemistry and energetics of the boundary. Ternary and higher-order alloys allow greater degrees of compositional freedom as two or more elements can segregate at the GB.
For example, the addition of certain secondary solutes is well known to enhance GB segregation of the primary solute due to mutually attractive interaction between the solutes at the GB \citep{gas1982interactive}. This is analogous to case VI.
Another example is the addition of a secondary solute which is also known to mitigate GB embrittlement caused by the undesired segregation of the primary solute--i.e. the solutes compete for GB sites and the strongly-segregating solute preferentially segregates by desegregating the weaker solute \citep{suzuki1991site}. This is analogous to case VII. Therefore, based on the nature of mutual interaction between the solutes, their combined addition in the ternaries causes GB segregation behaviors that are fundamentally different from that of the individual solutes in their respective binaries.
Xing et al. \citep{xing2018solute} presented a classification of the GB segregation/de-segregation behaviors in ternary single-phase alloys. The segregation mechanisms observed in current study are analogous to those observed for GBs.} 

The case studies presented in Sec.~\ref{sec:parametric} are qualitatively representative of the observed mechanisms in the literature. As mentioned in 
Sec.~\ref{sec:intro}, numerous cases of IB co-segregation has already been reported.
% : Mg+Ag at Al(matrix)/$\Omega$-Al$_2$Cu; Si+Mg and Mn+Zr at Al(matrix)/$\theta^{\prime}$-Al$_2$Cu
Several experimental and first-principles studies demonstrate enhanced high-temperature stability of the quaternary alloys due to IB co-segregation and its associated reduction in IB energy $\gamma$.  
Moreover, some studies suggest the need to desegregate certain solute from the IB: for example, Si at Al(matrix)/$\theta^{\prime}$-Al$_2$Cu in (Al-Cu)-Mn-Zr-Si is found to be deleterious to the microstructural stability in larger amounts \citep{shyam2019elevated}; Ca at Mg(matrix)/Al$_{11}$La$_3$ in (Mg-Al)-La-Ca-Mn is found to destabilize the strengthening precipitate \citep{yang2020interphase} by by nucleating Al$_2$Ca. Therefore, it is also necessary to understand the de-segregation mechanisms (cases VII and VIII) so that alloys could be designed to favor beneficial solutes to segregate in preference to the undesired impurities. 

The diffuse-interface thermodynamic formulation was shown to be consistent (see Sec.~\ref{app:GA}) with the classical Gibbs interface phase rule \citep{frolov2015phases}. At a given temperature, the number of components ($n_\mathcal{N}$) and the number of coexisting bulk phases ($n_\psi$) dictates the number of compositional degrees of freedom ($d_\theta^f = n_\mathcal{N} - n_\psi$). Thereby, two compositional degrees of  freedom in solute $3$ and $4$ concentrations were available to modulate the properties of the quaternary IB in Sec.~\ref{sec:parametric}. 
Analogously, if the model is applied to a ternary GB (where the common tangent constraint between phases is removed), two compositional degrees of  freedom in solutes $2$ and $3$ will be available. Therefore, in addition to multicomponent IB segregation, the current approach has the potential to be employed for recently observed GB segregations in multi-principal element/high entropy alloys \citep{li2020grain}.

\section{Summary and conclusions} \label{sec:conclusions}

We developed a diffuse-interface thermodynamic model for interphase boundary (IB) segregation in multicomponent alloys.
In this description, the diffuse IB consists of the IB phase and its compositional gradients with the adjoining bulk phases. Analytic solutions for the planar, stationary IB showed that the equilibrium between IB and bulk phases is set by the parallel tangent conditions. The model is consistent with the classical multicomponent segregation isotherms and the generalized Gibbs adsorption isotherm. Therefore, the resulting expressions for the relative solute excess and the excess IB energy are ideally suited for parameterization using solute segregation energies from atomistic calculations (like DFT), and for direct comparison of with experimental chemical characterizations (like atom probe tomography).

Due to the unavailability of comprehensive data on IB segregation and thermodynamic parameters, we performed a parametric study for a hypothetical quaternary alloy. The regular solution mixing behavior was employed to provide physical significance to the case studies. We demonstrated a variety of co-segregation and desegregation mechanisms in the quaternaries---these arise from unique interactions between the solutes in the IB and the bulk. The case studies resemble the experimentally observed segregation behaviors at quaternary IBs and ternary GBs. However, we assumed a simplified parameteric space for the purpose of illustration. A more thorough exploration of the parametric space will be necessary for understanding matrix-precipitate alloy.

The current nanoscopic model will require a fine descretization of the interface to resolve segregation. However, mesoscopic phase-field simulations are typically performed at the microstructure level with multiple precipitates or grains. Incorporating the nanoscopic model for mesocopic simulations will require further study to determine the possibility of using increased computational interface width to reduce computational expense, while also maintaining the required quantitative IB energy and solute excess. 
In addition to chemical thermodynamics, IB segregation in many experimental alloys are governed by the interface crystallography \cite{patala2019understanding, homer2015grain}, structure \cite{banadaki2018efficient, zhang2016faceted}, elastic energy contribution \cite{heo2011phase} and kinetic factors.
Therefore, further model development is expected to enable quantitative modeling that can provide design rules for segregation-engineered stable alloys. 
% The model in its current form can be readily adapted for grain boundary segregation in multicomponent alloys, including multi-principal element alloys. 

\appendix
\renewcommand{\thesection}{\Alph{section}}
\renewcommand{\theequation}{\Alph{section}\arabic{equation}}
\section{Equilibrium Phase-field} \label{app:profile}

We present the derivation for the equilibrium phase-field $\phi_e(x)$ in a one-dimensional system. Considering the phase concentrations $c_{2\psi}$, $c_{3\psi}$,\ldots, $c_{\mathcal{N}\psi}$ ($\psi=m,i,p$) as functions of ($c_2$,$\phi$) and ($c_3$,$\phi$), respectively, and differentiating $c_{\theta}(x)$ ($\theta=2,3,\ldots,\mathcal{N}$) in Eq.~\ref{eq:conc}, we get
\begin{flalign} \label{aeq:diff_conc_c} 
    M(\phi) \frac{\partial c_{\theta m}}{\partial c_\theta} + I(\phi) \frac{\partial c_{\theta i}}{\partial c_\theta} + P(\phi) \frac{\partial c_{\theta p}}{\partial c_\theta} = 1 &&
\end{flalign}
and
\begin{flalign} \nonumber
    &M(\phi) \frac{\partial c_{\theta m}}{\partial \phi} + I(\phi) \frac{\partial c_{\theta i}}{\partial \phi} + P(\phi) \frac{\partial c_{\theta p}}{\partial \phi} \\ \label{aeq:diff_conc_phi}
    &= \frac{d M(\phi)}{d \phi}\,(c_{\theta i}-c_{\theta m}) + \frac{d P(\phi)}{d \phi}\,(c_{\theta i}-c_{\theta p}). &&
\end{flalign}
Eqns.~\ref{aeq:diff_conc_c} and \ref{aeq:diff_conc_phi} are used to obtain the derivatives of $f(c_2,c_3,\ldots,c_\mathcal{N},\phi)$ (Eq.~\ref{eq:f_local}) as
\begin{flalign} \label{aeq:non-eq_dfdc}
    &\frac{\partial f}{\partial c_\theta} = M(\phi) \frac{\partial f^m}{\partial c_{\theta m}}\frac{\partial c_{\theta m}}{\partial c_\theta} \\ \nonumber
    & \hspace{1cm}+ I(\phi) \frac{\partial f^i}{\partial c_{\theta i}}\frac{\partial c_{\theta i}}{\partial c_\theta} + P(\phi) \frac{\partial f^p}{\partial c_{\theta p}}\frac{\partial c_{\theta p}}{\partial c_\theta} \\ \nonumber
    &= \left(M(\phi)\frac{\partial c_{\theta m}}{\partial c_\theta} + I(\phi)\frac{\partial c_{\theta i}}{\partial c_\theta} + P(\phi)\frac{\partial c_{\theta p}}{\partial c_\theta} \right){\mu}_{\theta 1}= {\mu}_{\theta 1} &&
\end{flalign}
and
\begin{flalign} \label{aeq:non-eq_dfdphi} \nonumber
    & \frac{\partial f}{\partial \phi} = \sum_{\theta=2:\mathcal{N}} \left[M(\phi)  \frac{\partial f^m}{\partial c_{\theta m}}\frac{\partial c_{\theta m}}{\partial \phi} + I(\phi) \frac{\partial f^i}{\partial c_{\theta i}}\frac{\partial c_{\theta i}}{\partial \phi} \right. \\ \nonumber
    & \hspace{1cm} \left.+  P(\phi) \frac{\partial f^p}{\partial c_{\theta p}}\frac{\partial c_{\theta p}}{\partial \phi} \right] \\ \nonumber
    &\hspace{1cm} +\frac{dM}{d\phi} f^m + \frac{dI}{d\phi} f^i + \frac{dP}{d\phi} f^p \\ \nonumber
    &= \sum_{\theta=2:\mathcal{N}} {\mu}_{\theta 1} \left(\frac{dM}{d\phi}(c_{\theta i}-c_{\theta m}) + \frac{dP}{d\phi}(c_{\theta i}-c_{\theta p})\right) \\ \nonumber
    &+ \frac{dM}{d\phi} f^m + \frac{dI}{d\phi} f^i + \frac{dP}{d\phi} f^p \\ \nonumber 
    =& \left(f^m - f^i - \sum_{\theta=2:\mathcal{N}} (c_{\theta m}-c_{\theta i}){\mu}_{\theta 1} \right) \frac{dM}{d\phi} \\
    &+ \left(f^p - f^i - \sum_{\theta=2:\mathcal{N}} (c_{\theta p}-c_{\theta i}){\mu}_{\theta 1}\right) \frac{dP}{d\phi}. && 
\end{flalign} 
where the condition (Eq.~\ref{eq:equilibrium_diff_pot}) of equal diffusion potentials, $\mu_{\theta 1}(x)$ ($\theta = 2:\mathcal{N}$), between the phases, and the identity $dI/d\phi = -dM/d\phi - dP/d\phi$ were used.

At equilibrium, $\phi_e(x)$ must satisfy the condition for stationary interface given by Eq.~\ref{eq:1D_eq_phi},
\begin{flalign} \label{aeq:eq_phase-field_begin}
    \frac{\partial f}{\partial \phi_e} = \varepsilon^2 \frac{d^2\phi_e}{dx^2}. && 
\end{flalign} 
The phase concentration fields must be constant value across the system,
$c_{\theta \psi}(x) = c^e_\theta$ ($\theta = 2:\mathcal{N}$) (Eq.~\ref{eq:equilibrium_diff_pot}). Using Eq.~\ref{aeq:non-eq_dfdphi} in Eq.~\ref{aeq:eq_phase-field_begin}, we get
\begin{flalign}
    -W^m_e\frac{dM(\phi_e)}{d\phi_e} - W^p_e\frac{dP(\phi_e)}{d\phi_e} = \varepsilon^2 \frac{d^2\phi_e}{dx^2}, &&
\end{flalign}
where $W^m_e \equiv f^i - f^m - \sum_{\theta=2:\mathcal{N}} (c_{\theta i}^e-c_{\theta m}^e){\mu}_{\theta 1}^e$ and $W^p_e \equiv f^i - f^p - \sum_{\theta=2:\mathcal{N}} (c_{\theta i}^e-c_{\theta m}^e){\mu}_{\theta 1}^e$ are spatially constant.
Multiplying both sides by $d\phi_e/dx$, integrating from $x = -\infty$ to $x = +\infty$, and changing the variable of integration to $\phi_e$, we get
\begin{flalign}
    {W^m_e \int_{\phi_e^m}^{\phi_e^i} \frac{dM(\phi_e)}{d\phi_e} d\phi_e + W^p_e \int_{\phi_e^i}^{\phi_e^p} \frac{dP(\phi_e)}{d\phi_e} d\phi_e = 0}, &&
\end{flalign}
where $\phi_e^m$, $\phi_e^i$ and $\phi_e^p$ are the values of the phase-field variable corresponding to exclusive phases $m$, $i$ and $p$, respectively. Accordingly, the interpolating functions must satisfy $M(\phi_e^m)=0$, $I(\phi_e^i)=0$ and $P(\phi_e^p)=0$ and $M(\phi_e) + I(\phi_e) + P(\phi_e) = 1$. The above equation reduces to
\begin{flalign}
    W^m_e \int_{\phi_e^m}^{\phi_e^i} {dI(\phi_e)} = - W^p_e \int_{\phi_e^i}^{\phi_e^p} {dI(\phi_e)}. &&
\end{flalign}
Therefore, $W^m_e = W^p_e \equiv W_e$ or $f^m_e-f^p_e-\sum_{\theta=2:\mathcal{N}} (c_{\theta m}^e-c_{\theta p}^e){\mu}_{\theta 1}^e=0$. This relation, together with Eq.~\ref{eq:equilibrium_diff_pot} constitutes the common tangent hyperplane between $f^m$ and $f^p$, and the parallel tangent hyperplane for $f^i$. $W_e$ defines the vertical distance between the common tangent and the parallel tangent hyperplanes.
Using $W^m_e = W^p_e \equiv W_e$ in Eq.~\ref{aeq:eq_phase-field_begin} and integrating after multiplying by $d\phi_e/dx$ gives the equation for equilibrium phase-field,
\begin{flalign} \label{aeq:phase-field_profile_final}
    \frac{\varepsilon^2}{2}\left(\frac{d\phi_e}{dx}\right)^2 = W_e I(\phi_e), &&
\end{flalign}
where the identities $-dM/d\phi_e = dI/d\phi_e$ for $\phi_e \in [\phi_e^m, \phi_e^i]$ and $-dP/d\phi_e = dI/d\phi_e$ for $\phi_e \in (\phi_e^i, \phi_e^p]$ were used.
\section{Interphase Boundary Energy} \label{app:IB_energy}

In this section, we derive the interphase boundary (IB) energy $\gamma$ from the equilibrium solution of the planar, diffuse interface in one dimension. Assuming equal molar volumes $v_m$ in each phase, the Gibbs solute excess $C_{xs}$ \citep{wheeler1993phase}, is evaluated with reference to the Gibbs dividing surface at $x=0$ as 
\begin{flalign} \label{aeq:solute_excess} \nonumber
    C^{xs}_\theta &= \int_{-l}^{+l} c^e_\theta (x) dx - \int_{-l}^0 c_{\theta m}^e dx - \int_0^{+l} c_{\theta p}^e dx \\
    &= {(2c_{\theta i}^e - c_{\theta m}^e - c_{\theta p}^e) \frac{\varepsilon}{3\sqrt{2W_e}}}, &&
\end{flalign}
where ${-l,+l}$ are layer bounds far from the diffuse interface. $\gamma$ is defined as the excess grand potential \citep{wheeler1992phase,mcfadden2002gibbs} of the diffuse interface, and is evaluated with reference to the physical mixture between the equilibrium bulk phases (geometrically, the common tangent hyperplane). For the multicomponent system,
\begin{flalign} \label{aeq:IB_energy_def}
    \gamma = (\mathcal{F}-\mathcal{F}_{o}) - \sum_{\theta=2:\mathcal{N}} C^{xs}_\theta {\mu}^e_{\theta 1}, &&
\end{flalign}
where $\mathcal{F}$ is the total free energy (Eq.~\ref{eq:F_functional}) of the diffuse-interface system and $\mathcal{F}_{o}$ is the free energy of the Gibbs reference system whose matrix and precipitate properties remain homogeneous up to the dividing surface at $x=0$. Therefore,
\begin{flalign} \label{aeq:IB_energy_expand} \nonumber
    \gamma =& \int_{-l}^{+l} \left[f(c^e_2,c^e_3,\ldots,c^e_{\mathcal{N}}, \phi_e) + \frac{\varepsilon^2}{2}\left(\frac{d\phi_e}{dx}\right)^2 \right] dx \\
    &- \int_{-l}^{0} f^m dx - \int_{0}^{+l} f^p dx - \sum_{\theta = 2:\mathcal{N}} C^{xs}_\theta {\mu}^e_{\theta 1}. &&
    % &- \int_{-l}^{+l} \left[M(\phi_e)c_m^e + P(\phi_e)c_p^e + I(\phi_e)c_i^e \right]\mu_e dx \\
    % &+ \int_{-l}^{0} c_m^e dx + \int_{0}^{+l} c_p^e dx  &&
\end{flalign}
Substituting the expressions for $f$ (Eq.~\ref{eq:f_local}) and $C_{xs}$ (Eq.~\ref{aeq:solute_excess}), and reorganizing the terms to evaluate the integrals piecewise over $[-l,0]$ and $(0,+l]$, we get
\begin{flalign} \label{aeq:IB_energy_piecewise_x} \nonumber
    \gamma 
    &= \int_{-l}^{+l} \frac{\varepsilon^2}{2}\left(\frac{d\phi_e}{dx}\right)^2 dx \\ \nonumber
    &+ \int_{-l}^{0} \left[\left(1-M(\phi_e)\right)\left(f^i_e - f^m_e - \sum_{2:\mathcal{N}} (c^e_{\theta i}- c^e_{\theta m}) {\mu}^e_{\theta 1} \right) \right] dx \\ \nonumber
    &+ \int_{0}^{+l} \left[\left(1-P(\phi_e)\right)\left(f^i_e - f^p_e - \sum_{2:\mathcal{N}} (c^e_{\theta i}-c^e_{\theta p}) {\mu}^e_{\theta 1} \right) \right] dx , &&
    % &+ \int_{-l}^{0} \left[{M(\phi_e)\left(f^m_e - c_{2m}^e \bar{\mu}_2^e - c_{3m}^e \bar{\mu}_3^e \right) +\left(1-M(\phi_e)\right) \left(f^i_e - c_{2i}^e \bar{\mu}_2^e - c_{2i}^e \bar{\mu}_3^e \right) - f^m_e - c_{2m}^e \bar{\mu}_3^e } \right] dx \\
    % % \nonumber
    % &+ \int_{-l}^{0} \left[{P(\phi_e)\left(f^p(c_p^e) - c_p^e \mu_e \right) +\left(1-P(\phi_e)\right) \left(f^i(c_i^e) - c_i^e \mu_e \right) - f^p(c_p^e) - c_p^e \mu_e } \right] dx &&
\end{flalign}
where the identities $I(\phi_e(x)) = 1-M(\phi_e(x))$ on $[-l,0]$ and $I(\phi_e(x)) = 1-P(\phi_e(x))$ on $(0,+l]$ were used. Simplifying the second and third terms, we get
\begin{flalign}  \nonumber
    \gamma &= \int_{-l}^{+l} \frac{\varepsilon^2}{2}\left(\frac{d\phi_e}{dx}\right)^2 dx + \int_{-l}^{+l} W_e I(\phi_e) dx \\
    &= \int_{-l}^{+l} \varepsilon^2 \left(\frac{d\phi_e}{dx}\right)^2 dx = \int_{\phi_e^m}^{\phi_e^p} \varepsilon^2 \frac{d\phi_e}{dx} d\phi_e ,&&
\end{flalign} \label{aeq:IB_energy_gradient}
where the definition of $W_e$ (Eq.~\ref{eq:parallel_tangent_We}) and the equality in Eq.~\ref{aeq:phase-field_profile_final} were used.

\section{Relative Excess Quantities} \label{app:GA}

In this section, we apply the layer treatment of \citep{cahn1979interfacial} to the diffuse-interface equilibirum system. This will allow the generalized Gibbs adsorption equation to be derived in terms of the relative solute excess $\Gamma^{(1,2)}_\theta$ and the relative entropy excess $\Gamma^{(1,2)}_S$ as contained in the current formulation. For the layer $[-l,l]$ containing the interfacial gradients, the following thermodynamic relation must hold at constant pressure \citep{cahn1979interfacial,frolov2015phases}
\begin{flalign} \label{aeq:GD_layer}
    d\gamma = -[S]dT - \displaystyle\sum_{\theta=1}^{\mathcal{N}} [N_
    \theta] d \bar{\mu}_\theta , &&
\end{flalign}
where $[S] = {S_l}/{A_i} = \int_{-l}^{l} s(x) dx$, $[N_\theta] = N_\theta/A_i = \int_{-l}^{l} \rho_\theta(x) dx $ are the layer contents of the respective physical quantities, per IB area $A_i$, and the terms in the integrand are the corresponding densities. For the bulk phases ($\psi = m,p$), the following Gibbs-Duhem equations must also hold
\begin{flalign} \label{aeq:GD_bulk}
    0 &= - S^\psi dT - \sum_{\theta=1}^{\mathcal{N}} N_\theta^\psi d \bar{\mu}_\theta .&&
\end{flalign}
% $[V] = {V_l}/{A_i} = \int_{-l}^{l} dx$. + V^\psi dP
% \begin{flalign} \label{aeq:GD_matrix}
%     0 &= - S^m dT + V^m dP - \sum_{\theta=1}^{3} N_\theta^m d \bar{\mu}_\theta \\ \label{aeq:GD_ppt}
%     0 &= - S^p dT + V^p dP -\sum_{\theta=1}^{3} N_\theta^p d \bar{\mu}_\theta %&&
% \end{flalign}
The two Gibbs-Duhem relations can be used to eliminate the variation in two intensive thermodynamic quantities among ${d\gamma,dT,d\bar{\mu}_1,\ldots,d\bar{\mu}_\mathcal{N}}$ from Eq.~\ref{aeq:GD_layer}. This can be done by applying Cramer's rule to Eqs.~\ref{aeq:GD_layer}--\ref{aeq:GD_bulk}. Following \citep{cahn1979interfacial}, and choosing the chemical potentials of $1$ and $2$ as the dependent variations to be eliminated, the generalized Gibbs adsorption equation is obtained as
% \begin{flalign}
%     d\gamma = -[S/XY] dT + [V/XY] dP - \sum_{\theta=1}^{3} [N_\theta/XY] d\bar{\mu}_\theta %&&
% \end{flalign}
% Here, $X$ and $Y$ are belong to $\{S,V,N_\theta\}$. Choosing $\{X,Y\}=\{N_1,N_2\}$, the variation $d\gamma$ with respect to the independent variations $dT$ and $d\mu^{\prime}_3$ is obtained as
\begin{flalign} \label{aeq:gen_GA} \nonumber
    d\gamma &= -[S/N_1 N_2]dT - \sum_{\theta=1}^{\mathcal{N}}[N_\theta/N_1 N_2]d\bar{\mu}_\theta \\
    &= -\Gamma^{(1,2)}_S dT - \sum_{\theta=1}^{\mathcal{N}} \Gamma^{(1,2)}_\theta d\mu_\theta , &&
\end{flalign}
where $\mu^e_\theta = \bar{\mu}_\theta/\rho_m$ is the equilibrium chemical potential of component $3$ in the bulk, and can be evaluated from the bulk phase free energy density (and its derivatives with concentrations $\mu^e_{\theta 1}$) as \citep{lupis1983chemical}
\begin{flalign} \label{aeq:chem_pot_3}
    \mu^e_\theta = f^m - \sum_{\xi=2:\mathcal{N}} \left(\delta_{\theta \xi} - c^e_{\xi m}\mu^e_{\xi 1} \right) , &&
\end{flalign}
where $\delta_{\theta \xi}$ is the Kronecker delta. The conjugate coefficients to the intensive quantities are defined as
% \begin{flalign} \label{aeq:determinant}
%     [Z/N_1 N_2] = \frac{\begin{vmatrix}
%         [Z] & [X] & [Y] \\
%         S^m & N_1^m & N_2^m \\
%         S^p & N_1^p & N_2^p
%     \end{vmatrix}}{
%     \begin{vmatrix}
%         N_1^m & N_2^m \\
%         N_1^p & N_2^p
%     \end{vmatrix}} %&&
% \end{flalign} % \frac{|[Z] X^m Y^p|}{|X^m Y^p|} =
\begin{flalign} \label{aeq:determinant}
    [Z/N_1 N_2] = {\begin{vmatrix}
        [Z] & [N_1] & [N_2] \\
        Z^m & N_1^m & N_2^m \\
        Z^p & N_1^p & N_2^p
    \end{vmatrix}} \div {
    \begin{vmatrix}
        N_1^m & N_2^m \\
        N_1^p & N_2^p
    \end{vmatrix}}, &&
\end{flalign} % \frac{|[Z] X^m Y^p|}{|X^m Y^p|} =
where $Z = \{S, N_3, \ldots , N_{\mathcal{N}}\}$. Unlike Eq.~\ref{aeq:GD_layer}, the generalized coefficients in Eq.~\ref{aeq:gen_GA} are invariant to the placement of the layer bounds. They are also independent of the Gibbs dividing surface convention.
% At constant $T$ and $P$, we have
% \begin{flalign}
%     \rho_o \left(\frac{\partial \gamma}{\partial {\mu}_3} \right)_{T,P} = -[N_3/N_1 N_2] %&&
% \end{flalign}
Assuming the molar densities ($\rho_m = 1/v_m$) of the components to be equal and constant throughout the system, $\rho_\theta(x) = \rho_m c_\theta(x)$, the determinants in Eq.~\ref{aeq:determinant} can be evaluated as 
% \begin{flalign}
%     [S/N_1 N_2] = \int_{-l}^{l} \left[(s(x) - s^m) - \frac{(s^p-s^m)}{(c_A^p-c_A^m)} (c_A(x)-c_A^m) \right] dx %&&
% \end{flalign}
\begin{flalign} \nonumber
    \Gamma^{(1,2)}_\theta &\equiv [N_\theta/N_1 N_2]v_m \\ \nonumber
    &= \int_{-l}^{l} \left[(c^e_\theta(x) - \frac{(c^e_{1m} c^e_{\theta p} - c^e_{\theta m} c^e_{1p})}{(c^e_{1m} c^e_{2p} - c^e_{2m} c^e_{1p})} c^e_2(x)\right. \\
    &- \left.\frac{(c^e_{\theta m} c^e_{2p} - c^e_{2m} c^e_{\theta p})}{(c^e_{1m} c^e_{2p} - c^e_{2m} c^e_{1p})} c^e_1(x) \right] dx &&
\end{flalign}
and
\begin{flalign} \nonumber
    \Gamma^{(1,2)}_S &\equiv [S/N_1 N_2] \\ \nonumber
    &= \int_{-l}^{l} \Bigg[(s_e(x) - \frac{(c^e_{1m} s^{p}_e - s^{m}_e c^e_{1p})}{(c^e_{1m} c^e_{2p} - c^e_{2m} c^e_{1p})} c^e_2(x) \Bigg. \\
    &- \Bigg.\frac{(s^{m}_e c^e_{2p} - c^e_{2m} s^{p}_e)}{(c^e_{1m} c^e_{2p} - c^e_{2m} c^e_{1p})} c^e_1(x) \Bigg] dx , &&
\end{flalign}
where the constraint $c^e_1(x)=1-\sum_{\theta=2}^{\mathcal{N}}c^e_\theta(x)$ is assumed and $\Gamma^{(1,2)}_{1/2}=0$ is realized. Analytic relations for the excess quantities can be obtained by evaluating the integrals with reference to the Gibbs dividing surface at $x=0$ as done in Eq.~\ref{aeq:solute_excess}. Therefore,
\begin{flalign} \label{aeq:entropy_excess_analytic} \nonumber
  \Gamma^{(1,2)}_S =& S^{xs} - \frac{(c^e_{1m} s^e_{p} - s^e_{m} c^e_{1p})}{(c^e_{1m} c^e_{2p} - c^e_{2m} c^e_{1p})} C^{xs}_2 \\
  &- \frac{(s^e_{m} c^e_{2p} - c^e_{2m} s^e_{p})}{(c^e_{1m} c^e_{2p} - c^e_{2m} c^e_{1p})} C^{xs}_1 , && 
\end{flalign} 
where
\begin{flalign} \label{aeq:Gibbs_ent_exs}
    {S^{xs} = \frac{\varepsilon}{3\sqrt{2W_e}} \left(2s^e_{i} - s^e_{m} - s^e_{p} \right)}. &&
\end{flalign}
Similarly Eq.~\ref{eq:solute_excess_analytic} is obtained for $\Gamma^{(1,2)}_\theta$ in terms of the analytic Gibbs solute excess Eq.~\ref{eq:Gibbs_exs}.

% \section{ACKNOWLEDGMENTS}
% This work was primarily supported by the U.S. National Science Foundation under Grant No. DMR-1554270, with partial support from North Carolina State University.

\section*{ACKNOWLEDGMENTS}
{SBK and SP were supported by the U.S. National Science Foundation under Grants DMR-1554270 and CMMI-1826173. The authors are grateful to Prof. Fadi Abdeljawad (Clemson University) for helpful discussions on phase-field modeling and multicomponent grain boundary segregation.}

\newpage

% \section*{References}
% \putbib
% \bibliographystyle{unsrt}
\bibliography{refs}

% \bibliographystyle{apsrev4-1} % Tell bibtex which bibliography style to use
% \bibliography{main} % Tell bibtex which .bib file to use (this one is some example file in TexLive's file tree)

% \end{bibunit}

% \clearpage
% \input{SupportingInfo.tex}

\end{document}